\documentclass[11pt,a4paper,openany,twoside,svgnames]{book}
\RequirePackage{etex}
\usepackage[utf8]{inputenc}
\usepackage{lmodern}
\usepackage{alphabeta}
\usepackage{setspace}
\usepackage{tikz}
\usepackage{verbatim}
\usepackage{etoc}
\usepackage{blindtext}
\usepackage[T1]{fontenc}
\usepackage[english,french]{babel}
\usepackage{tcolorbox}
\usepackage{chngpage}
\usepackage[absolute]{textpos}
\usepackage{mdframed}
\usepackage{ragged2e}
\usetikzlibrary{fit}
\newlength\tocrulewidth
\usepackage{lscape}
\usepackage{multirow}
\usepackage{caption}
\usepackage[ruled,linesnumbered,resetcount,algochapter]{algorithm2e}

\SetCommentSty{mycommfont}
\usepackage{graphicx,subcaption}
\usepackage{capt-of}
\usepackage{lineno,hyperref}
\usepackage{url}
\usepackage{rotating}
\usepackage{appendix}
\usepackage{amssymb}
\usepackage{breakcites}
\usepackage{enumerate}   
\usepackage{letltxmacro}

\LetLtxMacro\oldenum\enumerate
\LetLtxMacro\oldendenum\endenumerate
\usepackage{pifont}
\usepackage{xcolor}
\newcommand{\cmark}{\textcolor{green!80!black}{\ding{51}}}
\newcommand{\xmark}{\textcolor{red}{\ding{55}}}
\usepackage{float}

\usepackage{fancyhdr, blindtext}
\newcommand{\changefont}{%
    \fontsize{8}{11}\selectfont
}
\newcommand{\FontTitle}{%
    \fontsize{19}{40}\selectfont
}
\usepackage{silence}
\usepackage[explicit]{titlesec}
\usepackage{tikz-network}

\usepackage{abstract}
\colorlet{grisclaire}{gray!30}
\colorlet{grisfonce}{darkgray}

\newcommand*\chapterlabel{}
\titleformat{\chapter}
  {\gdef\chapterlabel{}
   \normalfont\sffamily\Huge\bfseries\scshape}
  {\gdef\chapterlabel{\thechapter\ }}{0pt}
  {\begin{tikzpicture}[remember picture,overlay]
    \node[yshift=-5cm] (0) at (current page.north west)
      {\begin{tikzpicture}[remember picture, overlay]
        \draw[fill=grisclaire] (0,0) rectangle
          (\paperwidth,5cm);
        \node[anchor=east,xshift=.9\paperwidth,rectangle,
              rounded corners=16pt,inner sep=11pt,
              fill=grisfonce, font=\large] (17) 
              {\color{white}#1};
       \end{tikzpicture}
      };
   \end{tikzpicture}
  }
\titlespacing*{\chapter}{0pt}{50pt}{-60pt}
\usepackage{lineno,hyperref}

\begin{document}
\begin{titlepage}
\centering
      {\large R\'epublique Alg\'erienne D\'emocratique et Populaire}\\
      {\large Minist\`ere de l'Enseignement Sup\'erieur et de la Recherche Scientifique}\\
      \vspace{0.5cm}
      \includegraphics[width=0.4\textwidth]{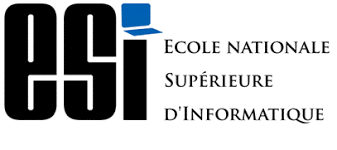}\\
      {\Large {\bfseries \'Ecole nationale Sup\'erieure d'Informatique}}\\
       \vspace{1cm}
      {\Large {\bfseries TH\`{E}SE}}\\
       \textit{Présentée pour obtenir le grade de }\\
      {\Large {\bfseries DOCTORAT EN SCIENCES}}\\

    	\textit{ Par}\\
    	{\Large {\bfseries Sou\^{a}ad \scshape{Boudebza}}}
    \noindent
{\color{black} \rule{\linewidth}{1mm} }\\
    {\huge\bfseries \color{black} \bfseries{Approche pour la D\'{e}tection de Communaut\'{e}s dynamiques dans les R\'{e}seaux Sociaux}} \\
       \noindent
{\color{black} \rule{\linewidth}{1mm} }\\

        Directeur de th\`ese : {\bfseries Omar \scshape{Nouali}}\\
        Co-Directeur de thèse : {\bfseries Fai\c{c}al \scshape{Azouaou}}\\
     \vspace{2cm}
    {\large \textbf{Soutenue le : 14/04/2022, devant le jury composé de:}}\\
	\vspace{0.5cm}
	\begin{tabular}{lll}
		M. Amar Balla, & Professeur, ESI, & Pr\'esident\\
        Mme Fatima Si Tayeb, & Professeur, ESI, & Examinatrice\\
        M. Hachem Slimani, & Professeur, Universit\'e de Béjaïa, & Examinateur\\
        Mme Malika Bessedik, & MCA, ESI, & Examinatrice\\
        M. Omar Nouali, & Directeur de Recherche, CERIST, & Directeur de thèse\\
	\end{tabular}
\newpage
\thispagestyle{empty}
\centering
      {\Large People's and Democratic Republic of Algeria}\\
      {\Large Ministry of Higher Education and Scientific Research}\\
      \vspace{0.5cm}
      \includegraphics[width=0.4\textwidth]{figures/Unknown.png}\\
   		{\bfseries {\large Higher National School of Computer Science}
   			}\\
   			 \vspace{1cm}
      {\Large {\bfseries THESIS}}\\
       \textit{To obtain the degree of }\\
      {\Large {\bfseries  DOCTOR OF SCIENCE}}\\

    	\textit{ By}\\
    	{\Large {\bfseries Sou\^{a}ad \scshape{Boudebza}}}
    \noindent
{\color{black} \rule{\linewidth}{1mm} }\\
       {\FontTitle\bfseries \color{black} \bfseries{An Approach for Detecting Dynamic Communities in Social Networks}} \\
       \noindent
{\color{black} \rule{\linewidth}{1mm} }\\
        Under the supervision of: {\bfseries Omar \scshape{Nouali}}\\
        and  {\bfseries Fai\c{c}al \scshape{Azouaou}}\\
    \vspace{2cm}
    {\large \textbf{Defended on: 14/04/2022, in front of a jury composed by:}}\\
	\vspace{0.5cm}
	\begin{tabular}{lll}
		{\large \textbf{Board of Examiners}}\\
	    Mr Amar Balla, & Professor, ESI, & Chairman\\
        Mrs Fatima Si Tayeb, & Professor,ESI, & Examiner\\
        Mr Hachem Slimani, & Professor, University of Bejaia, & Examiner\\
        Mrs Malika Bessedik, & MCA, ESI, & Examiner\\
        Mr Omar Nouali, & Director of Research, CERIST, & Advisor\\
	\end{tabular}
\end{titlepage}
\pagenumbering{roman}
\chapter*{Dedication}
\vspace{4cm}
\begin{flushright}
In loving memory of my Grandmother...
\end{flushright}

\chapter*{Acknowledgement}
\vspace{2cm}

I would like to express my deepest appreciation to everybody who inspired me, helped me, and contributed directly or indirectly to accomplishing this thesis. 

First, i would like to thank my supervisors: Dr. Omar Nouali and Pr. Faiçal Azouaou for their guidance, advice, and support.

I am thankful to the members of my thesis committee for accepting to read, evaluate and comment on this thesis.

I am deeply indebted to Dr. Rémy Cazabet at the University of Lyon, without him this thesis would not have been possible. Thank you for welcoming me several times to your laboratory.
Thank you for your invaluable contribution and your unwavering support. You stand amongst the kindest and most helpful professors I have known. 


Finally, my warmest and most sincere thanks go to my family, especially my beloved husband Antar who has been a great partner and supporter of my academic research. If not for them, I would not have completed this thesis.
\chapter*{Abstract}
\thispagestyle{empty}
\vspace{2cm}
 
Recent developments in the internet and technology have made major advancements in tools that facilitate the collection of social data, opening up thus new opportunities for analyzing social networks. Social network analysis studies the patterns of social relations and aims at discovering the hidden features embedded in the structure of social networks. One of the most important features in social networks is community structure: densely knit groups of individuals. 
The dynamic nature of interaction in social networks often challenges the detection of such community structures. 
The contributions in this thesis fall into two categories.
The first category highlights the problem of identifying overlapping communities over time. To carry out such analysis, a framework called OLCPM (Online Label propagation and Clique Percolation Method)  is proposed. It is an online algorithm based on clique percolation and label propagation methods. OLCPM has two main features: the first one is its ability to discover overlapping communities, while the second is its effectiveness in handling fine-grained temporal networks.
As for as the second category is concerned, it 
emphasizes on the problem of analyzing communities that are embedded at different temporal scales. For example, in networks of interaction such as e-mails or phone calls, individuals are involved in daily as well as occasional conversations. We propose a first method for analyzing communities at multiple temporal scales. 
Hence, the dynamic network (link streams) is studied at different temporal granularities, and coherent communities (called stable communities) over a period of time are detected at each temporal granularity.  
The two proposed approaches are validated on both synthetic and real-world datasets.
\chapter*{Résumé}
\selectlanguage{french} 
\vspace{2cm}

Le d\'eveloppement r\'ecent d'internet et technologie a fait un grand progr\`es en mati\`ere des outils qui facilitent la collection des donn\'ees sur les r\'eseaux sociaux, ouvrant ainsi de nouvelles opportunit\'es pour l'analyse de ces derniers. 
Cette analyse s'interesse \`a l'\'etude des relations sociales.
Elle permet de d\'evoiler les propi\'et\'es caract\'erisant les structures sociales.
Une des propi\'et\'es importantes des r\'eseaux sociaux est la pr\'esence des groupes denses applel\'es communaut\'es. La nature dynamique des interactions au sein des r\'eseaux sociaux représente un grand challenge lors de la d\'etection des communaut\'es.
Les contributions dans cette th\`ese s'articulent autour de deux axes. Le premier axe aborde le probl\`eme de d\'et\'ection de communaut\'es dynamiques et recouvrantes. Nous proposons un framework appel\'e OLCPM, bas\'e sur les m\'ethode de percollation de cliques et de propagation de lables. OLCPM permet de d\'ecouvrir les communaut\'es recouvrantes et il est capable de traiter des r\'eseaux dynamiques \`a granularit\'e tr\`es fine.
Le deuxi\`eme axe aborde le probl\`eme de d\'et\'ection de commuanut\'es \`a plusieurs \'echelles temporelles. Nous proposons une premi\`ere m\'ethode pour l'analyse des communaut\'es \`a multiples \'echelles temporelles.
Le r\'eseau dynamique est \'etudi\'e \`a diff\'erentes \'echelles temporelles. Les communaut\'es stables sur une p\'eriode de temps sont d\'et\'ect\'ees pour chaque granularit\'e. Les deux contributions sont valid\'ees et test\'ees sur des r\'eseaux synth\'etiques et r\'eels.

\selectlanguage{english} 
\tableofcontents
\clearpage\phantomsection\addcontentsline{toc}{chapter}{List of Figures}
\listoffigures
\clearpage\phantomsection\addcontentsline{toc}{chapter}{List of Tables}
\listoftables
\newpage
\pagenumbering{arabic}
\clearpage
\titleformat{\chapter}[display]
  {\gdef\chapterlabel{}
   \normalfont\sffamily\Huge\bfseries\scshape}
  {\gdef\chapterlabel{\thechapter\ }}{0pt}
  {\begin{tikzpicture}[remember picture,overlay]
    \node[yshift=-5cm] (0) at (current page.north west)
      {\begin{tikzpicture}[remember picture, overlay]
        \draw[fill=grisclaire] (0,0) rectangle
          (\paperwidth,5cm);
          \tikzstyle{GN}=[circle,fill=grisfonce, draw,text=white,font=\bfseries,font=\Huge,
          minimum size=3cm]
           \tikzstyle{PN}=[circle,fill=white, draw,text=black,scale=0.4]
            \path 
 node[GN] (0) at (3.3,3) {\thechapter}
 node[PN] (1) at (5.3,3.3) {}
 node[PN] (2) at (6,2) {} 
 node[PN] (3) at (7.1,2.75) {}
 node[PN] (4) at (6.2,4.25) {}
 node[PN] (5) at (7.75,4) {} 
 node[PN] (6) at (8.2,2.5) {} 
 node[PN] (7) at (9,3.3) {} 
 node[PN] (8) at (10,4) {} 
 node[PN] (9) at (10.8,2.4) {}
 node[PN] (10) at (12.9,3.25) {}
 node[PN] (11) at (12.3,3.8) {}
 node[PN] (12) at (13.9,3) {}
 node[PN] (13) at (16,2.5) {}
 node[PN] (14) at (17,3) {}
 node[PN] (15) at (18.3,2.5) {}
 node[PN] (16) at (17.1,2.25) {};
        \node[anchor=east,xshift=.9\paperwidth,rectangle,
              rounded corners=16pt,inner sep=11pt,
              fill=grisfonce,font=\large] (17) 
              {\color{white}#1};
              \draw[line width=0mm,   black] (0)--(1)--(2)--(3)--(4)--(1)--(3)--(6)--(7)--(5)--(4);
 \draw[line width=0mm,   black] (7)--(8)--(9)--(7)--(11)--(10)--(9)--(11)--(8);
  \draw[line width=0mm,   black] (10)--(12)--(13)--(16)--(15)--(14)--(12)--(15);
  \draw[line width=0mm,   black] (13)--(17);
       \end{tikzpicture}
      };
   \end{tikzpicture}
  }
\chapter{Introduction}\label{chp1:introduction}
\thispagestyle{empty}
\section{Thesis context}
The study of complex networks, referred to as network science, has become a highly active field in the last few decades with a broad variety of applications. The latter ranges from technological systems like the Internet and World Wide Web, biological systems such as the nervous system or protein interactions to transportation infrastructures such as roadways, airlines, power grids, waterways, pipelines, and others. One of the most prominent applications in the field falls within the domain of social network analysis.

The study of social networks has a deep root in sociology, featuring pioneering work on sociometry by Moreno in 1934
\cite{Moreno1934}. Social networks have seen spectacular growth in recent years, mainly due to the advent of the information age and the internet, which have made the collection of enormous amounts of social data possible. The latter paved the way for promising perspectives on the study of social networks. The aim of social network analysis is to analyze relationship patterns among social entities and to understand the general properties and features of the whole network. The graph theory is at the heart of the research conducted in this field. A graph consists of a set of nodes representing social actors within the network (people, organizations, groups, or any other entities) and a set of edges between pairs of nodes representing interactions between those actors (friendship, collaboration, influence, idea, etc.). The graph theory has been successfully drawn upon to identify and characterize hidden patterns, often non-trivial, in social networks. 

Over the past years, researchers have studied different structural properties of social networks. Among the most important and revolutionary findings are the Small World property by \cite{watts1998collective},  the Scale Free by \cite{barabasi1999emergence} and the Community Structure by \cite{girvan2002community}. The Small World property means that the average distance between any pairs of nodes in the network is small, due to the existence of few long-distance connections ("Six degrees of separation" concept by \cite{milgram1967small}).
The Scale Free property means that nodes in the network have heterogeneous link connections (degree): while a few nodes have high degrees, most nodes have a very low degree. Community Structures are believed to be one of the most prominent features of social networks. A community is characterized by the existence of a collection of nodes, where nodes within a collection tend to interact more with each other than
with the rest of the network \cite{radicchi2004defining}. For instance, in social networks, individuals within the same community often
share similar properties such as interests, social ties, location, occupation, etc.

The ability to detect such community structures could be of great importance in a number of research areas, such as recommender systems \cite{boratto2009,deng2014}, email communication \cite{moradi2012}, epidemiology \cite{kitchovitch2011}, criminology \cite{ferrara2014}, marketing and advertising \cite{mckenzie1999, fenn2009}, etc.
In collaboration networks, where nodes represent researchers and edges represent co-authorship links between researchers, community detection can discover groups of researchers working in the same area and may, thus, help to find researchers with expertise in a given area. In online networking sites like Facebook -- an example of a network representing acquaintances for a particular user--, the community detection can find user's social circles, such as: family, work colleagues, or college friends, rendering it useful for recommender systems in Facebook. 
\section{Statement of the Problem}\label{chpt1:problem}

The present thesis falls within the domain of social network analysis. Therefore, special attention is paid to the analysis of community structures in social networks. This field offers interesting yet challenging problems. This thesis addresses the essential issues facing community detection, in particular, in the context of social networks. We aim to investigate the following issues: 


\begin{itemize}
\item \textbf{Overlapping community detection} :
Early work focused on the simplest form of the community detection problem which is the partitioning of the networks into disjoint communities, where each node belongs to a unique community. However, a more realistic form seems to be overlapping community structure.
In real-world networks, notably in social networks, communities are not always disjoint from each other. In fact, nodes in social networks tend to be part of several groups at once. For instance, individuals often belong to familial and professional circles; scientists collaborate with several research groups, etc. Such shared nodes, called overlapping nodes, play a crucial role in the network. It servers as a bridge between different groups. 
The application of disjoint community detection methods on such networks may lead to misleading characterization of their overlapping community structure. This problem prompts the urgent need to consider the overlap feature for discovering community structures in social networks.
\item \textbf{Dynamic community detection:}
Social networks are dynamic by nature; their social entities and interactions evolve constantly. This evolution is characterized either by adding or removing nodes or edges from the network. For instance, in online social networks like Facebook, changes are introduced by users joining or withdrawing from the network, or by people adding each other as "friend". As the network evolves over time, the community structure may undergo various changes, also known as critical events. \cite{palla2007quantifying} proposed six types of events that may occur during the evolution of communities: birth, growth, shrink, merge, split, and death. The communities can grow or shrink as members are added or removed from an existing community. As time goes by, new communities can be born and old communities may disappear. Two communities can become closely related and merged into a single one, or conversely, a single community can split into two or more distinct ones. 
Even though many methods have been proposed to deal with the problem of community discovery in dynamic networks, this problem remains a serious challenge.

\item \textbf{Detecting community structures at multiple temporal scales characterizing the network evolution:} Several algorithms have been proposed in recent years to discover evolving  community structures, but no method has yet been proposed to deal with the multi-scale temporal evolution property of social networks.  In fact, fluctuations in social networks can be observed at yearly, monthly, daily, hourly, or even smaller scales. For instance, if one were to look at interactions among workers in a company or laboratory, one could expect to discover clusters of people corresponding to meetings and/or coffee breaks, interacting at a high frequency (e.g., every few seconds) for short periods (e.g., few minutes), project members interacting at a medium frequency (e.g., once a day) for medium periods (e.g., a few months), coordination groups interacting at low frequency (e.g., once a month) for longer periods (e.g., a few years), etc. Communities may exist, therefore,  at different temporal scales: short, medium, and large periods.  The question that may arise here is how to detect such community structures at the different scales characterizing the network evolution?
\end{itemize}

\section{Contributions}\label{chp1:contributions}
The contribution of this thesis is twofold.
First, the first two issues about finding overlapping and evolving community structures are addressed. Hence, OLCPM \cite{boudebza2018}, an online algorithm based on clique percolation and label propagation methods is proposed to carry out the analysis. OLCPM can detect overlapping communities and works on temporal networks with a fine granularity. By locally updating the community structure, OLCPM delivers significant improvement in running time compared with previous clique percolation techniques. The experimental results on both synthetic and real-world networks illustrate the effectiveness of the method. 
Second, as a response to the third challenge about the  multi-scale temporal evolution aspect, we propose an algorithm to detect stable community structures by identifying change points within meaningful communities \cite{boudebza2019detecting}. Unlike existing dynamic community detection algorithms, the proposed method is able to discover stable communities efficiently at multiple temporal scales. The effectiveness of the method is tested on synthetic networks as well as on high-resolution time-varying networks of contacts drawn from real social networks.
\newpage
\section{Structure of the thesis}\label{chp1:structure}
This thesis is organized as follows:

Chapter \ref{chp2:BackgroundSN} introduces the essential background information on social network analysis (SNA). It provides the key concepts, notations and measurements used in SNA. The purpose of this chapter is to ease the reading of the rest of the manuscript.  

Chapter \ref{chp3:BackgroundCD} presents a literature review about community detection. The concept of community and the main approaches for static community detection are first presented. The problem of dynamic community detection is addressed thereafter, reviewing in the process the existing methods for dynamic community detection.

Chapter \ref{chp4:OLCPM} presents our first contribution which pertains to the detection of dynamic overlapping community. First, the rational basis of this proposal is expressed. Then, the fully dynamic network formalism we proposed to model evolving graph is introduced. After that, the proposed framework OLCPM for detecting overlapping dynamic communities is thoroughly explained. The last part in this chapter reports the experimental process conducted to assess the effectiveness of the proposed framework.

Chapter \ref{chp:Stable} is devoted to the description of our second contribution on the temporal multi-scale detection of stable communities in link streams. Initially, we provide a view to both Link stream analysis and change-point methods which are the roots of the proposed method. Moreover, the proposed method is further described. The experimental results on both synthetic and real-world networks are then discussed. The main outcomes of this study are discussed to the end of this chapter.

Finally, Chapter \ref{chp:conclusionfuturework} summarizes the key findings of this thesis and sheds light on similar potential research paths.
\chapter{Social Network Analysis}\label{chp2:BackgroundSN}

\thispagestyle{empty}
\vspace{1cm}

\parindent=0em
\etocsettocstyle{\rule{\linewidth}{\tocrulewidth}\vskip0.5\baselineskip}{\rule{\linewidth}{\tocrulewidth}}
\localtableofcontents

\clearpage
\section{Introduction}\label{chp2:Introduction}
This chapter presents background information on social network analysis which is intended to facilitate the reading of the manuscript. 
At first, we outline the key concepts, definitions and main representations behind the social network perspective. Then, we provide an overview of social network analysis measurements. At the end of this chapter, we present the common features of social networks. 
The expert reader can skip this chapter and go directly to the next one, which is a literature review on dynamic community detection.

\section{Related concepts}
This section introduces the basic concepts used around social network analysis and aims to enable the reader to make link between these different concepts.

\textbf{Network or graph}. 
A mathematical structure to model pairwise relations between objects. It is composed of a set of nodes (or vertices) representing objects and a set of edges (or links) representing relations between pairs of nodes. 
In network science, the terminology: network, node and link refers to real systems. In mathematics (graph theory), the terms: graph, vertex, and edge refer to the mathematical representation of these networks \cite{barabasi2013network}. In this manuscript, the two terminologies are used interchangeably.

\textbf{Complex networks.}
Networks (or graphs) to model complex systems in the real world. A complex system is a system made up of a large number of components interacting with each other in a nontrivial way \cite{simon1991architecture}, reflecting a complex pattern that is neither completely regular nor completely random \cite{watts1998collective}. Examples of complex systems include biological systems (such as nervous system or protein interactions), technological systems (like the Internet and World Wide Web),  
social systems (such as acquaintance or collaboration patterns between people), etc.

\textbf{Social networks. }
A subset of complex networks 
in which the vertices are social actors (e.g., people, groups of people, organizations, nations, etc.), and the edges represent some form of social interaction between them, such as familial, friendship and collaboration ties between people, trade relations between countries, etc.  \cite{Newman:2010:NI:1809753}.

\textbf{Network science.} 
An interdisciplinary field that has emerged in the 21st century, focusing on the study (understanding and modeling) of patterns of connections within complex systems in many areas such as biology, computer science, and social science \cite{barabasi2013network}.

\textbf{Social network analysis.}
A sub-part of Network Science focusing on the relationships and the interconnected behavior of social actors.

\section{What are social networks?}\label{sec:WSN}

Social network refers to the articulation of a social relationship, ascribed or achieved, among social entities \cite{laumann2013networks}. Two main components can clearly be distinguished:  

\begin{itemize}
\item The social entities, termed \textit{actors} are most commonly persons, but in principle, it could be any entity that can be connected to another entity, such as organizations, countries, web pages, scientific papers, and so on. 
\item Social relationships or \textit{ties} could include, for example, friendships, collaborations, trade, Web links, citations, resource flows, information flows, exchanges of social support to name a few \cite{wasserman1994social}.
\end{itemize}

\section{Types of social network}\label{sec:typeSN}

Social networks can be categorized either by the nature of interacting entities (known as network mode) or according to the properties of ties among these entities.  

\subsection{Network mode}\label{sec:Nmode}

The term "mode" refers to a distinct set of social actors or nodes (e.g. individuals or places) on which ties of a specific kind are measured between pairs of actors \cite{wasserman1994social}. Ties can be measured on one, two, or even more sets of actors. The number of distinct sets of actors a network contains refers to the number of modes. Ties measured on a single type of actor result in one-mode networks, as an example student's friendship networks. Two-mode networks contain ties between two distinct sets of actors, a typical example is the affiliation networks (nodes are individuals and events (e.g. clubs or organizations), and each individual is linked to the event he attends to). 

Two-mode networks can easily be converted to one-mode networks, but it often involves a loss of information. 
One simple transformation of the previous two-mode network example can be done as follows: we keep only one set of actors, individuals for example. Then, we link every two individuals if they attend at least one common event. Links can be weighted to indicate the number of common events between individuals.

One or two-mode networks are the most extensively studied while three (or highly)-mode networks are rarely studied because of their complex structure.

\subsection{Type of ties}

The ties between actors can be of different types. \cite{borgatti2009network} identify four broad categories of ties: similarities, social relations, interactions, and flows.
\begin{itemize}
\item Similarities occur when actors share common characteristics and properties, such as spatial and temporal proximity, co‐membership in groups and events, or demographic characteristics. 
\item Social relations include kinship, role relations (e.g., friend, student); sentimental ties (e.g., like, hate); or cognitive awareness (e.g., knowing). 
\item Interactions refer to behavior-based ties such as speaking with, helping.
\item Flows are those tangible and intangible things that are transmitted through interactions such as resources, information, or influence.
\end{itemize}
\section{Representations for social network}\label{sec:notations}
There are many ways to describe social networks. In this section, we will introduce two representations: the sociometric representation, which is considered as the ancestor of all representations, and the graph--theoretic representation which is nowadays one of the most useful ways for representing networks. 

\subsection{Sociometric representation}\label{sec:sociometric}

Sociometry was first proposed by \cite{Moreno1934} in his famous book "Shall we survive?". This approach has generated a great deal of interest among psychologists and sociologists and has opened the door to a completely new way for studying interpersonal relationships structure among groups. Sociometry is concerned with the study of positive and negative emotional relations, such as liking/disliking and friendship/enemies among a set of people. The sociometric is referred to as the network data set containing people and their emotional relations. It has two main representations: sociogram and sociomatrix.

\begin{figure}[ht]
\centering
\includegraphics[scale=1]{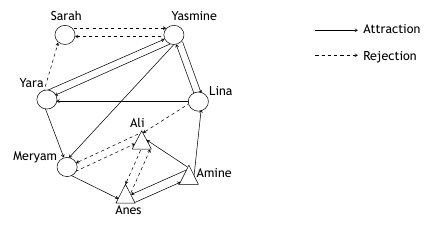}
\caption{Example of sociogram representing attraction/rejection relations between a group of students (three boys represented by triangles and five girls represented by circles).}
\label{fig:sociogram}
\end{figure}

The sociogram is a visual display for depicting interpersonal relationships structure of groups, where social entities are represented as points and relationships among pairs of entities are represented as lines linking the corresponding points. Figure \ref{fig:sociogram} depicts an example of sociogram. 

Sociomatrices are two-way matrices in which the rows and the columns are respectively indexed by the sending actors and the receiving actors. The sociomatrix is associated to one kind of relationship, and the entries of this matrix decode the values of ties between pairs of actors.   
In one-mode networks, the rows and columns of the sociomatrix represent the same set of actors. Two-mode networks, like ‘affiliation’ networks, are represented by a rectangular sociomatrix in which the rows represent one type of actors and the columns represent the other type. Table \ref{tab:sociomatrices} gives the two sociomatrices for the two relations in the previous example of sociogram (Figure \ref{fig:sociogram}).

\begin{table}[!ht]
\resizebox{\textwidth}{!}{%
\small
\begin{tabular}{lllllllll}
\multicolumn{9}{c}{\textbf{Attraction relationship}}                  \\
         & Sarah & Yassmine & Yara & Ali & Lina & Amine & Anes & Maya \\
         \hline
Sarah    & 0     & 0        & 0    & 0   & 0    & 0     & 0    & 0    \\
Yassmine & 0     & 0        & 1    & 0   & 1    & 0     & 0    & 1    \\
Yara     & 0     & 1        & 0    & 0   & 0    & 0     & 0    & 1    \\
Ali      & 0     & 0        & 0    & 0   & 0    & 0     & 0    & 0    \\
Lina     & 0     & 1        & 1    & 0   & 0    & 0     & 0    & 0    \\
Amine    & 0     & 0        & 0    & 1   & 1    & 0     & 1    & 0    \\
Anes     & 0     & 0        & 0    & 0   & 0    & 1     & 0    & 0    \\
Meryam     & 0     & 0        & 0    & 0   & 0    & 0     & 1    & 0   
\end{tabular}}
\resizebox{\textwidth}{!}{%
\begin{tabular}{lllllllll}
\multicolumn{9}{c}{\textbf{Rejection relationship}}                  \\
         \hline
         & Sarah & Yassmine & Yara & Ali & Lina & Amine & Anes & Maya \\
Sarah    & 0     & 1        & 0    & 0   & 0    & 0     & 0    & 0    \\
Yassmine & 1     & 0        & 0    & 0   & 0    & 0     & 0    & 0    \\
Yara     & 1     & 0        & 0    & 0   & 0    & 0     & 0    & 0    \\
Ali      & 0     & 0        & 0    & 0   & 0    & 0     & 1    & 1    \\
Lina     & 0     & 0        & 0    & 1   & 0    & 0     & 0    & 0    \\
Amine    & 0     & 0        & 0    & 0   & 0    & 0     & 0    & 0    \\
Anes     & 0     & 0        & 0    & 1   & 0    & 0     & 0    & 0    \\
Maya     & 0     & 0        & 0    & 1   & 0    & 0     & 0    & 0   
\end{tabular}}
\caption{Sociomatrices for the two relations of the sociogram in Figure \ref{fig:sociogram}.}
\label{tab:sociomatrices}
\end{table}

The sociomatrix encodes the same information as a sociogram, but with a great advantage to deal with large networks for which it is often difficult to draw readable sociograms.  
 
\subsection{Graph--theoretic representation}\label{sec:graph}

Since the late 1940s, the graph--theoretic representation is used as a key formalism to study social networks. This representation provides a straightforward yet powerful way to represent actors and their relations. A graph consists of a set of points (also called vertices or nodes) representing actors and a set of lines (or edges) connecting pairs of actors (points). 

\textbf{Definition of graph.}

Suppose we have a set of actors. In the graph--theoretic representation, these actors are represented by vertices or nodes.  We denote \(V=\{v_1,v_2,v_3,...,v_n \}\) the set of nodes containing $n$ actors. We denote $E$ the set of all existing ties between each pair of actors in $V$. If a tie exists between two actors $v_i$ and $v_j$, we denote it by the tuple $(v_i,v_j)$. 

Mathematically, a graph can be described by the two sets $V$ and $E$, we denote the graph as \(G=(V,E)\). This notation represents one set of actors of the same kind and one kind of relationship between the set of actors. 

The graph can be also represented through an adjacency matrix. This representation has the advantage of being more useful for computation.


\textbf{Graph models.}

The concept of the graph can be extended to take into account different properties of nodes and edges 
(direction, intensity, multiplicity, etc.), and thus several graph based-models are proposed:

\begin{itemize}
\item \textit{Directed graphs :} if the relation between pairs of actors is directional, the tie $(v_i,v_j)$ is distinct from the tie $(v_j,v_i)$. In such a case, the graph is referred to as a directed graph and the ties are refereed as arcs (or directed lines). 
Note that at most there can be $n(n-1)$ ties, and at least $0$ ties.
Directed graphs are adopted for representing social networks with symmetric relationships such as follow influence relationships or interactions in phone call networks.
\item \textit{Undirected graphs:} in non-directed graphs, the order of actors in a pair of ties 
no longer matters. When one actor relates to the second, the second relates to the first, hence, we cannot distinguish between $(n_i,n_j)$ and $(n_j,n_i)$. In this case, at most there can be $n(n-1)/2$ ties. Non-directed graphs are well suited for representing asymmetric relationships, like the neighborly relationship, co-authorship,  kinship, and marriage links.
\item  \textit{Weighted graphs:} graphs can be extended to take into account the intensity or strength of a tie by assigning to it values called weights. Weighted graphs are often used to model communication networks. In such networks, weights on edges denote the occurrence of interactions (e.g. number of messages, or comments) between people. 
\item  \textit{Labeled graphs:} these graphs are well suited to model social networks with different types of relationships. Between each pair of nodes, there can be multiple labels each of which represents a type of relationship. In a social network like Facebook, for example,  the labels: friend,  family,  favorite,  etc. can be used to type relationships. 
\item  \textit{Attributed graphs:} graph models have also been extended for representing the attributes of the actors (nodes) by associating discrete or continuous valued attributes on nodes.
\item \textit{Bipartite graphs:} are commonly used to model two-mode networks like affiliation networks (e.g.,  attendance at the same events) using two types of nodes.  
\end{itemize}
\section{Network analysis}
A whole body of research has been devoted to characterizing and analyzing social networks properties since the graph theory was introduced. Therefore, a variety of measurement methods are developed. These methods are often tailored according to their levels of analysis and are mainly classified into three levels: element-level, group-level, and network-level \cite{Brandes:2005:NAM:1062400}. In this section, we discuss some graph based-measurements for each level of analysis. In order to formulate these measurements, we assume that we have a network represented by the graph $G= (V, E)$ containing $|V|$ nodes and $|E|$ edges. 

The following section presents preliminary concepts that are on the basis of most of the network measurements we are going to introduce. We assume that the graph is simple, i.e., unweighted, non-labeled, one-mode, and non-attributed.   

\subsection{Preliminaries}

\textbf{Degree.} We define the degree of a node $i \in V$, denoted by $deg(i)$, 
the number of edges that are incident with it, or equivalently as the number of its adjacent nodes. The degree of a node goes from $0$ if the node has no incident edges (the node is called isolate) to $|V|-1$ if the node has edges with all nodes in the graph. 

In a directed graph, a node $i$  has two variants of degrees, the \textit{out-degree} denoted $deg^{+}(u)$ which describes the number of edges that have their origin in $i$ and the \textit{in-degree }denoted $deg^{-}(u)$, which represents the number of edges that have their destination in $i$.

\textbf{Paths and distances.} A \textit{path} among two nodes $u$, $v$ $\in V$ is defined as the sequence of edges that are crossed during a visit starting from $u$ and ending in $v$.  The \textit{length} of a path is the number of edges it contains. We denote $|P|$ the length of the path $P$. Furthermore, the \textit{geodesic distance} is the shortest-length path connecting a given pair of nodes. It is defined as:  
\[d_G(u, v) = min \{ |P|, P \text{ is a path from $u$ to $v$}\}.\]
The \textit{eccentricity} is the greatest geodesic distance between a given node $v$ and any other in the network. It is defined as:
\[E(v) = max_{i\in V \backslash \{v\}} d_G(v, i).
\]
\subsection{Element--level measurements}\label{sec:ElementMeasurement}

The key question addressed in the element-level analysis is how to assess the relevance of nodes or edges. It is generally dealt with centrality measurements.

\textbf{Centrality.} Centrality indices are used to quantify the most important nodes or edges. Node centrality is often used as an indicator of power, influence, popularity, and prestige of actors. On edges, the centrality is typically used to measure how much communication or flow passes through a link. There is a large number of different centrality measures that have been proposed over the years. The most important are:
\begin{itemize}
\item \textit{Degree centrality: } this centrality, also referred to as neighborliness centrality, ranks nodes according to the number of neighbors (or degree) \cite{freeman1978centrality}: \(C_D(v)=deg(v)\). The higher the degree, the more central the node is. The degree centrality is often used to 
find people with many social connections (popular people, individuals who can quickly spread information, etc.)
\item \textit{Closeness centrality 
(or distance centrality):} it measures the distance of a node to all other nodes in the graph. The node with a small total distance is considered to be the most important. The most accepted closeness distance definition is that formulated by \cite{freeman1978centrality}. The author defined  centrality  as the reciprocal of the total geodesic distance from a given node to all other node: \(C_C(v)=\frac{1}{\sum_{ i \in V} d_G(v,i)} \). In the context of information diffusion, this measure is often used to find individuals who can quickly spread information to all the other people in the network. 
\item \textit{Betweenness centrality:} is another well-known measure proposed by \cite{freeman1978centrality}. It quantifies the number of times a node acts as a bridge along the shortest path between two other nodes in the network, it is defined as : \(C_B(v)=\sum_{i\neq j \neq v \in V}\frac{\sigma_{ij}(v)}{\sigma_{ij}}\), where  $\sigma_{ij}$ is the total number of shortest paths between the nodes $i$ and $j$, and $\sigma_{ij}(v)$ is the total number of shortest paths between the nodes $i$ and $j$ that passe through the node $v$. A node is the more central the more shortest paths run through it. Betweenness centrality highlights actors which are well connected to the rest of the nodes within the network. 
Such actors are called bridges, they serve as a liaison between different graph regions. In the diffusion of information, bridges enable information to spread into unconnected parts of the network. 

\end{itemize}

The centrality measures discussed above are designed for nodes, but most of them can be easily 
adapted to measure edge centrality.

\subsection{Group--level measurements}

The main question here is to identify cohesive groups in the network, i.e., groups having strong linkages among its members. Such groups are typically founded by common goals, interests, preferences or other similarities. Components and cliques are among the most common ways to conceptualize group cohesiveness.   

\textbf{Component.}
Also called connected component which is a maximal connected sub-graph, i.e., a sub-graph in which every node can be reached from every other node and which does not contain any other connected sub-graph. 

\textbf{Clique.} 
Cliques are a typical example of cohesive groups. 
This concept originated as early as 1949 \cite{luce1949method} in sociology. 
A clique is defined as a set of nodes such that there is an edge between every pair of nodes in this set. In other terms, a clique is a complete sub-graph. 

The discovery of cohesive groups is a fundamental issue in social network analysis, it is known as community detection. The latter will be discussed extensively in Chapter \ref{chp3:BackgroundCD}.

\subsection{Network--level measurements}\label{chpt2:networMeasures}

The question addressed at this level is to characterize the global properties of the network. The most important network measures are: diameter, average degree, mean geodesic distance, density, clustering coefficient, and degree distribution.

\textbf{Diameter.} The diameter of a network is the maximal geodesic distance between any pair of nodes in the network, or in other terms, it is the maximum of the eccentricity of the set of nodes in the network. 
\[Diam(G) = max_{u,v\in V } d_G(u, v).\]
In the context of social network analysis, this metric gives an idea about the proximity of pairs of actors in the network indicating how far two nodes are, in the worst of cases. 

\textbf{Average degree.} Also called graph degree is simply the mean of the degrees of all nodes in a network. The average degree can be used to measure the global connectivity of a network \cite{costa2007}.  More precisely:
\[ deg(G)=\frac{\sum_{i\in V} deg(i)}{|V|}=\frac{2|E|}{|V|}.\]

\textbf{Mean geodesic distance. } This metric measures how far apart, on average, any pair of nodes lie in the network. It represents the average shortest path distances for all pairs of nodes in the network, as follows:
\[dist_G=\frac{1}{|V|(|V|-1)}\sum_{u,v\in V} d_G(u,v).\]

\textbf{Density.} The density measures the network connectedness. It is defined as the ratio between the number of edges actually present in the graph and the maximum possible number of edges that can be present in the graph:
\[D=\frac{|E|}{|V|(|V|-1)/2}.\]
Its value ranges from $0$, if no edges are present, to $1$, if all possible edges are present, i.e., the graph is complete.    
The density may give insights into certain phenomena such as information spread through the network. 

\textbf{Transitivity.} 
The transitivity, also known as clustering coefficient, was proposed by \cite{luce1949method}. 
It measures the extent to which two nodes adjacent to any node are adjacent to each other. 
In other words, if there is a tie from $x$ to $y$ and also from $y$ to $z$, then there is probably a tie from $x$ to $z$.  
It is defined as the ratio of the number of closed triples to the total number of triples. 
\[ T=\frac{\text{number of closed triples}}{\text{number of triples}}=\frac{3 \times \text{number of triangles}}{\text{number of triples}},\] where a triple is a connected sub-graph composed of three nodes, i.e.,  nodes are connected either by two edges (opened triple) or by three edges (closed triple). A triangle is a complete sub-graph formed by 3 nodes, it is composed of three closed triples.  

The graph is transitive if every triple it contains is a closed triple, i.e., transitive.
Real-world networks, notably, social networks, exhibit a high clustering coefficient (transitivity).

\textbf{Degree distribution.} The degree distribution $p_k$ is the probability that a randomly selected node has degree $k$.  In other words, it is the fraction of nodes with degree $k$. More formally:

For $k= 1,2,..., $ let we pick a node $v$  uniformly and randomly, $P[deg(v) =k] =p_k$

In a real-world network, most nodes have a relatively small degree, but a few nodes have a very large degree, being connected to many other nodes. 
\section{Social network properties}

Like most complex networks, social networks share some common features. 
Among the most well-known properties are: the small-world property, the scale-free feature, and the presence of community structures.

\subsection{Small world}

\begin{figure}[!ht]
\centering
\includegraphics
[scale=0.7]
{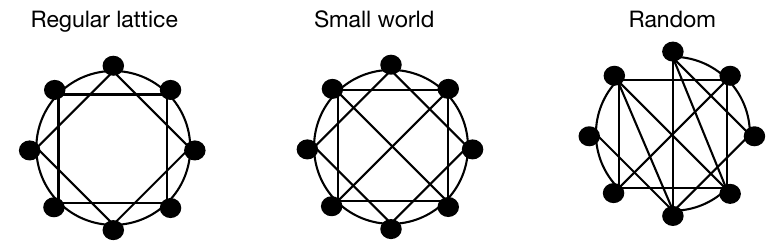}
\caption{Small world feature \cite{watts1998collective}.}
\label{fig:SmallWorld}
\end{figure}

Small-world networks are networks that exhibit simultaneously two properties: the small world effect and the high clustering. 
\begin{itemize}
\item \textit{Small world effect: }the small world effect was outlined since the seminal experiments by \cite{milgram1967small}.     
It means that the mean geodesic distance between any pair of nodes is relatively small (small network diameter) \cite{watts1998collective}. As noticed in Milgram's experiment that actors in social networks were separated by six degrees of social contacts on average. This characteristic is quantified using the average shortest path distance over all nodes in the network.  This distance scales logarithmically with the number of nodes, meaning that between any two nodes, the expected distance is $O(log(n))$ (where $n$ is the network size).
The small-world phenomenon is common in 
, even in sparse networks, i.e., those in which the number of links is much smaller than the maximum number of links the network can have.  
\item \textit{High clustering:} Most large real-world networks, and especially social networks,  exhibit a high clustering coefficient. It has been observed that two nodes having a common neighbor are more likely to be connected to each other. In a friendship network, for example, people tend to be friends with the friends of their friends. This property can be quantified by the clustering coefficient (see Section \ref{chpt2:networMeasures}).  
\end{itemize}
Small-world networks are different from regular lattices and random networks. As noticed by \cite{watts1998collective}, they are halfway between a regular lattice and a totally random network (see Figure \ref{fig:SmallWorld}). Regular lattices are highly clustered but do not exhibit the small-world effect in general, while, random graphs show the small-world effect, but do not show a high clustering.

\subsection{Scale free}

\begin{figure}[!ht]
\centering
\includegraphics[scale=0.5]
{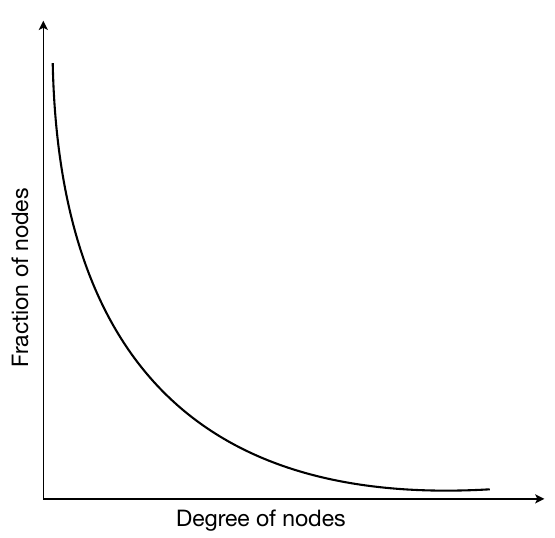}
\caption{Power-law distribution of node linkages \cite{barabasi2003scale}.}
\label{fig:ScaleFree}
\end{figure}

Social networks also exhibit a highly heterogeneous degree distribution (few nodes with higher degree and a majority of nodes with small degree), which follows a  power-law \cite{albert2002statistical} (see Figure \ref{fig:ScaleFree}). The term \textit{'scale free'} means that whatever the scale at which we observe, the network looks the same, i.e., the power law is preserved regardless of the scale.

\subsection{Community structure}

Community structure is one of the most prominent observed features in social networks. A community is a sub-graph that has high connectivity within its members and comparably fewer connections with the rest of the network. Extracting the community structure of a network is called community detection. It is of great importance since it offers insight into the network's structure and functionalities. This feature will be discussed in detail in the following chapter.   

\section{Conclusion}

This chapter introduced the concepts, representations, and measurements of social network analysis which provide the necessary background to understand the problems addressed in the rest of the thesis. It also presented the social network's features. One of the most important of these features is the community structure. The latter represents the main subject of this thesis and it will be discussed in detail in the next chapter.

\chapter{Community Detection}\label{chp3:BackgroundCD}

\thispagestyle{empty}
\vspace{1cm}

\parindent=0em
\etocsettocstyle{\rule{\linewidth}{\tocrulewidth}\vskip0.5\baselineskip}{\rule{\linewidth}{\tocrulewidth}}
\localtableofcontents

\clearpage
\section{Introduction}\label{chp3:Introduction}
This chapter presents a literature review on community detection. 
At first, the concept of community is introduced along with an overview of the classical static methods for community detection. Following this, the issue of dynamic community detection in time-evolving networks is then addressed. In the light of that, each of the concept of dynamic community and dynamic network models are presented followed by a review of the current literature on dynamic community detection methods.

\section{Community definition}
A community (also called cluster or module) is traditionally defined as a sub-graph that consists of a set of nodes that are more densely connected to each other than to the other nodes in the rest of the network \cite{fortunato2010community,porter2009communities}. A generally accepted definition is still lacking since it may not always be straightforward to give a precise definition of what "more densely connected" means. As a matter of fact, the definition is subjective and may depend on the context of the application. 
Researchers in many fields, e.g., social science, computer science, and physics, have drawn up several definitions. These definitions can broadly be classified into two main categories \cite{fortunato2010community}, namely local definition and global definition. 

\subsection{Local community definition}

To some extent, the community can be viewed as an autonomous entity, which is separated from the whole graph. It is therefore examined independently from the rest of the graph, focusing only on its nodes and maybe its direct neighbors. Many local criteria are used to identify communities: complete mutuality, reachability, vertex degree and the comparison of internal versus external cohesion \cite{tang2010graph}. 

Communities can be defined as a perfect cohesive group, where all its members are connected to each other (complete mutuality) \cite{luce1949method}. In graph theory, this corresponds to a clique - a maximal complete subgraph in which all nodes are adjacent to each other. Such a criterion is too strict especially for social networks which are known to have many triangles (the simplest cliques), but few larger cliques. 
More relaxed definitions of cliques have been proposed in order to comply with the general characteristics observed in real-world social networks. Some methods use the reachability property, i.e., the existence (and the length) of a path between nodes.
In the k-clique based community, two nodes can be considered as belonging to the same community if there exists a path between the two nodes of length no more than $k$. 

Another criterion to define a cohesive group uses the nodal degree and requires a relatively large number of adjacent nodes within the group. In the context of social network analysis, two complementary definitions were proposed: $1)$  a k-plex is a maximal subgroup in which each node is adjacent to all other nodes of the subgroup except at most k of them \cite{seidman1978graph}; $2)$ a k-core is a maximal subgraph in which each node is adjacent to at least k nodes of the sub-graph \cite{seidman1983network}. These two definitions impose conditions on the minimal number of absent or present edges. 

Communities can be defined by comparing the internal and external cohesion of a sub-graph. \cite{radicchi2004defining} proposed the definition of strong community which is a sub-graph that requires that the internal degree of each node is greater than its external degree.  This stringent condition can be relaxed into the definition of weak community \cite{radicchi2004defining}, for which it suffices that the internal degree of the sub-graph exceeds its external degree.

\subsection{Global community definition} 

The community can be defined by considering the whole network. This is particularly appropriate in the case in which sub-graphs cannot be taken apart without seriously affecting the functioning of the system \cite{fortunato2010community}. There exist many global criteria to identify communities, the most well-known of which is the modularity proposed by \cite{girvan2002community}. This criterion is based on the idea that a 
random graph has no meaningful community structure, and thus, the null model is used as a term of comparison, to verify whether the graph displays a community structure. The null model is a randomized version of the original graph, where edges are rewired at random, under the constraint that the expected degree of each vertex matches the degree of the vertex in the original graph \cite{newman2004finding}. The modularity compares the partition of real networks with their randomized part in the null model. A sub-graph is a community if the fraction of edges inside the sub-graph exceeds the expected fraction of edges that the same sub-graph would have in the null model.  The community structure can be detected by optimizing the modularity to find the optimal partition. This criterion will be discussed more thoroughly in Section \ref{chpt2:Modularity}.  
Another important measurement is the Map equation by \cite{rosvall2008maps}. It is based on the principle of the Minimum Description Length (MDL) \cite{grunwald2000model}, whereby any regularity in the data can be used to compress it. By considering the community structure as a set of regularities in the network and the path description of the random walk on the network as the data to compress, communities can be detected during the compression of the path description. 

\section{Static methods for community detection}

Nowadays, there is a vast literature in the field of community detection. A broad variety of algorithms have been developed for the identification of static communities. The approaches for the latter can be roughly classified into: graph partitioning, hierarchical clustering, modularity optimization, Clique percolation and label propagation community detection algorithms. In this section, we provide a brief overview of these approaches. The comprehensive overview of the community detection techniques can be found in \cite{fortunato2010community,fortunato2016community} 

\subsection{Traditional methods}
\subsubsection{Graph partitioning}
The community detection has its roots in graph partitioning. 
The latter consists in partitioning the graph into a predefined number of subgraphs, so that the number of edges between the subgraphs (called cut size) is minimal.
The representative algorithm of graph partitioning methods is Kernighan-Lin (KL) algorithm \cite{kernighan1970efficient}. The KL algorithm is a heuristic optimization method which introduces a gain function $Q$ in the process of dividing communities. The value of $Q$ represents the difference between the number of edges inside the communities and the number of edges connecting between them. 
Spectral bisection method by \cite{barnes1982algorithm} is also one of the famous graph partitioning algorithms. This method is based on the spectral properties of the Laplacian matrix.

Algorithms for graph partitioning are of limited use because it is necessary to provide as input the number of communities which is almost impossible to know.

\subsubsection{Hierarchical clustering}

Graphs may exhibit hierarchical organization which displays several levels of clusters, i.e., clusters at a higher level can contain several lower level clusters. 
In such cases, hierarchical clustering algorithms can be used to reveal the multilevel community structure of the graph. The basic idea behind hierarchical clustering is to define a similarity measure between vertices and to form communities containing most similar vertices. Hierarchical clustering algorithms can be classified in two categories. The first category is called Agglomerative algorithm, in which similar clusters are iteratively merged. It starts from the vertices as separate clusters (singletons) and ends up with a unique cluster. The seconde category is called Divisive algorithm in which clusters are split by eliminating links joining nodes with low similarity. This algorithm follows the opposite direction of agglomerative algorithm. It starts by the whole network as a single cluster and ends up with clusters containing similar vertices. 

\begin{figure*}[h!]
\centering
\includegraphics[scale = 1.7]{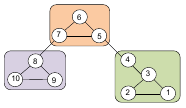}
\includegraphics[scale = 1.7]{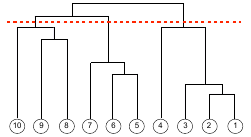}
\caption{Example of the dendrogram (or hierarchical tree). The circles represent nodes in the graph. The horizontal cut in red corresponds to a partition of the graph into three main communities.} 
\label{fig:dendrogram}
\end{figure*}

The hierarchical clustering results can be better represented  as a tree diagram, named dendrogram, like the one in Figure \ref{fig:dendrogram}. 

One of the most popular divisive algorithms is the one by Girvan and Newman \cite{girvan2002community}. The authors introduced the edge betweenness centrality, which refers to the number of the shortest paths that go through an edge in a graph (see Section \ref{sec:ElementMeasurement}). 
The algorithm follows the following process:
\begin{enumerate}
\item Compute  edge betweenness for all edges in the graph
\item Remove edges with the highest edge betweenness (when ties exist with other edges, one edge is to be chosen at random), 
\item Recompute edge betweenness on the remaining links, 
\item Iterate from Step \textit{2} until the graph has no more edges,
\end{enumerate}
Each iteration of this process that increases the number of communities corresponds to a hierarchical level.  

Unlike graph partitioning, hierarchical clustering does not require to specify the number of clusters to be generated as input. However, one of the problems with hierarchical clustering is that it produces many partitions from which we need to identify the appropriate one.

\subsubsection{Partitional clustering}
In this approach, the community detection is formulated as a data clustering problem that aims to separate the set data points into $K$ disjoint clusters such as to minimizing/maximizing a given cost function based on distance measure between nodes. 
One of the most popular partitional clustering algorithms is the $K$-means by \cite{macqueen1967some}.  The algorithm starts from $K$ initial centroids (cluster centers) where $K$ is the desired number of clusters. Each data point is then assigned to the closest centroid based on Euclidean distance. Then, the mean of the resulted collection of data points (clusters) is calculated and centroids are updated to mean value. The assignment and update steps are repeated until the centroids stop changing (convergence).

\subsubsection{Spectral clustering}
Spectral clustering refers to the class of methods using the eigenvectors of a matrix to find graph partitions. It was introduced in the early 1970s with the work of \cite{donath2003lower} who first used similarity matrices, and of  \cite{fiedler1973algebraic} who proposed using the normalized Laplacian matrix. The most common form of spectral clustering involves three main steps: $i)$ constructing a matrix representation of the graph, $ii)$computing of eigenvalues and eigenvectors of the matrix and mapping of each point to a lower-dimensional representation based on one or more eigenvectors, and $iii)$ finally, clustering of points based on the new representation. 
 The main difference between spectral clustering algorithms lies in the matrix representation (Adjacency matrix, Normalized/Unnormalized Laplacian matrix, Gaussian kernel, etc.) 


\subsection{Modularity-based approaches} \label{chpt2:Modularity}

A large number methods has been suggested to find optimal community structures. Modularity 
is one of the most widely used technique in optimization based methods. The present subsection first introduces Modularity and then presents the most popular techniques for modularity optimization. It discusses at the end the resolution limit problem which modularity suffers from.

\subsubsection{Modularity definition}

Modularity has been initially introduced as a stopping criterion in the algorithm of Girvan and Newman \cite{newman2004finding} to select the best cut in the dendrogram. Since then, it has become one of the the most used and the most significant quality measure for communities. It quantifies the difference between the fraction of edges within communities and the expected of such fraction in the random graph with the same number of nodes and the same node degrees as the original graph (null model). The idea behind this definition is that the communities should have more internal links than what is expected in a random graph. 

More formally, let us consider a graph $G=(V,E)$ comprising a set of nodes (or vertices) $V$ connected by set of links (or edges) $E$. $|V|$ and $|E|$ denotes the number of elements in $V$ and $E$ respectively. 
$C$ denotes the community structure of the graph $G$. In a  random graph, the probability of linking two nodes $i$ and $j$ with degrees $d_i$ and $d_j$ respectively is : 
\(\frac{d_{i}d_{j}}{{(4|E|)}^{2}}\)
 and hence the expected fraction for links in a  community $c$ is given as : 
 \({\left(\frac{d_{c}}{2|E|}\right)}^{2}\), where  $d_{c}=\sum_{i\in c} d_{i}$(sum of node degrees in $c$ ).  
Modularity can be defined as: 
\begin{equation}
\label{eqn:eq1}
Q=\sum_{c\in C} Q_{c}=\sum_{c\in C}\left( \frac{|E_{c}^{in}|}{|E|}-{\left( \frac{2|E_{c}^{in}|+|E_{c}^{out}|}{2|E|}\right)}^{2}\right),
\end{equation}

where \(\frac{|E_{c}^{in}|}{|E|}\) is the number of links inside the community $c$. Accordingly, \(\left(2|E_{c}^{in}|+|E_{c}^{out}\right)\)is the fraction of links within the community $c$. 
The equation \ref{eqn:eq1} can also be expressed in equation \ref{eqn:eq2} and \ref{eqn:eq3}: 
\begin{equation}
\label{eqn:eq2}
Q=\frac{1}{2|E|}\sum_{c\in C} \sum_{i,j\in c}\left( A_{ij} -\frac{d_{i}d_{j}}{2|E|}\right),
\end{equation}

\begin{equation}
\label{eqn:eq3}
Q=\frac{1}{2|E|}\sum_{i,j\in V}\left( A_{ij} -\frac{d_{i}d_{j}}{2|E|}\right)\delta(c_{i},c_{j}),
\end{equation}

where, $A_{ij}$ is an element of the adjacency matrix between nodes $i$ and $j$ ($ A_{ij} =1 $ if i and j are connected, otherwise $ A_{ij} =0 $), $\delta(c_{i},c_{j})$ is the Kronecker delta symbol which indicates whether the nodes $i$ and $j$ belong to the same community (if ($c_{i}=c_{j}$) $\delta(c_{i},c_{j})= 1$, otherwise $\delta(c_{i},c_{j})= 0$). 

Higher modularity usually means better community structure. The modularity can be either positive or negative, it takes values between $-1$ and $1$. A positive value indicates the possible presence of community structure. If each node is a community itself, the modularity is always negative and it is zero when taking the whole graph as a single community. 

Modularity is the basis of many methods for community detection. It is often used as a quality function to be optimized. Modularity optimization is known to be NP-hard problem, so one usually employs heuristics or approximation algorithms. The most popular of these algorithms  will be presented in the following section. 
\subsubsection{Greedy optimization}

Newman's greedy algorithm \cite{newman2004fast} was the first algorithm to maximize modularity. It is an agglomerative hierarchical clustering method, where initially each node in the graph is considered as a single community, then they are merged iteratively in order to get the greatest value increase of modularity. Only those communities sharing one or more edges can be merged at each iteration.


Another well-known method is the Louvain algorithm by \cite{blondel2008fast} that is a heuristic greedy algorithm for detecting communities in weighted graphs. It is also based on modularity optimization. Louvain method has two phases
\begin{enumerate}
\item First, it assigns each node to a different community as singleton communities. Then, it tries to reassign the node to the community of its neighbor that resulted in the greatest modularity increases. If no increase is possible, then the node stays in its own original community. The process is repeated until no further increase in modularity can be achieved.
\item The algorithm then builds a new network, called \textit{super-graph}, in the way that communities identified from the first step are contracted into super-nodes, edges between super-nodes are weighted with the sum of the weights of the edges between the represented communities at the previous step, self-loop represents edges between nodes of the same community. 
\end{enumerate}

\begin{figure}[!htbp]
\centering
\includegraphics[scale=1.5]
{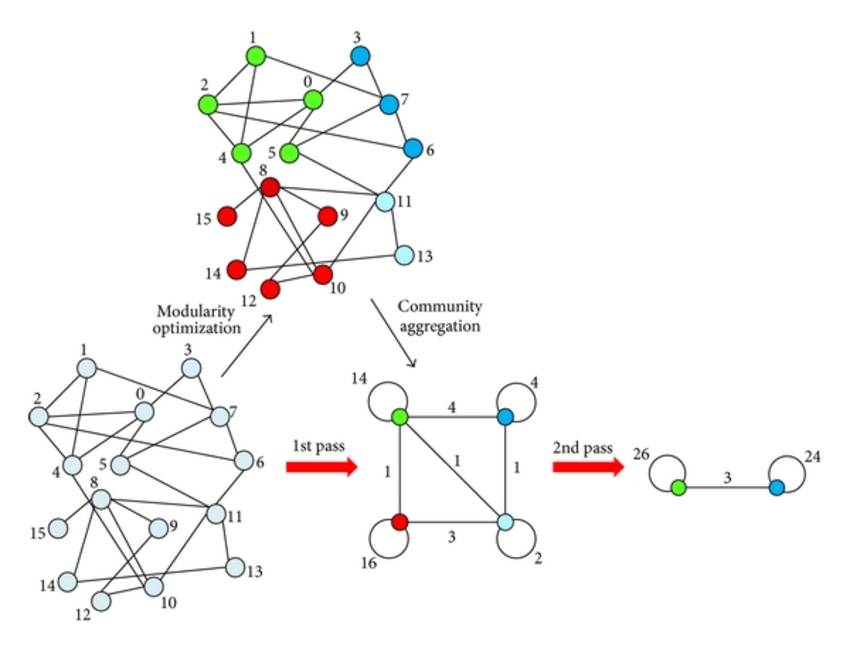}
\caption{Calculation process for Louvain algorithm \cite{blondel2008fast}}
\label{fig:louvain}
\end{figure}

The two phases 
of the algorithm are then repeated iteratively, until the contraction does not reduce the number of nodes (when modularity cannot increase anymore). Note that the modularity gain at the first step is always computed for the original network (not for the super-graph). 
The Louvain algorithm is illustrated in Figure \ref{fig:louvain}.

The algorithm is extremely fast and produces good solutions in practice\cite{fortunato2010community}. 

\subsubsection{Simulated Annealing}
Simulated annealing is a probabilistic meta-heuristic for the global optimization problem which avoids the risk of getting trapped in a local minimum. It was first employed for modularity optimization by \cite{guimera2004modularity}. At first, the algorithm starts by partitioning the network into random partitions. Then, in each iteration, both random local and global moves take place based on modularity gain. Local moves shift a node randomly from one module to another, while global moves consist of splitting and merging modules. 

\subsubsection{Extremal optimization}
Extremal optimization is a meta-heuristic technique for combinatorial optimization problems that first appeared in the field of statistical physics by \cite{boettcher1999extremal}. It basically operates on optimizing a global variable by improving extremal local variables. This technique was applied for modularity optimization problem by \cite{duch2005community}. The authors used Modularity as the global variable to optimize, and define node fitness (the ratio of the local modularity of the node by its degree) as a local variable in the extremal optimization process. The proposed heuristic evolves as follows: it starts by randomly splitting the network into two partitions of equal number of nodes. Then, at each iteration, it moves the node with the lowest fitness from its own community to another community. The shift changes the community structure, so the fitness of many other nodes needs to be recalculated. 
The process repeats until it cannot increase modularity. After that, it generates sub-community networks by deleting the inter-community edges and proceeds recursively on each sub-community network until an "optimal state" with a maximum value of modularity is reached.
\subsubsection{Spectral optimization}
Spectral optimization applies the classical sepctral clustrening approach for modularity optimization. For instance, in the popular work by  \cite{newman2006modularity} modularity is reformulated in terms of eigenvectors of a new representation matrix for the graph called modularity matrix.
\subsubsection{Resolution limit}

\begin{figure*}[h!]
\centering
\includegraphics[scale=0.3]
{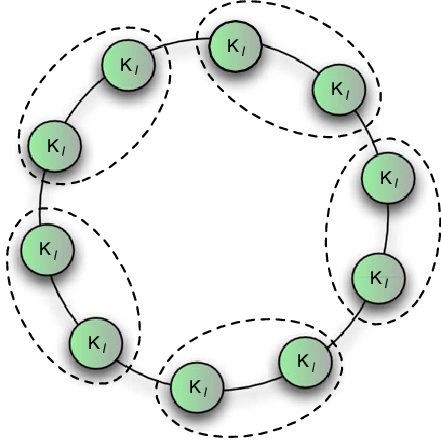}
\caption{Resolution limit of modularity optimization \cite{fortunato2007resolution}. A network made of identical cliques of size $m$ connected by single links. If the number of cliques is larger than about $\sqrt{L}$ ($L$ represents the total number of links in the network), the partition with Modularity optimization corresponds to clusters containing two or more cliques (represented by dotted lines).}
\label{fig:resolution}
\end{figure*}

Despite the huge success of modularity optimization, one drawback is that it suffers from the problem of resolution limit. This problem has been discussed in the study by \cite{fortunato2007resolution}. 
The authors have proved that there is some scale depending on the network size beyond which smaller communities cannot be detected, even when they are well defined (like cliques) and loosely connected to each other (see Figure \ref{fig:resolution}). This problem has a great impact in practice since most real world networks contain communities of very different sizes.

The resolution limit stems from the definition of modularity, and in particular from the null model which presupposes that each node can interact with every other node. This, however, is not the case for large world networks, in which every node interacts only with a portion of the network. 

In order to resolve this problem, many researchers proposed modified versions of modularity: multiresolution modularity \cite{reichardt2006statistical,arenas2008analysis}, modularity desnsity \cite{li2008quantitative}, etc.   

\newpage

\subsection{Clique Percolation Method}

\begin{figure*}[h!]
\centering
\includegraphics[scale=0.8]{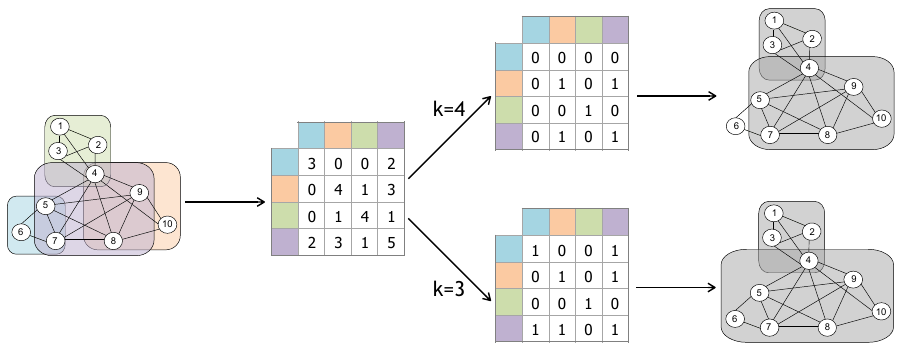}
\caption{Example illustrating the principle of CPM to find k-clique communities on an example of the graph at $k = 3$ and $k = 4$.}
\label{fig:cpmExp}
\end{figure*}

Most community detection algorithms are designed to identify disjoint communities and therefore are not suitable for detecting overlapping communities. However, in many real-world networks, it is natural to find nodes that belong to more than one
community at the same time. The Clique Percolation Method (CPM) by \cite{palla2005uncovering} was among the first methods for detecting overlapping communities. In this method, a k-clique is rolled over the network to other cliques sharing k-1 nodes. In this way, a community is composed of the union of all k-cliques that can be reached from each other by rolling on the network. The algorithm works as follows:
\begin{enumerate}
\item Find all maximal cliques, a clique is maximal if it is not included in a larger clique.
\item Create clique-clique overlap matrix. Each entry in the matrix indicates the number of nodes shared between the respective cliques.
\item Erase every off-diagonal entry smaller than k-1 and every diagonal element smaller than k in the matrix and replace the remaining elements by one.
\item The resulting components from the matrix are equivalent to k-clique communities.
\end{enumerate}
Figure \ref{fig:cpmExp} illustrates the principal of CPM  to find k-clique communities on an example of graph for $k = 3$ and $k = 4$.

\subsection{Label Propagation method}
The label propagation method is a simple and fast community detection method which was originally introduced by \cite{raghavan2007near}. In the initial stage, the algorithm assigns a unique label to each node. Then, an iterative process is followed, where each node updates its label to the one shared by the largest number of its neighboring nodes. The process is repeated until convergence, i.e., the label of each node in the network no longer changes. Communities are then obtained by considering groups of nodes with the same label.

\section{Dynamic communities}\label{chpt3:Communityevolution}

\begin{figure*}[h!]
\centering
\includegraphics[scale=1.3]{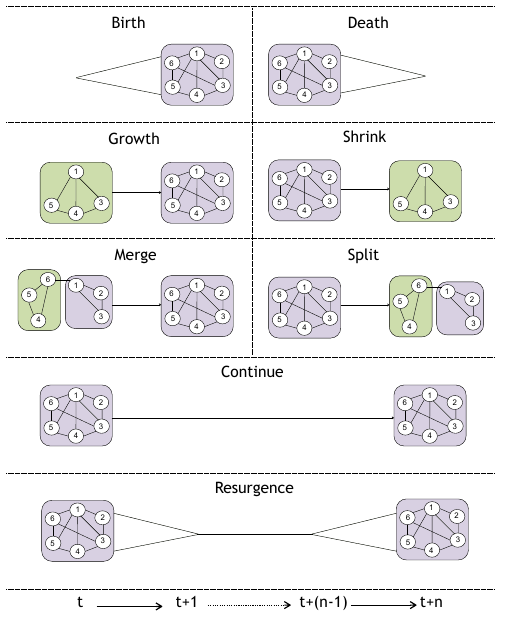}
\caption{Events characterizing evolution of communities 
\cite{rossetti2018community}.} 
\label{fig:life}
\end{figure*}

\cite{palla2007quantifying} have introduced six types of events to characterize the evolution of communities: birth, growth, shrink, merge, split, and death. \cite{cazabet2014dynamic} proposed a new operation: resurgence. The continue event is often considered in the community life cycle. 
In the following, we describe these events in detail (see Figure \ref{fig:life} ):
\begin{itemize}
\item Birth: a new community emerges if it has never been observed before.
\item Death: The dissolution of a community occurs when it does not appear in the next times.
\item Growth: a community grows when new nodes join the community, making its size larger than in the preceding time.
\item Shrink: a community shrinks when it loses some of its nodes.  
The size of this community is thus reduced compared to that in the previous time. 
\item Merge: a merge occurs if two distinct communities or more are combined into a single community at the next time. 
\item Split: it may occur that a single community splits into two or more communities. 
\item Resurgence: A community vanishes for a period, then comes back without perturbations as if it has never stopped existing. 
\end{itemize}

It should be noted that not all these events may be handled in the same community detection method. Also, the way how to deal with these events may differ from one method to another, depending on their purpose and their area of application. Several methods proposed formal definitions of these events. 


\cite{asur2009event} formalized five of the eight events described above (see Table \ref{tab:CommunityEvents}). Let $C_i$ and $C_{i+1}$ denoting the set of clusters observed at two consecutive times $i$ and $i+1$ respectively.

\begin{table}[!htbp]
\resizebox{\textwidth}{!}{%
\small
\centering
\begin{tabular}{|p{6 cm}|p{9cm}|}
  \hline
  $Continue(C_i^k, C_{i+1}^j) = 1$ if $V_i^k=V_{i+1}^j$ & A cluster is a  continuation of another cluster if the two clusters share the same nodes.  \\
  \hline
 $k$-$Merge(C_i^k, C_i^l, k)$= 1 if $\exists$ $C_{i+1}^j$ such that :

\begin{itemize}
\item  $\frac{|(V_i^k \cup V_i^l) \cap V_{i+1}^j |}{Max(|V_i^k \cup V_i^l|, |V_{i+1}^j |)} > k \%$.

\item $|V_i^k \cap V_{i+1}^j|> \frac{|C_i^k|}{2}$

\item $|V_i^k \cap V_{i+1}| > \frac{|C_i^l|}{2}$
\end{itemize} & two clusters merge if there exists a cluster in the next time step that contains at least $k\%$ of the nodes belonging to their union and the renewal of these two clusters is at least $50\%$.\\
  \hline
  $K$-$Split(C_{i}^j, k)=1$ if $\exists$ $C_{i+1}^k , C_{i+1}^l$ such that :
\begin{itemize}
\item  $\frac{|(V_{i+1}^k \cup V_{i+1}^l) \cap V_{i}^j |}{max(|V_{i+1}^k \cup V_{i+1}^l|, |V_{i}^j |)} > k \%$.

\item $|V_{i+1}^k \cap V_{i+1}^j|> \frac{|C_{i+1}^k|}{2}$

\item $|V_{i+1}^l \cap V_{i+1}^j| > \frac{|C_{i+1}^l|}{2}$
\end{itemize} & A cluster splits if $k\%$ of its nodes are present in two different clusters in the next time step.\\
  \hline
  $Form(C_{i+1}^k)=1$ if $\exists$ no $C_i^j$ such that : $V_{i+1}^k \cap V_i^j > 1$ & a new cluster appears if none of the nodes in the cluster were grouped together at the previous time step. \\
\hline
$Dissolve(C_i^k)=1$ if $\exists$ no $C_{i+1}^j$ such that : $V_i^k \cap V_{i+1}^j > 1$ & A cluster vanishes if none of the vertices in the cluster are in the same cluster in the next time step\\
\hline
\end{tabular}}
\caption{Formal definition of events characterizing community evolution \cite{asur2009event}}
\label{tab:CommunityEvents}
\end{table}

The definition has some weaknesses:
\begin{itemize}
\item The authors consider only events between consecutive snapshots. They do not discuss basic events like contraction or growth.  
\item The definition of events is restrictive. In some cases, the community may continue, even it loses some of its nodes. A community may form not only with one overlapping node.  
\item The definition of continue and form events considers only nodes. The authors do not impose any restrictions on edges. This may allow the continuation of a community that loses all its edges when it should rather be a death or the emergence of a new community without any edges.  
\item The dependency on the parameter $k$ to define merge and split events. The final result depends on the value of $k$.  
\end{itemize}
Several other definitions have been proposed: similarly to \cite{asur2009event}, \cite{greene2010tracking} and \cite{brodka2013ged} proposed descriptions based on matching techniques. \cite{chen2010detecting} characterize community dynamics by tracking community core evolution. 

\section{Dynamic network models}
Like many networking systems, social networks are dynamic by nature. Friendships, communications, collaborations between social entities may shift over time. For instance, not all friendships last forever, some of them fade away over time. In a research laboratory, some researchers only have short-term collaborations, while some others have rather long and sustainable research collaborations. This dynamic is of much importance to provide a complete understanding of the network system. Researchers have proposed many representations to deal with time-varying networks. We can distinguish three broad approaches: aggregated graphs model, series of snapshots model, and temporal networks. In this section, each of these models will be discussed.

\subsection{Aggregated graphs model}
A straightforward way to handle a dynamic network is to simplify it into a single static network by aggregating all contacts between each pair of nodes in a single edge. One can build a binary static network where nodes are only linked or not, but obviously, a lot of information is lost. Another alternative 
is to incorporate the frequency of interactions between nodes by constructing weighted networks. All these approaches can never retain all temporal network evolution information. It does not allow longitudinal analysis, for instance tracking the evolution of communities.
\subsection{Series of snapshots model}
The evolving network is modeled through a series of snapshots, each of which is a static network 
representing the state of the network at a given time. There are two basic ways to construct a snapshot, either by capturing the contacts that exist at 
a given time step (e.g. every hour, week, year, etc.) or by aggregating all contacts during a given period of time (window time). 
The main issue of this approach is to determine the temporal granularity, i.e., the 'right' number of temporal steps or time windows. Tracking communities across network sequences can be difficult if important temporal information is lost between snapshots.
\subsection{Temporal networks}
Temporal networks conserve all known temporal information. There are two main models: series of contact and interval graph \cite{holme2012temporal}. In a sequence of contact, each interaction is represented as a triple $(i,j,t)$ where $i$ and $j$ are the interacting entities and $t$ is the time when the relationship is activated. In an interval graph, interaction is represented as a quadruplet $(i,j,t,\delta t)$ which means that $i$ is involved in contact with $j$ from $t$ to  $t+\delta t$. In these models, only the temporal information about interactions is represented, there is no temporal information about nodes.
\section{Literature review on dynamic community detection}\label{chpt2:DCDM}

The main goal in this section is to review relevant methods and algorithms proposed so far for the problem of dynamic community detection. 
This section is organized as follows:
\begin{enumerate}
\item As a first step, we will present an overview of the main existing surveys on dynamic community detection. The different classifications proposed by four surveys \cite{aynaud2013communities, hartmann2016clustering, rossetti2018community, dakiche2019tracking} will be discussed.
\item The survey proposing the most reliable classification of those surveys will then be extended and enriched by reviewing many diverse methods and algorithms, especially recent ones.
\item Then, we will present the advantages and drawbacks of each class of approaches. 
\item At the end of this section, we will discuss the main findings and observations taken from this survey.
\end{enumerate}



\subsection{Existing surveys}
Given the growing interest in dynamic community detection and the increasing number of methods being proposed in this topic, several related surveys have been conducted by different researchers \cite{aynaud2013communities, hartmann2016clustering, dakiche2019tracking, rossetti2018community}, proposing different classifications. 

\cite{aynaud2013communities} distinguished three classes: 

\begin{enumerate}
\item Two-stage approaches which first discover time-independent communities at each time step and then identify the community evolution by matching communities across different time steps.
\item Evolutionary clustering methods that simultaneously optimize the clustering quality and the temporal smoothness, thus requiring the current network topology and the communities found in the previous time step.
\item Coupling graph clustering methods which detect community structure on a graph built by adding links between instances of nodes at different time steps.
\end{enumerate}

The survey by \cite{hartmann2016clustering} identify two categories: 
\begin{enumerate}
\item Online approaches which use information about the network topology and community structure in the previous time steps.
\item Offline approaches which  use information from both previous and subsequent time steps. This survey focuses only on online methods. The latter are divided into two subcategories: $i)$ Temporal smoothness methods which compute communities from scratch at each time step and $ii)$ Dynamic update approaches which update communities found in previous time steps.
\end{enumerate}

The survey by \cite{cazabet2014dynamic}, extended by \cite{rossetti2018community} proposed a taxonomy of three classes, corresponding to different dynamic community definitions: 
\begin{enumerate}
\item Instant-optimal approaches which only consider the current state of the network.
\item Temporal Trade-off approaches which only consider past and present clustering and past network topology.
\item Cross-time approaches which consider the entire evolution in both network topology and clustering. 
\end{enumerate}
Each of these classes is divided into subcategories, corresponding to different techniques used to find communities matching the underlying definition.

\cite{dakiche2019tracking} adopted a classification inspired from the one proposed by \cite{cazabet2014dynamic}. They distinguished four classes: 
\begin{enumerate}
\item Independent community detection and matching methods which correspond to Two-stage approaches.
\item Dependent community detection methods which require the current network topology and the community structure found in the previous time step. 
\item Simultaneous community detection on all snapshots which corresponds to Coupling graph methods.
\item Methods working on temporal graphs which update the network at each network change.
\end{enumerate}

Regarding these literature reviews, most of them are not comprehensive. The proposed classifications do not always cover all the techniques of dynamic community detection, since some methods, particularly recent ones, do not fit into any category. Also, some of the classifications are overlapped, since some methods can be classified under more than one category.
The survey by \cite{rossetti2018community} seems to be more exhaustive and reliable. It provides an in-depth overview of dynamic community discovery approaches. 

\subsection{Approaches for dynamic community detection}

Given the robustness and the exhaustiveness of the survey by \cite{rossetti2018community}, we will opt for their classification in our review. Our goal is not to rewrite the original survey, but rather to extend it by adding some relevant methods which are not included in the survey, especially recent ones. 

\begin{figure*}[!h]
\centering
\includegraphics[scale=.4]
{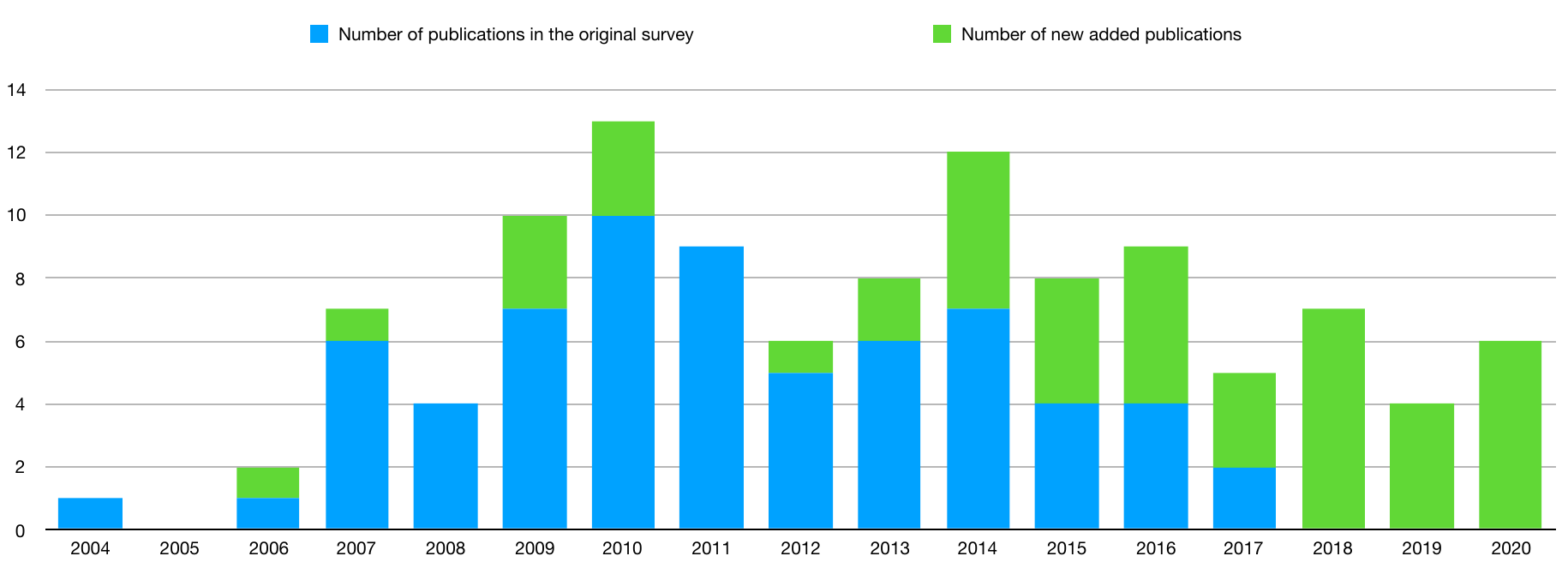}
\caption{Analyzed papers per year}
\label{fig:pub}
\end{figure*}

Figure \ref{fig:pub} illustrates the relevance of such an extension, as it includes  more recent methods. 

\begin{figure*}[!htbp]
\centering
\includegraphics[scale=.8]
{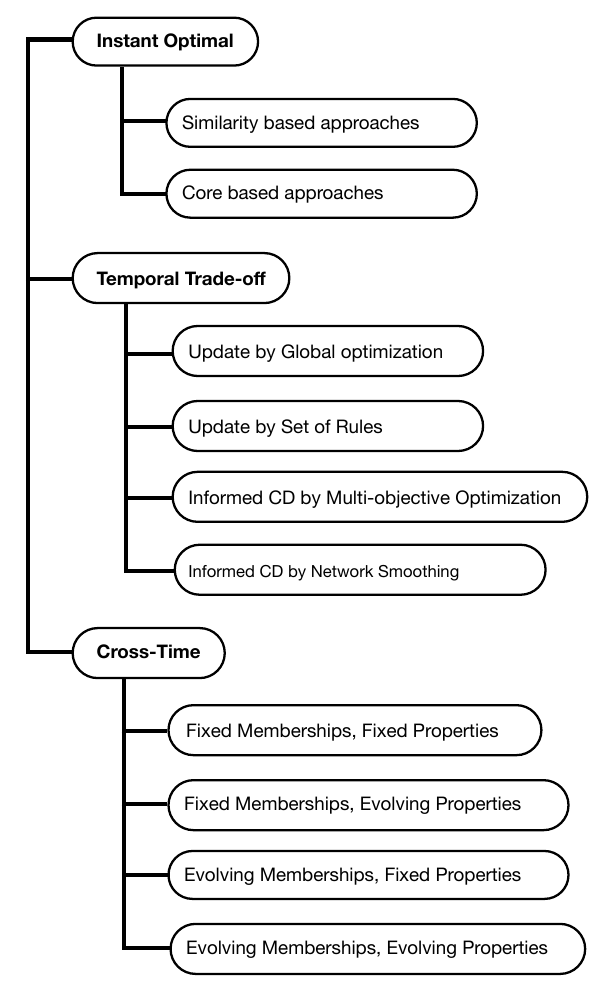}
\caption{Classification of dynamic community detection algorithms by \cite{rossetti2018community}.}
\label{fig:tax}
\end{figure*}

The following methodology is adopted in this review.
\begin{itemize}
\item We will use the taxonomy by \cite{rossetti2018community}, a two-level classification (see Figure \ref{fig:tax}). The high-level distinguishes three classes corresponding to different dynamic community definitions. Each of these classes is divided into subcategories, corresponding to different techniques used to find communities. In this survey, for each class, we will present its dynamic community definition and we will outline its subcategories.   
\item We will not describe all the methods presented in the survey by \cite{rossetti2018community}.  For each subcategory of approaches, we will present only some representative methods and the rest of the methods will just be cited (for further detail, the reader can refer to \cite{rossetti2018community}). 
\item We will extend the survey by \cite{rossetti2018community}, by categorizing and reviewing new relevant methods which are not already presented in the survey, with a particular emphasis on recent ones. 
\item A table will be provided to summarize all the reviewed methods for each category of approaches.
\item For ease of reading, each subcategory of approaches will be marked by two horizontal lines. The listed methods will be distinguished by formatting their citations with bold font. The newly added methods will be marked by a star "$^{*}$". 
\end{itemize}

\subsubsection{Instant-optimal}
This category of approaches 
proposes to adapt static community detection methods to the dynamic case. In such a category, the network evolution is modeled as a snapshot model, i.e., a sequence of temporal steps. At first, an optimal partition is calculated for each snapshot using any static algorithm. Then, the evolution of the community structure is tracked over time by comparing the obtained optimal partitions either between consecutive snapshots or even between far apart snapshots. 
An example of Instant-optimal methods are community matching approaches, also called "Two-Stage Approaches" (or identify and match) 
\cite{aynaud2013communities}, which comprise two main steps: a static community detection step and an iterative matching step to align communities found in successive snapshots.



Based on the matching technique used to compare optimal partitions, methods in this class can be classified into similarity-based approaches and core-nodes based approaches. Based on the scope of their matching, each of these two subcategories, in turn, can be divided into Iterative matching and  Multi-step matching.
\newline
\noindent\rule{\linewidth}{0.4pt}\par
\vspace{-0.25em} 
\noindent\textit{\textbf{Similarity-based approaches}}\par 
\vspace{-0.75em} 
\noindent\rule{\linewidth}{0.4pt}\par

Similarity-based approaches are the most popular in the Instant-optimal class. These approaches 
use a quality function to measure the similarity between communities in adjacent time steps (Iterative similarity-based methods) or in far apart time steps (Multi-step similarity-based methods).
Communities with the highest similarity are considered as part of the same dynamic community.

The method by \textbf{\cite{hopcroft2004tracking}} is one of the earliest methods in this subcategory. It starts by identifying "natural communities" that remain stable under multiple clustering runs. Iterative clustering algorithm by \cite{jain1988algorithms} is used at this step. Then it tracks them over time by matching similar successive communities.  The authors proposed the following matching function:
\(match(C_{1},C_{2})=Min(\frac{|C_{1}\cap C_{2}|}{|C_{1}|},\frac{|C_{1}\cap C_{2}|}{|C_{2}|})\), Where $C_{1}$ and $C_{2}$ are the communities (set of nodes) to match.

\textbf{\cite{palla2007quantifying} } extended the original clique percolation method \cite{palla2005uncovering} to the dynamic case. First, the CPM \cite{palla2005uncovering} is applied on each time step to find communities. Then, it is applied to the joining graphs for pairs of consecutive time steps. Finally, the resulting communities are matched with communities between consecutive time steps while finding the community-centric events (birth, growth, merge, split, and death). 
The authors used the Jaccard index : \(Jaccard(X,Y)= \frac{|Y\cap Y|}{|X \cup Y|)}\)  to measure relative overlap value between two communities. The cluster $X$ at time $t+1$ is matched to the cluster $Y$ which has the largest overlap at time $t$. 
\begin{figure*}[h!]
\centering
\includegraphics[scale=1.1]{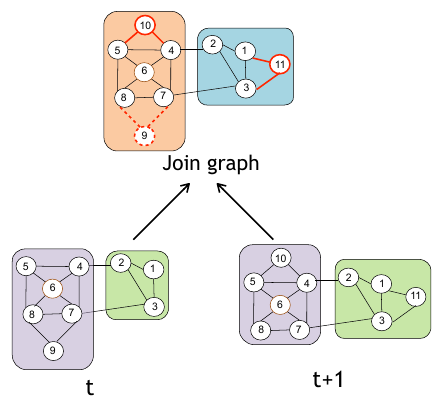}
\caption{Example illustrating the principle of D-CPM. Red dashed lines represent nodes/ edges that appear only at time $t$. Red solid lines denote new nodes/edges at time $t+1$.}
\label{fig:DCPM}
\end{figure*}

\textbf{\cite{asur2009event} } proposed a an event-based method. At first, they apply the MCL algorithm \cite{vandongen2000cluster} (a well-known modularity-based clustering algorithm) to obtain the clusters in each snapshot. Then, they characterize the transformation of these clusters by defining five basic events (see Table \ref{tab:CommunityEvents} in Section \ref{chpt3:Communityevolution}). The matching in this method is simply to compute events between consecutive snapshots. It is implemented as bit operations on timestamp cluster membership matrices. Several works with the same rationale have been proposed by different researchers, notably by \cite{greene2010tracking}, \cite{sun2015matrix}, \cite{bota2011dynamic}, \cite{brodka2013ged}, \cite{zhu2016reconstructed}, \cite{ilhan2015predicting}, \cite{rossetti2020angel}.

In the method by \textbf{\cite{greene2010tracking}}, the static algorithm MOSES \cite{mcdaid1detecting} is applied to the individual snapshot graphs. Then, the Jaccard coefficient is used to match successive snapshots. The key events in the community life cycle are formulated as a set of rules characterizing the evolution of communities.

\textbf{\cite{sun2015matrix}$^{*}$} proposed a matrix-based algorithm to identify community evolution events. First, they applied the Louvain algorithm \cite{blondel2008fast} to detect communities in each snapshot. Then, they built two correlation matrices to describe the relationship between communities in each pair of adjacent snapshots. Based on these matrices, they defined decision rules to detect the evolution of events.


\textbf{\cite{bota2011dynamic}} proposed a method capable of handling communities found by any non-monotonic community detection algorithm. The authors extended the basic community events described in \cite{palla2007quantifying} by introducing five events: Grow-merge, Contraction-merge, Grow-split, Contraction-split, and Obscure case (combinations of possibly more than two events). They adopt the same methodology of \cite{palla2007quantifying}:  The $N^{++}$ algorithm \cite{bota2010community} is used to find communities and the community matching is performed on union graphs between consecutive snapshots according to a set of rules based on sizes of communities. 

\textbf{\cite{brodka2013ged} }suggested the GED method to discover group evolution, i.e., events (changes) which may occur in social groups (communities): continuing, shrinking, growing, splitting, merging, dissolving, and forming. To enable the analysis of group evolution, the authors introduced a new measure called inclusion, which respects both the quantity (the number of members) and quality (the importance of members) of the group. 

The authors in \textbf{\cite{zhu2016reconstructed}$^{*}$} reconstructed the event framework by \cite{asur2009event} and proposed the concept of community attributes to analyze the difference of communities between consecutive snapshots.


\textbf{\cite{ilhan2015predicting} }proposed an event-based framework to track similar communities and identify events over time. Louvain algorithm \cite{blondel2008fast} is applied to discover communities. The set of communities at consecutive snapshots are matched with each other using a custom measure called community similarity: two communities are similar if the ratio of their mutual members exceeds a threshold. Significant events of the communities are identified based on this matching measure. To detected events, such as survive, grow and shrink, a metric, called fluctuation is proposed to compute the percentage of increase/decrease in the number of community members. 
The authors proposed also a time series model to predict community features.

The Angel algorithm by \textbf{\cite{rossetti2020angel}$^{*}$} is a node-centric method for detecting overlapping communities in both static and dynamic networks. The dynamic variant of Angel works as follows: first, communities in the different snapshots are extracted using the static variant of Angel. Then, the precision score: \(Precision(X,Y)= \frac{|X\cap Y|}{|X |}\)is used to match communities between consecutive snapshots in both directions. This makes it possible to identify the different events in community life-cycle.

Other methods use the same approach based on similarity to match communities between different snapshots
, not only between consecutive time steps. The works by \cite{falkowski2007users, falkowski2006mining}, \cite{goldberg2011tracking}, \cite{tajeuna2015tracking} and  \cite{mohammadmosaferi2020evolution} fall in this category.

In the work by \textbf{\cite{falkowski2007users, falkowski2006mining}}, the authors used the concept of community survival graph, a graph built by representing communities detected at different time steps as nodes which are linked via edges based on their overlap similarity: \(Overlap(X,Y)= \frac{|Y\cap Y|}{Min(|X|, |Y|)}\).  Then, the hierarchical edge betweenness clustering algorithm \cite{newman2004finding} is run on this graph to find final communities. 

\textbf{\cite{goldberg2011tracking} }suggested a framework for tracking community evolution by identifying evolutive chains of communities. 
For each snapshot, a set of communities are extracted using any static algorithm. Communities are then linked (matched) via a strength measure (any set intersection measure can be used). For each community, a valid maximal chain of communities is constructed. 
A chain is valid if the strength of its weakest link is above a threshold and it is considered as maximal if it is not a proper subchain of any other valid chain. 


\textbf{\cite{tajeuna2015tracking}$^{*}$} proposed a method which initially identifies the community structure at each time step by using Infomap algorithm \cite{rosvall2008maps} and then proceeds to build a matrix that counts the number of nodes shared between each pair of communities. 
The individual rows of the obtained matrix are then used to 
capture the trace of the communities that should be compared over all time points. 
The authors proposed a new similarity measure, named mutual transition, to compare communities and a set of rules to capture significant transition events a community can undergo. 


The recent work by \textbf{\cite{mohammadmosaferi2020evolution}$^{*}$} put forward a mapping-based method to identify community evolution. First, communities are detected for each time step (using Infomap \cite{rosvall2008maps} and Leiden \cite{traag2019louvain}). Members of each community are then mapped to a pair which includes a time window and community index. Community mapping is defined using two kinds of similarities: partially similar communities and very similar communities. It is implemented using Python's hash-map.

Table \ref{tab:methods1} lists similarity-based methods and the key elements describing their rationale.

\begin{table}[!htbp]
\resizebox{\textwidth}{!}{%
\small
\begin{tabular}{|p{2cm}|p{3.5cm}||p{7.5cm}|}
\hline
\textbf{Category}                    & \textbf{Reference} & \textbf{Key elements} \\ \hline \multirow{10}{*}{\textbf{Iterative}}
& \cite{hopcroft2004tracking}                  & Natural communities, static iterative clustering by \cite{jain1988algorithms}, matching function                     \\ \cline{2-3} 
                                     & \cite{palla2007quantifying}                  & Clique percolation method(CPM)\cite{palla2005uncovering}, joining graph, community events, Jaccard index, overlapping communities
                                     \\ \cline{2-3} 
                                                    & \cite{asur2009event}                  & Event based method, bit operation, cluster membership matrices                     \\ \cline{2-3} 
                                                    & \cite{greene2010tracking}                 & Event-based, MOSES Static algorithm by \cite{mcdaid1detecting}, Jaccard index, rules characterizing community evolution                     \\ \cline{2-3} 
                                                    & \cite{sun2015matrix} $^{*}$                  & Matrix based algorithm, Louvain algorithm, correlation matrix, decision rules                     \\ \cline{2-3} 
                                                    & \cite{bota2011dynamic}                  & Basic and obscure community events, $N^{++}$ algorithm \cite{bota2010community}, union graph, rules                     \\ \cline{2-3} 
                                                    & \cite{brodka2013ged}                  & Group evolution, inclusion measure, event-based method                      \\ \cline{2-3}  
                                                    & \cite{zhu2016reconstructed}$^{*}$                  & Event framework, community attributes                     \\ \cline{2-3} 
                                                    & \cite{ilhan2015predicting}                  & Event-based framework, Louvain algorthim \cite{blondel2008fast}, community similarity measure, flactuation measure, time series model                     \\ \cline{2-3}
                                                   & \cite{rossetti2020angel} $^{*}$                 & Node-centric method method, overlapping communities, precision score, community life-cycle events                       \\ \hline \multirow{5}{*}{\textbf{Multi-step}} & \cite{falkowski2007users, falkowski2006mining}                  & Community survival graph, overlap similarity, hirachical edge betweenness  clustering algorithm \cite{newman2004finding}
                                                    \\ \cline{2-3} 
                                                    & \cite{goldberg2011tracking}                  & Evolutive chain of communties, strength measure, valid maximal chain                     \\ \cline{2-3} 
                                                     & \cite{tajeuna2015tracking}$^{*}$                   & Infomap algorithm \cite{rosvall2008maps}, Matrix of common nodes, mutual transition measure                     \\ \cline{2-3} 
                                                    & \cite{mohammadmosaferi2020evolution}$^{*}$                   & Mapping communities, Infomap \cite{rosvall2008maps} and Leiden algorithms\cite{traag2019louvain}, partially similar communities, very similar communities, hash-map                     \\ \hline
\end{tabular}}
\caption{Matching-based methods and the key elements describing their rationale. New references added to the survey in \cite{rossetti2018community} are marked with "$^{*}$".}
\label{tab:methods1}
\end{table}

Other methods in \textbf{\cite{dhouioui2014tracking, rosvall2010mapping, takaffoli2011modec, bourqui2009detecting}} fall also in this category (their rationale can be found in the original survey by \cite{rossetti2018community}).
\newpage
\noindent\rule{\linewidth}{0.4pt}\par
\vspace{-0.25em} 
\noindent\textit{\textbf{Core-based approaches}}\par 
\vspace{-0.75em} 
\noindent\rule{\linewidth}{0.4pt}\par

In this subcategory, the matching is reduced to only specific nodes, called core-nodes, instead of all members in communities. Core-nodes represent the most stable subset of nodes in a community and they should not move through time. Core nodes can be identified in different ways, by using centrality criterion, k-core decomposition, etc. Two communities are matched if they share the same core nodes.

\textbf{\cite{wang2008community}} proposed a method called CommTracker in which core nodes are distinguished from ordinary nodes based on both the community topology and the node weight. The authors defined community core, i.e., a significant group of nodes in a community, as each node $v$ in the community that satisfies \(\sum_{u\in neighbors} d_{v}-d_{u}>0\) (nodes with non-negative centrality). Community core nodes are then used to establish the evolving relationships among communities at successive snapshots. Two clusters are matched if their community cores share the highest similarity value.




\textbf{\cite{beiro2010visualizing}$^{*}$} apply their static submodular algorithm \cite{albert2002statistical}  for each snapshot, and to evaluate the similarity between communities in successive snapshots they count the coincident nodes in the central hub of both communities. To find a community’s central hub they used the $k$-core decomposition which consists of identifying the largest subgraph induced by the community nodes, in which each node has a degree equal or bigger than $k$.

\textbf{\cite{chen2010detecting} } introduced the notion of graph representatives and community representatives to detect and track community dynamics. Representatives of a graph at $t$ are nodes that also appear in the graph at $t-1$. Community representative is the node that has the minimum number of appearances in other communities of the same graph.
First, they find graph representatives in each snapshot, and they enumerate the communities in each graph using the graph representatives as the seeds
. They then use community representatives to establish the relationship between the communities from different time steps. At final, decision rules are used to decide the type of community dynamics.

Table \ref{tab:methods2} lists core-based methods and the key elements describing their rationale.

\begin{table}[!htbp]
\resizebox{\textwidth}{!}{%
\small
\begin{tabular}{|p{2cm}|p{3.5cm}||p{7.5cm}|}
\hline
\textbf{Category}                    & \textbf{Reference} & \textbf{Key elements} \\ \hline
\multirow{2}{*}{\textbf{Iterative}}  & \cite{wang2008community}                  & Core community, non negative centrality                     \\ \cline{2-3} 
                             & \cite{beiro2010visualizing}$^{*}$                   & Community central hubs, k-core decomposition, static submodular algorithm \cite{albert2002statistical}                     \\ \hline \multirow{2}{*}{\textbf{Multi-step}} & \cite{chen2010detecting}  & Graph representatives, community representatives, decision rules
                   \\ \hline
\end{tabular}}
\caption{Core-based methods and the key elements describing their rationale. New references added to the survey in \cite{rossetti2018community} are marked with "$^{*}$".}
\label{tab:methods2}
\end{table}

The methods by \textbf{\cite{morini2017revealing}} falls in this category (please refer to the original survey \cite{rossetti2018community} for more detail)

\subsubsection{Temporal Trade-off}\label{chp3:tradeoff}

In these approaches, communities at an instant $t$ are defined as a trade-off between optimal solution at $t$ and known past. In other words, communities at $t$ depend on the actual network topology at that time and the past network topology and/or past partition; It does not depend on future topological perturbations. The community detection process in such approaches can be summarized as follows:  
\begin{enumerate}
\item Detect communities on the initial network state;
\item For each temporal step $t$ that follows, detect communities at $t$ using graph at $t$ and \textit{past} information.
\end{enumerate}

In such a process, both dynamic network models (snapshot model and temporal network) can be used. When handling temporal networks, communities are usually updated by considering local modifications in the network. On the contrary, when handling snapshot models, communities are often calculated for the whole network.

Approaches in this category may be classified into two main categories according to how the communities are calculated: 
\begin{itemize}
\item Updating approaches which update communities found previously at each network evolution. Based on the strategy used to update communities, the authors in \cite{rossetti2018community} distinguished two subcategories: updating by using global optimization and updating by using local rules.  
\item Informed community detection (ICD) approaches which run communities from scratch at each network evolution while considering information in the previous steps. Two main strategies are used in these approaches: using multi-objective optimization or using network smoothing. 
\end{itemize}


\rule{\linewidth}{0.4pt}\par
\vspace{-0.25em} 
\noindent\textit{\textbf{Updating by Global Optimization}}\par 
\vspace{-0.75em} 
\noindent\rule{\linewidth}{0.4pt}\par


These approaches work as follows: first, the community structure at the initial network state is calculated. Then, the partition found at $t$ is used as a seed to initialize a global optimization process at $t+1$. Quality functions (like modularity, conductance, etc.) and heuristics (such as greedy methods, simulated annealing, spectral methods, etc.) used in the static case can be used here.

The method by \textbf{\cite{aynaud2010static}} is a typical example in this category. The authors proposed the Stabilized Louvain Method, a dynamic variant of Louvain \cite{blondel2008fast} to find coherent communities over time. First, they run the original version of Louvain on the first time step. Then, for the next time steps, they used the modified version of Louvain which consists of initializing the Louvain with the community structure found at the previous time step. Methods in \cite{he2015fast,chong2013incremental,wang2010mining} are also based on the same principle. 
All these methods use the Louvain algorithm to detect communities on each snapshot, they differ in the way to initialize this algorithm.

In the method by \textbf{\cite{he2015fast}$^{*}$}, 
the community structure of the first snapshot is initialized by using the Louvain algorithm. Then, at each time step $t$ a small network is constructed according to the network structure at $t$ and the community information at $t$-1, and then the Louvain algorithm is used to detect communities in this newly constructed graph.

\textbf{\cite{chong2013incremental}$^{*}$} proposed an incremental batch method that can handle large and complex network changes. Communities are derived from the initial network using the Louvain technique. For subsequent snapshots, the following technique is applied: first, nodes that are directly affected by network changes are initialized to singleton communities and all other nodes retain their previous community memberships, then the Louvain algorithm is applied to the resulting structure.

\textbf{\cite{wang2010mining}$^{*}$} reused the idea of core nodes to reduce the instability problem. Core vertices are defined as those that do not change communities if the same algorithm is repeatedly run on the same slightly modified network. The authors use the Louvain algorithm to detect communities on each snapshot, and they initialize this algorithm with the core vertices found in the previous step.

\textbf{\cite{aktunc2015dynamic}$^{*}$} extended the Smart Local Moving (SLM) algorithm defined by \cite{waltman2013smart} (a well-known static modularity based algorithm) to be incremental and dynamic. 
The proposed dynamic SLM (dSLM) modifies the initial community assignments of nodes in SLM by using the historical results of community detection: for each new node, singleton new communities are constructed and added to the previously found communities.

\textbf{\cite{ning2010incremental}$^{*}$} suggested an incremental spectral clustering method, based on the normalized cut. The method is capable of handling two kinds of dynamics: similarity change and insertion/deletion of data points, which are represented as incidence vector/matrix. The proposed algorithm is initialized by a standard spectral clustering. Then, it continuously updates the eigenvalue system and generates instant cluster labels, as the data set is evolving. 

\textbf{\cite{dinh2009towards}$^{*}$} proposed a graph-based method in which the modular structure found in the previous network state is used as a guide to adaptively identify modules in the next state. The method comprises two main steps: $(1)$ compress the network into a compact representation while preserving the modular structure and $(2)$ run module identification algorithm on the compact representation to update the modular structure. Modularity is used to quantify the strength of the modular structure. The CNM \cite{clauset2004finding} algorithm is used to detect the initial modular structure of the network. 

Table \ref{tab:methods3} lists updating methods using global optimization and the key elements describing their rationale.

\begin{table}[!htbp]
\resizebox{\textwidth}{!}{%
\small
\begin{tabular}{|p{4cm}||p{9cm}|}
\hline
\textbf{Reference} & \textbf{Key elements} \\ \hline
\cite{aynaud2010static}                  & Dynamic variant of Louvain algorithm \cite{blondel2008fast}, initialization with previous communities                     \\ \hline
\cite{he2015fast}$^{*}$                  & Louvain algorithm \cite{blondel2008fast}, constructed network based on the actual network state and previous communities                      \\ \hline
\cite{chong2013incremental}$^{*}$                  & Incremental batch method, large and complex changes, Louvain algorithm \cite{blondel2008fast}, singleton communities for nodes affected by changes                      \\ \hline
\cite{wang2010mining}$^{*}$                  & Core nodes, Louvain algorithm \cite{blondel2008fast}                     \\ \hline
\cite{aktunc2015dynamic}$^{*}$                 & An incremental and dynamic variant of SLM Algorithm \cite{waltman2013smart}, singleton communities for new nodes                     \\ \hline
\cite{ning2010incremental}$^{*}$                  & Incremental spectral method, Normalized cut, similarity change, insertion/deletion of data points, incidence vector/matrix, eigenvalue system                     \\ \hline
\cite{dinh2009towards}$^{*}$                  & Graph-based method, compact representation, Modularity, CNM algorithm \cite{clauset2004finding}                     \\ \hline
\end{tabular}}
\caption{Updating methods using global optimization and the key elements describing their rationale. New references added to the survey in \cite{rossetti2018community} are marked with "$^{*}$".}
\label{tab:methods3}
\end{table}

Other methods falling in this category can be found the original survey \cite{rossetti2018community} are: \textbf{\cite{alvari2014community,bansal2011fast,gorke2010modularity,miller2009continuous,shang2014real}}.
\newpage
\noindent\rule{\linewidth}{0.4pt}\par
\vspace{-0.25em} 
\noindent\textit{\textbf{Updating by Set of Rules}}\par
\vspace{-0.75em} 
\noindent\rule{\linewidth}{0.4pt}\par

These approaches define a set of rules to update the community structure according to each network change (node/edge apparition/vanishing).

\textbf{\cite{nguyen2011adaptive}} proposed QCA the Quick Community Adaptation, modularity based optimization algorithm for dynamic community detection. QCA requires basic communities which can be obtained at the first time step by running any static community detection method. For subsequent time steps, QCA uses the basic structure found at the previous step for updating the local optimal communities to adapt to network changes (node/edge addition/removal). The same authors introduced AFOCS \cite{nguyen2011overlapping} as an extension to QCA to deal  with overlapping communities. 

The algorithm by \textbf{\cite{cordeiro2016dynamic}$^{*}$} shares the same principle of QCA and AFOCS. The authors proposed 
a modified version of the Louvain algorithm to support incremental community structure changes when nodes or edges are added or removed from the network.
The proposed algorithm performs a local modularity optimization that maximizes the modularity gain function only for communities which are affected by the network change, keeping most of the communities of the previous snapshot unchanged.

HotTracker is a framework proposed by \textbf{\cite{bhat2014hoctracker}$^{*}$} to track community evolution in dynamic social networks. It identifies a preliminary community structure for the initial network state through a novel density-based overlapping community detection algorithm. 
Then for every new network state, it uses only active nodes (those that have caused the network to change) to adapt the preliminary community structure. It uses a log-based approach to map the evolution between communities at two successive snapshots. 

\textbf{\cite{marquez2020overlapping}} suggested a framework, called ADIS, for revealing overlapping communities. ADIS first locates a basic community partition on the first network snapshot, and then updates the community structure in each snapshot according to the changes of the network structure. The authors define updating strategies based on the optimization of conductance for each type of changes.

An incremental label propagation method for detecting the structures of communities in real time is propopsed by \textbf{\cite{pang2009realtime}$^{*}$}. The method attempts to deal with the network changes incrementally. First, each node will be allocated the label (group) number randomlly. Then the label will be changed based on the neighbors'labels. The node will be given the label which the majority of its neighbors have. 

\textbf{\cite{zakrzewska2016tracking}} extended their work in \cite{zakrzewska2015dynamic} 
and proposed an algorithm for dynamic seed set expansion, which maintains a local community over time by incrementally updating as the underlying graph changes. At the first phase, a static seed set expansion is applied to the initial graph. At the second phase, a stream of graph update is applied, and with each graph update, the algorithm updates the community.

\textbf{Wang et al. \cite{wang2018tracking}$^{*}$} proposed DOCET (Dynamic Overlapping Community Evolution Tracking) method. DOCET first detects the initial overlapping community structure based on node location analysis in the peak-valley structure of the topology potential field. Then it incrementally updates the dynamic community structure based on influence scope analysis in the topology potential field. Finally it tracks community evolution events based on the variation of core nodes in the topology potential field.

\textbf{\cite{samie2018change}$^{*}$} introduced a change-aware community detection framework for community detection which first detects the change type (gradual or abrupt), and then decides how to discover the communities of the current snapshot, with or without considering the information of the previous one(s). 

DEMON (Democratic Estimate of the Modular Organization of a Network) is a local-first method to community discovery \textbf{\cite{coscia2012demon} $^{*}$}. The method adopts a democratic  appraoch, in which each node votes for the existing communities in its local view of the network, i.e., its ego neighborhood, using a label propagation algorithm; and then, the local communities are merged into a global collection. DEMON runs in a streamed fashion considering incremental updates of the graph as they arrive in subsequent batches: as batches of new nodes and new links arrive, the new communities can be found by considering only the ego networks of the nodes affected by the updates (both new nodes and old nodes reached by new links).  

The authors in \textbf{\cite{hu2016local}$^{*}$} proposed a local method for discovering communities and their evolutionary behaviors in dynamic networks, called Local Dynamic Method for Community Evolution Track (LDM-CET). To discover the community structure for each timestep, they represent different types of change such as deleted edges and new created edges as nodes of change, and use the approximate personalized PageRank community finder to explore the local views of the nodes of change. The dynamic communities are obtained by combing the local views of the nodes of change so as to update the community structure. To track the evolutionary behaviors of communities, they construct a partial evolution graph which only contains the communities involved in evolution. 

The ARTISON algorithm by \textbf{\cite{cheraghchi2017toward}$^{*}$} is an incremental community detection method, which is inspired by the Adaptive Resonance Theory technique- a famous adaptive clustering model in neural networks. 
The proposed method can handle both low and abrupt change in network.

IncOrder \textbf{\cite{sun2014incorder}$^{*}$} is an incremental density-based method which contains two separate stages: online and offline stages. Based on a symmetric measure core-connectivity-similarity between pairs of adjacent nodes, the online stage builds an index structure called core-connected chain for dynamic networks. As the network changes, the method maintains the chain incrementally online. The offline stage extracts the community structure from the maintained chain.

\textbf{\cite{asadi2018incremental}$^{*}$} introduced an unsupervised machine learning algorithm for incremental detection of communities using a label propagation method, called incremental speaker listener propagation algorithm ISLPA. ISLPA can detect both overlapping and non overlapping communities incrementally after removing or adding a batch of nodes or edges over time.

\textbf{\cite{zhao2019incremental}$^{*}$} proposed an incremental method to detect communities in dynamic social networks.
The main idea is to detect communities at the initial network state, then to collect and analyze the incremental dynamic changes, i.e., all changes between two time steps, and finally to update communities incrementally. 
The method handles subgraph addition (including nodes and edges). Updating strategies are defined to derive communities at the current state according to the relationship between subgraph including incremental changes and the communities at previous time step.

\textbf{\cite{xu2020superspreaders}$^{*}$} suggested a two-stage method called EAS (error accumulation sensitive) for incrementally detecting communities in dynamic social networks. 
In the first stage, an error accumulation sensitive (EAS) incremental clustering algorithm is proposed.
The EAS algorithm first calculates the error accumulation degree. If it does not exceed a pre-defined threshold, the community structure of the current snapshot will be obtained by partially updating the community of the previous snapshot. Otherwise, the current snapshot will be totally re-partitioned with the method used for the initial snapshot (any algorithm can be used here).
In the second stage, a superspreaders and superblockers (SAS) based algorithm for identification of critical evolution events is developed. The SAS algorithm first identifies superspreaders and superblockers for each individual snapshot, and then uncovers the birth, merge, and growth of dynamic communities with the identified superspreader nodes, and the death, split, and shrink of dynamic communities with the detected superblocker nodes.

\textbf{\cite{nath2019detecting}$^{*}$} proposed InDEN algorithm to detect both intrinsic and disjoint communities from evolving networks. They introduced two concepts, (i) detection of density variation with time and (ii) intra-community strength between nodes and their neighbors to calculate the Affinity score. InDEN adopts an incremental process to discover community. Initially, the process starts with zero community. Then, with the arrival of the first edge, a community seed is formed with its nodes. Affinity score is used to assign any new incoming node to the community with the maximum score. A new community is formed if there exists community or communities with low or no Affinity score.

\textbf{\cite{shang2014real}$^{*}$} proposed a modularity-based algorithm to detect and track communities over time in incremental networks. This algorithm considers the network change as a sequence of new edges. It comprises two steps: First, the Louvain algorithm is applied to obtain an initial community structure. Then, according to the edges’ type, incremental updating strategies are used to track the dynamic communities. Each strategy should be able to increase the modularity of the community structure, if not, it should make the lost in modularity as low as possible.

Table \ref{tab:methods4} lists updating methods using a set of rules and the key elements describing their rationale.

\begin{table}[!htbp]
\resizebox{\textwidth}{!}{%
\small
\begin{tabular}{|p{4cm}||p{9cm}|}
\hline
\textbf{Reference} & \textbf{Key elements} \\ \hline
\cite{nguyen2011adaptive}                  & Modularity, community adaptation, local optimal communities                      \\ \hline
\cite{cordeiro2016dynamic}$^{*}$                  & Louvain extension, incremental changes, local modularity optimization                      \\ \hline
\cite{bhat2014hoctracker}$^{*}$                 & Density-based, overlapping communities, active nodes, log-based mapping                     \\ \hline
\cite{marquez2020overlapping}$^{*}$                  & Overlapping communities, conductance optimization                      \\ \hline
\cite{pang2009realtime}$^{*}$                  & Incremental Label propagation                     \\ \hline
\cite{zakrzewska2016tracking}                  & Seed expansion, stream graph updates                     \\ \hline
\cite{wang2018tracking}$^{*}$                  & Overlapping communities, node location analysis, influence scope analysis, core nodes variation                     \\ \hline
\cite{samie2018change}$^{*}$                  & Change-aware model, gradual and abrupt changes                     \\ \hline
\cite{coscia2012demon}$^{*}$                  & Local-first approach, democratic approach, ego-neighborhood, label propagation, streams                     \\ \hline
\cite{hu2016local}$^{*}$                 & Local method, approximate personalized PageRank, partial evolution graph                     \\ \hline
\cite{cheraghchi2017toward}$^{*}$                  & Adaptive resonance theory, low and abrupt changes                     \\ \hline
\cite{sun2014incorder}$^{*}$                  & Density-based method, core-connectivity measure, core connected chain, online and offline stages                     \\ \hline
\cite{asadi2018incremental}$^{*}$                  & Unsupervised  machine learning, speaker listener propagation, overlapping, batch changes                      \\ \hline
\cite{zhao2019incremental}$^{*}$                  & Incremental changes, subgraph addition                     \\ \hline
\cite{xu2020superspreaders}$^{*}$                  & Error accumulation sensitive degree, superspreader and superblocker nodes                     \\ \hline
\cite{nath2019detecting}$^{*}$                 & Intrinsic communities, density variation, intra-community strength, affinity score                     \\ \hline
\cite{shang2014real}$^{*}$                  & Modularity, Louvain algorithm \cite{blondel2008fast}                     \\ \hline
\end{tabular}}
\caption{Updating Methods using a set of rules and the key elements describing their rationale. New references added to the survey in \cite{rossetti2018community} are marked with "$^{*}$".}
\label{tab:methods4}
\end{table}

Other works falling in this category can be found in the original survey: \textbf{\cite{agarwal2012real,cazabet2011simulate,cazabet2010detection,duan2012incremental,falkowski2008studying,gorke2009dynamic,lee2014incremental,ma2013cut,nguyen2011overlapping,nguyen2011adaptive,rossetti2017tiles,xie2013labelrankt,zakrzewska2015dynamic,sun2010community}}

\noindent\rule{\linewidth}{0.4pt}\par
\vspace{-0.25em} 
\noindent\textit{\textbf{ICD by Multi-objective Optimization}}\par 
\vspace{-0.75em} 
\noindent\rule{\linewidth}{0.4pt}\par

Methods in this subcategory try to balance both partition quality and temporal partition coherence at the same time for each snapshot so that a partition found at $t$ represents the natural evolution of the one found at $t-1$. 
 
The work by \textbf{\cite{chakrabarti2006evolutionary}$^{*}$} is a typical example of methods in this category. The authors introduced the first evolutionary clustering method which aims to 
optimize two criteria: the snapshot quality which means that the clustering at any point in time should be of high quality; and the history cost which means that the clustering should not shift dramatically from one time step to the next. To find an optimal cluster sequence, they suggested to minimize the difference between the snapshot quality and the history cost at each time step $t$, with a relative weight $cp$: \(SQ(C_{t},M_{t}) - cp .HQ(C_{t-1},C_{t})\), where $C_{t}$ and $C_{t-1}$ are respectively partitions at $t$ and $t-1$, $M_t$ is the adjacency/similarity matrix at time $t$. Two instances of this framework are proposed: k-means and agglomerative hierarchical clustering.

FaceNet \textbf{\cite{lin2009analyzing} } is a probabilistic model which captures the community evolution by optimizing the cost function:  \(cost=\alpha CS + (1 - \alpha ) CT\), where $CS$ is the snapshot cost which measures the goodness of communities found at a given snapshot,  and $CT$ is the temporal cost which measures the coherence of the actual communities with respect to the previous ones. At each time step, the community structure is identified by using the mixture model proposed in \cite{yu2006soft}, and the cost function is then used to regularize the community structure at the current time based on the community structure at the previous snapshot.

\textbf{\cite{chi2007evolutionary}$^{*}$} proposed two evolutionary spectral clustering frameworks. The cost function is defined as a linear combination of two costs: 1) the snapshot cost ($SC$) which measures the quality of the current partition, where a higher snapshot cost means worse snapshot quality; 2) the temporal cost ($TC$)  which measures the temporal smoothness in terms of the goodness-of-fit of the current partition with respect to either historic network state or historic partition, where a higher temporal cost means worse temporal smoothness. The cost function is given as : \(cost=\alpha . CS +\beta . CT\), where \(0\leqslant\alpha, \beta\leqslant 1\) 
and \(\beta (=1-\alpha)\). In both frameworks, the snapshot cost $CS$ is measured by the clustering quality used to obtain the partition of the current network. The two frameworks differ in the definition of the temporal cost $TC$. In the first framework named preserving cluster quality (PCQ), the current partition is applied to historic data and the resulting cluster quality determines the temporal cost. In the second named preserving cluster membership (PCM), the temporal cost is expressed as the difference between the current partition and the historic partition.


\textbf{\cite{messaoudi2019multi}$^{*}$} proposed a multi-objective bat algorithm for discovering communities in dynamic networks optimizing Modularity and Normalized Mutual Information (NMI) as objective functions. The algorithm uses the Mean Shift algorithm to generate the initial population and avoid the random process by defining a new mutation operator. 

A similar work is proposed by \textbf{\cite{liu2020detecting}$^{*}$}. The authors proposed a multi-objective evolutionary algorithm, denoted as DECS, mainly to capture the evolving patterns of communities in dynamic social networks. They also used Modularity and Mutual information as objective functions. They developed a migration operator cooperating with the classic genetic operators (selection, crossover, and mutation operators) to search for inter-community connections, and adopt the genome representation in\cite{li2014community} to represent networks. 

EvoLeaders \textbf{\cite{gao2016evolutionary}$^{*}$} is an evolutionary community discovery algorithm based on leader nodes. 
First, the top leader algorithm by \cite{khorasgani2010top} is used to find the initial leader nodes and their communities at the first snapshot. This algorithm regards each community as a set of follower nodes congregating close to a potential leader, and a leader node as the most central node in the corresponding community. 
Then, for each subsequent snapshot an updating strategy, which is incorporated with temporal information, is used at first to get the initial leader nodes. Furthermore, a community splitting algorithm is used to isolate the nodes in the initial communities that are not contained in the corresponding communities at the last snapshot. Therefore, the resulted small communities are merged so that to improve the community quality. The quality measure used is a trade-off between Snapshot Modularity, which computes modularity for the communities detected at the current network, and History Modularity, which evaluates the current communities with respect to the previous snapshot. At the final step, the leader nodes are updated for each community.

\textbf{\cite{rozenshtein2014discovering}$^{*}$} proposed an alternating optimization method for finding dense dynamic communities in interaction networks. The goal of the proposed method is to find communities that satisfy the two requirements: dense interactions that occur within a number of short time intervals. The first requirement is formulated as the densest subgraph problem for finding the optimal set of nodes given a set of time intervals,  and the second is formulated as the maximum-coverage with multiple budgets (mcmb) problem i.e., finding the optimal set of time intervals for a given set of nodes. The method works in an alternating fashion, it starts from an initial time interval, and obtains a solution by iteratively solving the two problems until convergence.

A multiobjective evolutionary algorithm based on structural and attribute similarities (MOEA-SA) is proposed by \textbf{\cite{li2017multiobjective}$^{*}$} to handle attributed graph clustering problems. In MOEA-SA,  two objectives are used to be maximized in the algorithm: modularity and attribute similarity. A hybrid representation based on locus-based adjacency and character string representations is used to make full use of the relationships between vertices. A multi-individual-based mutation operator is used with a neighborhood correction strategy. The hill-climbing strategy is applied to MOEA-SA to speed up the convergence of modularity.

An intimacy evolutionary clustering algorithm is suggested by \textbf{\cite{chen2020community}$^{*}$}. Firstly, the time-weighted similarity matrix is utilized and calculated to grasp time variation during the community evolution. Secondly, the differential equations are adopted to learn the intimacy evolutionary behaviors. During the interactions, intimacy between two nodes would be updated based on the iteration model. Nodes with higher intimacy would gather into the same cluster and nodes with lower intimacy would get away, then the community structure would be formed in dynamic networks. 

\textbf{\cite{jiao2018constrained}$^{*}$} proposed a constrained common cluster-based model (C3 model) to analyze and explore community structure and common cluster structure hidden in the temporal or multiplex networks.  They first construct the Markov steady-state matrices of each snapshot of the temporal network or each slice of the multiplex network. Next, they propose the object function of C3 model by combining the Markov steady-state matrices, similarity matrices, and community membership matrices of the network in a theoretical way. Finally, a gradient descent algorithm based on non-negative matrix factorization is proposed for optimizing the objective function. 

\textbf{\cite{said2018cc}$^{*}$} proposed a  Clustering Coefficient-based Genetic Algorithm (CC-GA) which is an evolutionary algorithm. The method generates the initial population based on the clustering coefficient. The Modularity measure is used as a fitness function to assess the quality of the population. Uniform crossover and random mutation are used as genetic operators.

Table \ref{tab:methods5} lists methods falling in the category of ICD by multi-objective optimization and the key elements describing their rationale.

\begin{table}[!htbp]
\resizebox{\textwidth}{!}{%
\small
\begin{tabular}{|p{4cm}||p{9cm}|}
\hline
\textbf{Reference} & \textbf{Key elements} \\ \hline
\cite{chakrabarti2006evolutionary}$^{*}$                 & Evolutionary  clustering, snapshot quality cost, history cost, adjacency/similarity matrix, k-means, agglomerative hierarchical clustering                    \\ \hline
\cite{lin2009analyzing}                  & Probabilistic model, snapshot cost, temporal cost, mixture model                      \\ \hline
\cite{chi2007evolutionary} $^{*}$                 & Evolutionary  spectral clustering, snapshot cost, temporal cost, historical network state, historical partition, preserving cluster membership, preserving cluster quality                     \\ \hline
\cite{messaoudi2019multi}$^{*}$                  & Bat algorithm, Modularity, Normalized Mutual Information, Mean Shift algorithm                    \\ \hline
\cite{liu2020detecting}$^{*}$                  & Modularity, Normalized Mutual Information, migration operator, genome representation                     \\ \hline
\cite{gao2016evolutionary}$^{*}$                  & Leader nodes, Snapshot Modularity, History Modularity                     \\ \hline
\cite{rozenshtein2014discovering}$^{*}$                  & Alternating optimization, densest subgraph, maximum coverage with multiple budget                     \\ \hline
\cite{li2017multiobjective}$^{*}$                 & Modularity, attribute similarity, Locus-based adjacency, character string, Multi-individual mutation operator, neighborhood correction strategy, hill-climbing strategy                       \\ \hline
\cite{chen2020community}$^{*}$                  & Intimacy evolutionary behaviours, weighted similarity matrix, differential equations, iteration model                     \\ \hline
\cite{jiao2018constrained}$^{*}$                  & Constraint  common cluster, Markov steady-state matrices, similarity matrices, community membership matrices, gradient descent algorithm, non-negative matrix factorization                     \\ \hline
\cite{said2018cc}$^{*}$                 & Genetic algorithm, clustering coefficient, Modularity, Uniform crossover, random mutation                     \\ \hline
\end{tabular}}
\caption{Methods falling in the category of ICD by Multi-objective Optimization and the key elements describing their rationale. New references added to the survey in \cite{rossetti2018community} are marked with "$^{*}$".}
\label{tab:methods5}
\end{table}

Other methods in this category can be found in the original survey: \textbf{\cite{crane2015community,folino2010multiobjective,gong2012community,gorke2013dynamic,kawadia2012sequential,lin2008facetnet,lin2009analyzing,tang2010graph,yang2009bayesian,zhou2007discovering}}.
\newpage
\noindent\rule{\linewidth}{0.4pt}\par
\vspace{-0.25em} 
\noindent\textit{\textbf{ICD by network smoothing}}\par 
\vspace{-0.75em} 
\noindent\rule{\linewidth}{0.4pt}\par

Methods in this subcategory first search for communities at $t$ by running the community detection algorithm, not on the graph as it at $t$, but on a version of it that is smoothed according to the past network evolution (e.g., by adding weights to track edges). Then, communities are usually matched between snapshots. Contrary to the Instant-optimal methods in which the matching is based on the previous partition, the matching here is based on the previous network state. 

\textbf{\cite{kim2009particle} }proposed a new particle-and-density based evolutionary clustering method. They model a dynamic network as a collection of lots of particles, and a community as a densely connected subset of particles. 
Each particle contains a small amount of information about the evolution of data or communities. 
They propose a density-based clustering method that efficiently finds temporally smoothed local clusters of high quality by using a cost embedding technique and optimal modularity. 
The proposed cost embedding technique performs smoothing at the data level instead of at the clustering result level.  
They also propose a mapping method based on information theory.  

The method by \textbf{\cite{xu2013analyzing,xu2013community}} detects the stable community core in mobile social networks. The main idea of this method is to find a partition based on stable links in a given network. Two main concepts are used: 

\begin{itemize}
\item \textit{Cumulative Stable Contact (CSC):} a CSC exists between two nodes if and only if their history contact duration is higher than a threshold.
\item \textit{Community core:} which is resulted in partitioning a reduced network containing only useful links.
\end{itemize}
The whole process is divided into timestamps. Nodes and their connections can be added or removed at each timestamp, and historical contacts are considered when detecting the community core. Also, community cores can be tracked through incremental computing which can help to recognize the evolution of community structure.

\textbf{\cite{guo2014evolutionary} }proposed an Evolutionary Community Structure Discovery (ECSD) algorithm for weighted networks. First, evolutionary matrices are built as the input which considers both the adjacency matrix and the community structure in previous time steps, and then an initial community with a node whose node strength is maximum is discovered. Afterward, the community is expanded by adding nodes that can improve the quality of the community. Finally, the communities whose numbers of nodes are smaller than a threshold are merged to improve the total quality.

The authors in \textbf{\cite{guo2016dynamic}$^{*}$} proposed a dynamic community detection algorithm based on distance dynamics. The network increments are treated as disturbances of the network. The idea behind this algorithm is to limit the range of disturbance inside a small neighbor in order to reduce the number of iterations and to speed-up the convergence of the local interaction model during the increment community detection in dynamic networks. The Attractor algorithm by \cite{shao2015community} is firstly used to detect communities at the first snapshot. When the network increments come, the disturbed edges are added to the candidate set. And the local interaction model is used to make iterative computations on the candidate set which includes added and removed nodes and edges. The distances are updated iteratively to achieve community detection.

\textbf{\cite{zeng2019consensus}$^{*}$} suggested a framework of consensus community based on PSO (Particle Swarm Optimization) called CCPSO. The consensus community is introduced by extracting the existing common communities in the current population and the population in the previous step. Then, based on the consensus community, the community structure of the current step can evolve toward the direction which is similar to the structure of the previous step for the sense of continuity.

Table \ref{tab:methods6} lists methods falling in the category of ICD by network smoothing and the key elements describing their rationale.

\begin{table}[!htbp]
\resizebox{\textwidth}{!}{%
\small
\begin{tabular}{|p{4cm}||p{9cm}|}
\hline
\textbf{Reference} & \textbf{Key elements} \\ \hline
\cite{kim2009particle}                  & Particle and density based clustering, information theory based mapping                     \\ \hline
\cite{xu2013analyzing,xu2013community}                  & Stable community core, cumulative stable contact, incremental computing                      \\ \hline
\cite{guo2014evolutionary}                  & Weighed networks, evolutionary matrices, node strength                      \\ \hline
\cite{guo2016dynamic}$^{*}$                  & Distance dynamic, network disturbance, Attractor algorithm \cite{shao2015community}, local interaction model, iterative computing                      \\ \hline
\cite{zeng2019consensus}$^{*}$                  & Consensus community, particle swarm optimization                      \\ \hline
\end{tabular}}
\caption{Methods falling in the category of ICD by network smoothing and the key elements describing their rationale. New references added to the survey in \cite{rossetti2018community} are marked with "$^{*}$".}
\label{tab:methods6}
\end{table}
\subsubsection{Cross-Time}

Approaches that fall in this category do not consider independently the different steps of the network evolution,  all steps of evolution are studied simultaneously. Communities are detected in a one-stage process, considering in a single pass all steps of network evolution and yielding a single temporal community decomposition.  

Methods in the Cross-time class differ on how to search communities. \cite{rossetti2018community} classify them based on two constraints: 
\begin{enumerate}
\item Evolution of node memberships: if nodes can switch between communities along time.
\item Evolution of community properties: if communities can appear or disappear along the studied period or not.
\end{enumerate}
Based on the above constraints, four categories can be distinguished: $(i)$ Fixed memberships; fixed properties $(ii)$ Fixed memberships; evolving properties $(iii)$ Evolving memberships; fixed properties $(iv)$ Evolving memberships; evolving properties.

\noindent\rule{\linewidth}{0.4pt}\par
\vspace{-0.25em} 
\noindent\textit{\textbf{Fixed memberships, fixed properties}}\par 
\vspace{-0.75em} 
\noindent\rule{\linewidth}{0.4pt}\par

Methods in this category assume that communities remain unchanged throughout the studied period: nodes can not change their memberships and communities can not appear or disappear. In doing so, they seek the best partition, on average, over a period of time. To improve the solution, most methods suggest splitting the network evolution into homogeneous periods of time by detecting dramatic changes in the networks. 

\textbf{\cite{aynaud2011multi}} proposed a framework for detecting a unique community partition that is relevant for almost every time step during a given period, called the time window. To detect such partition, the authors proposed two modularity-based methods. The first method consists of building a sum network, which is a weighted graph representing the union of all snapshots in a given time window $T$. Each edge of the sum graph is weighted by the total time during which this edge exists in $T$. Then, they apply Louvain a static community detection method on that network. The second method consists in defining the average modularity over a set of time steps (time window $T$), as follows: \(Q_{avg}(G,\pi,T) =\frac{1}{n}\sum_{t \in T} Q(G_{t},\pi)\). This average modularity is then optimized by modifying the Louvain method as follows : 

\begin{enumerate}
\item Redefine the quality gain in the first phase as the average of the static gains for each snapshot of $T$.
\item Change the way to build the super graph in the second one as follows: given a partition $\pi$, the same transformation as for the Louvain is applied on every snapshot of $T$ independently (with different weights for each snapshot) to obtain a new evolving network between the communities of $\pi$.   
\end{enumerate}

The authors in \textbf{\cite{sun2007graphscope} }proposed GraphScope method for discovering communities and monitoring their changes in directed bipartite stream graphs. The method is based on the Minimum Description Length (MDL) principle by \cite{grunwald2007minimum}. 
The primary idea underlying this work is to find the minimum encoding cost for the description of a time sequence of graphs (called graph segments) and their partitions in communities. To do so, an incremental process is proposed which first constructs graph segments by grouping similar consecutive snapshots so that the encoding cost is minimized. Then, within each segment, finds the best partition of source/destination nodes that yields a smaller encoding cost. The change point remarks the beginning of the new graph segment.

Using the same principle of MDL, \textbf{\cite{tan2014online}$^{*}$} proposed a method to discover the community transition for individual users in dynamic networks. It starts by constructing a trajectory to represent the evolution of communities, then a trajectory segmentation approach is proposed to discover the best partition that yields minimum encoding cost. 

Table \ref{tab:methods7} lists methods in the category of "fixed memberships, fixed properties" and the key elements describing their rationale.

\begin{table}[!htbp]
\resizebox{\textwidth}{!}{%
\small
\begin{tabular}{|p{4cm}||p{9cm}|}
\hline
\textbf{Reference} & \textbf{Key elements} \\ \hline
\cite{aynaud2011multi}                  & Time window, Average Modularity, sum network, Louvain algorithm \cite{blondel2008fast},                        \\ \hline
\cite{sun2007graphscope}                  & Directed bipartite stream graph, Minimum Description Length(MDL), graph segments, change points                      \\ \hline
\cite{tan2014online}$^{*}$                  & Minimum Description Length(MDL), community trajectory, segmentation                      \\ \hline
\end{tabular}}
\caption{Methods in the category of "fixed memberships, fixed properties" and the key elements describing their rationale. New references added to the survey in \cite{rossetti2018community} are marked with "$^{*}$".}
\label{tab:methods7}
\end{table}

Another work which falls in this subcategory is proposed by \textbf{\cite{duan2009community} }(the principal of this work can be found in the survey by \cite{rossetti2018community})

\newpage
\noindent\rule{\linewidth}{0.4pt}\par
\vspace{-0.25em} 
\noindent\textit{\textbf{Fixed memberships, evolving properties}}\par 
\vspace{-0.75em} 
\noindent\rule{\linewidth}{0.4pt}\par


These methods do not allow nodes to switch communities. They also assume that communities are not homogeneous throughout the studied period of time. The activity within a community can increase or decrease over time, for example, nodes can interact more actively during some recurrent periods. They assign to each community a temporal profile that corresponds to the evolution of their activity. 

\textbf{\cite{gauvin2014detecting}}  used non-negative tensor factorization to extract the community-activity structure of temporal networks. The method allows to simultaneously identify communities and to track their activity over time. The temporal network is represented as a time-ordered sequence of adjacency matrices, each one describing the state of the network at a given point in time. The adjacency matrices are combined in a three-way tensor. They propose a non-negative factorization technique to the tensor. This technique takes as input the desired number of components (communities) and returns as outputs two matrices: the first gives the membership weight of nodes to the different components and the second gives the activity level of components at different snapshots.

The recent work by\textbf{ \cite{sarantopoulos2018timerank}$^{*}$}
proposed a similar method, called TimeRank. 
The dynamic network is represented as a three-dimensional tensor where the first two dimensions refer to the nodes and the third dimension to the snapshots as relation. 
Then,  "inter-timeframe edges" are introduced to connect a node with its image in other snapshots. The reasoning to add this kind of edge is to allow dynamic patterns to be more easily unveiled. A random walk clustering is performed on the tensor to discover a predefined number of dynamic communities. 

Table \ref{tab:methods8} lists methods in the category of "fixed memberships, evolving properties" and the key elements describing their rationale.

\begin{table}[!htbp]
\resizebox{\textwidth}{!}{%
\small
\begin{tabular}{|p{4.1cm}||p{8.9cm}|}
\hline
\textbf{Reference} & \textbf{Key elements} \\ \hline
\cite{gauvin2014detecting}                  & Non-negative factorization, time adjacency matrix, three-way tensor                     \\ \hline
\cite{sarantopoulos2018timerank}$^{*}$                  & Three dimensional tensor, inter-timeframe edges, random walk clustering                     \\ \hline
\end{tabular}}
\caption{Fixed memberships, evolving properties methods and the key elements describing their rationale.}
\label{tab:methods8}
\end{table}

Other methods in this subcategory can be found in the original survey:\textbf{\cite{gauvin2014detecting,matias2017statistical}}.

\noindent\rule{\linewidth}{0.4pt}\par
\vspace{-0.25em} 
\noindent\textit{\textbf{Evolving memberships, Fixed properties}}\par 
\vspace{-0.75em} 
\noindent\rule{\linewidth}{0.4pt}\par


These methods permit nodes to switch between communities. Because they use stochastic block models, the number of communities and their density is fixed for the whole studied period.

\textbf{\cite{yang2009bayesian}} used a dynamic stochastic block model for modeling communities and their evolutions in a unified probabilistic framework. 
In the static SBM model, a network is generated in the following way: first, each node is assigned to a community following a probability \(\pi\); then, links between nodes are generated following a Bernoulli distribution with parameter $P$. The Dynamic Stochastic Block Model (DSBM) and SBM differ in how the community assignments are determined. In the DSBM model, instead of following a prior distribution π, the community assignments at any time are determined by those at the previous time through a transition matrix that aims to capture the dynamic evolutions of communities. For parameter estimation, Bayesian inference is used to compute the posterior distributions for all the unknown parameters. Two versions of the inference method are introduced, i.e., an online learning version (a Temporal Trade-off method) that iteratively updates the probabilistic model over time, and an offline learning version (Cross-time method) that learns the probabilistic model with network data obtained at all time steps.

\textbf{\cite{ludkin2018dynamic}$^{*}$} proposed a dynamic extension to the stochastic block model which includes autoregressive terms, named the autoregressive stochastic block model (ARSBM). To allow block membership to evolve in time, the authors assume that the community membership of a node follows a continuous-time Markov Chain (CTMC) which means that a node spends an exponentially distributed time in the community before moving to a new community. A reversible jump Markov chain Monte Carlo (RJMCMC) algorithm was proposed to infer the changing block membership of nodes in the ARSBM. 

Table \ref{tab:methods9} lists methods in the category of "evolving memberships, fixed properties" and the key elements describing their rationale.

\begin{table}[!htbp]
\resizebox{\textwidth}{!}{%
\small
\begin{tabular}{|p{4cm}||p{9cm}|}
\hline
\textbf{Reference} & \textbf{Key elements} \\ \hline
\cite{yang2009bayesian}                  & Dynamic Stochastic Block Model, transition matrix, Bayesian inference, online learning, offline learning                     \\ \hline
\cite{ludkin2018dynamic}$^{*}$                  & Autoregressive Stochastic Block Model, Continuous Time Markov Chain                      \\ \hline
\end{tabular}}
\caption{Evolving memberships, fixed properties methods and the key elements describing their rationale.}
\label{tab:methods9}
\end{table}

Other methods in this category can be found in the original survey:\textbf{\cite{ghasemian2016detectability,herlau2013modeling,ishiguro2010dynamic,Matias2015EstimationAC,xu2014dynamic,yang2009bayesian,yang2011detecting}}.
\newline
\noindent\rule{\linewidth}{0.4pt}\par
\vspace{-0.25em} 
\noindent\textit{\textbf{Evolving memberships, Evolving properties}}\par 
\vspace{-0.75em} 
\noindent\rule{\linewidth}{0.4pt}\par

Methods in this category do not set out any restrictions on the way to search communities: nodes can change their memberships and communities can appear or disappear and they can change their activity in time. 

\textbf{\cite{jdidia2007communities}} introduced the first method in the class of Cross-time approaches. The evolving network is viewed as a single evolving graph where edges are defined as follows:

\begin{itemize}
\item There is an edge between the node $(i, t)$ and the node $(j,t+1)$ if they have a common neighbour $k$ ($i$ and $k$ are neighbours at  $t$ and $k$ and $j$ are neighbours at $t + 1$). 
\item There is an edge between $(i, t)$ and $(j, t)$ if $i$ and $j$ are neighbours at time $t$; 
\item There is an edge between the same node $(i, t)$ and $(i, t + 1)$  if $i$ is present at $t$ and $t + 1$.
\end{itemize}
Communities are identified by applying the WALKTRAP algorithm (a static method based on random walk clustering )  \cite{pons2005computing} on the constructed graph.

\begin{figure*}[h!]
\centering
\includegraphics[scale = 1.5] {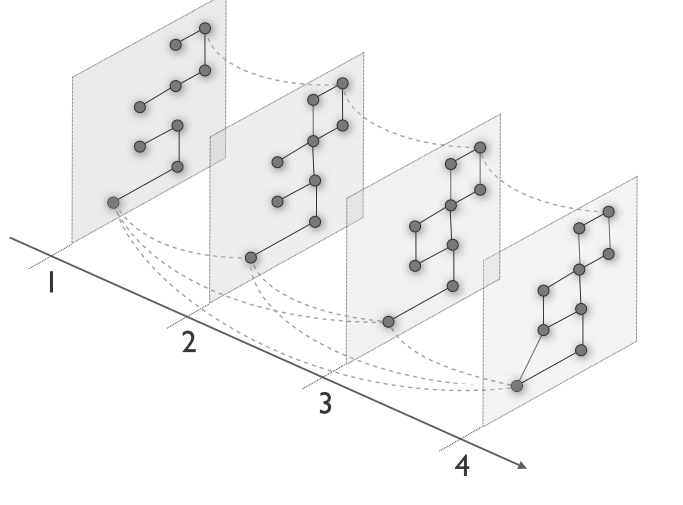}
\caption{Example of a multi-layer network. Four slices $S=\{1, 2, 3, 4\}$ represented by adjacencies $A_{ijs}$ encode intra-slice connections (solid). Inter-slice connections (dashed) are encoded by $C_{jrs}$, specifying coupling of node j to itself between slices $r$ and $s$. For clarity, inter-slice couplings are shown for only two nodes and depict
two diﬀerent types of couplings: $(1)$ coupling between neighboring slices, appropriate for
ordered slices; and $(2)$ all-to-all inter-slice coupling, appropriate for categorical slices. Reprinted from \cite{mucha2010community}}
\label{fig:slices}
\end{figure*}

\textbf{\cite{mucha2010community}} adopted a modularity based 
method for multi-layer networks. The method is general, and it can handle different multi-layer networks: multi-relational networks, dynamics networks, etc. The idea is as follows: 
\begin{enumerate}
\item First, they build a multi-slice network that encompasses the variation of connections (e.g. dynamic interactions) (see Figure \ref{fig:slices}). Each layer has an adjacency matrix describing connections between nodes belonging to the previously considered slice.
\item Then, they use the multi-slice generalization of modularity on the resulting network. The authors introduced a coupling parameter that links nodes across network slices.
 The new modularity is given as:   
\[Q_{multislice}=  \frac{1}{2\mu} \sum_{i j s r} \left( \left( A_{i j s} - \gamma _s  \frac{d_{i s} d_{j s}}{2m_s}\right) + \delta _{i j} C_{jsr}\right)\delta (c_{is},c_{jr}),  \]
\end{enumerate}
where: $2\mu= \sum_{jr} d_{jr} $ is a normalization factor, $A_{i j s}$  is the adjacency between $i$ and $j$ in the slice $s$ of the network, $\gamma _s $ the resolution parameter of the slice $s$, $d_{i s}=\sum_j A_{ijs}$ is the strength (or the degree) of node $i$ in slice $s $, $m_s=\sum_{ij} A_{ijs}$ is the total strength (or links) in slice $s$, $C_{jsr}$ is the inter-slice couplings that connect node $j$ in slice $r$ to itself in slice $s$. which is supposed to take binary values $\{0,\omega\}$ indicating absence/presence of inter-slice links, $c_{is}$ indicates that community assignment of node $i$ in slice $s$.

This method considers flexible constraints on node memberships and community properties. Nodes can switch communities 
, because nodes in different layers are assigned to separate communities. Communities can evolve through network slices.

The method by \textbf{\cite{Viard:2016:CMC:2853249.2853730}} used the stochastic block models to discover communities in link streams. Its principle is to find groups of nodes that interact similarly between themselves or with other groups of nodes during one or several periods of time, without constraint on the frequency of those interactions, but only on their similarity. 

\textbf{\cite{ma2017evolutionary}$^{*}$} proposed the sE-NMF by combing Evolutionary Nonnegative Matrix Factorization (E-NMF) and spectral clustering in which semi-supervision is incorporated into the objective of ENMF. The dynamic network is represented by a 3-dimensional adjacency matrix: the two first dimensions represent vertices and the third one represents time. sE-NMF consists of two major components: $(i)$ discovering local clusters based on the temporal smoothness framework using a priori information, and $(ii)$ mapping local clusters to identify dynamic communities using a $T$-partite graph ($T$ is the number of snapshots). The number of local clusters(communities) is given as input of the sE-NMF algorithm.

Table \ref{tab:methods10} lists methods in the category of "evolving memberships, evolving properties" and the key elements describing their rationale.

\begin{table}[!htbp]
\resizebox{\textwidth}{!}{%
\small
\begin{tabular}{|p{4cm}||p{9cm}|}
\hline
\textbf{Reference} & \textbf{Key elements} \\ \hline
\cite{jdidia2007communities}                  & Single evolving graph, Walktrap algorithm \cite{pons2005computing}                     \\ \hline
\cite{mucha2010community}                  & Multi-layer networks, multi-slice network, multi-slice modularity                     \\ \hline
\cite{Viard:2016:CMC:2853249.2853730}                  & Stochastic block models, link stream                               \\ \hline
\cite{ma2017evolutionary}$^{*}$                  & Evolutionary  non-negative matrix factorization, spectral clustering,  3-dimensional adjacency matrix, local clusters                     \\ \hline
\end{tabular}}
\caption{Evolving memberships, evolving properties methods and the key elements describing their rationale. New references added to the survey in \cite{rossetti2018community} are marked with "$^{*}$".}
\label{tab:methods10}
\end{table}

Other methods in this subcategory can be found in the original survey:\textbf{\cite{Himmel2016EnumeratingMC,bassett2013robust}}.

\subsection{Advantages and drawbacks of dynamic community detection approaches}

In this survey, we have listed three main classes of approaches for dynamic community detection: Instant-optimal approaches, Temporal Trade-off approaches, and Cross-time approaches. 
Each class of approaches has its own advantages and drawbacks. 
A summary of the major strengths and limitations of each class is provided in Table \ref{tab:CompareClasses}, and the details will be discussed hereafter.

\begin{table}[!htbp]
\resizebox{\textwidth}{!}{%
\small
\begin{tabular}{|p{0.25\linewidth} || p{0.5\linewidth}| p{0.5\linewidth}|}
\hline
\textbf{Class}               & \textbf{Advantages}                                                                                                                                                                                                                                                                        & \textbf{Drawbacks}                                                                                                                                                                                                                             \\ 
\hline
\textbf{Instant-optimal}     & \begin{itemize} 
            \item Reuse of static algorithms and matching techniques
            \item Ease of parallelization 
        \end{itemize} & \begin{itemize} 
            \item Instability of static algorithms
            \item High complexity
            \item It works only on snapshot model
        \end{itemize}                               \\ 
\hline
\textbf{Temporal Trade-off } & \begin{itemize} 
            \item Stability of the community detection 
            \item Updating approaches have low complexity
            \item Approaches working on temporal networks can handle all the network evolution
        \end{itemize}& \begin{itemize} 
            \item Methods calculating communities from scratch have high complexity
            \item It is difficult to parallelize
            \item Community drift
        \end{itemize} \\ 
\hline
\textbf{Cross-time}          & \begin{itemize} 
            \item It copes with the instability and the community drift problems
        \end{itemize}& \begin{itemize} 
            \item It works only in an offline fashion
        \end{itemize} \\
\hline
\end{tabular}
}
\caption{A comparison of the advantages and drawbacks of the three classes.}
\label{tab:CompareClasses}
\end{table}

\textbf{\textit{Instant-optimal approaches: }}The key strength of Instant-optimal approaches is that it reuses existing static community detection methods and matching techniques. 
The main drawback of these approaches is related to the instability of static algorithms. Indeed, most of them are often non-deterministic, so the same algorithm run on the same (or a slightly modified) graph can yield different communities. Thus, it is not possible to distinguish between changes due to the evolution of the community structure and changes due to the instability of algorithms. Besides, these approaches have a very high complexity due to the high number of runs of the detection and matching. However, since the communities at each step are detected independently, it is possible to run the detection for multiple steps in parallel and this could significantly reduce their temporal complexity. Methods in this category work only on the snapshot network model which fails to capture the full temporal evolution of the network.

\textbf{\textit{Temporal Trade-off approaches:}}
These approaches tackle the issue of instability that affects Instant-optimal approaches while keeping a similar principle: searching partitions at each time step. Approaches updating communities at each network evolution has the advantage of being very fast. Some approaches in this category enable handling all the evolution steps in the network. This could be very important when handling highly dynamic networks.
Despite all these advantages, these approaches still have some weaknesses. For example, methods calculating communities from scratch at each network evolution have high complexity. Another shortcoming is the difficulty of parallelization since the communities at each step are based on communities found in a previous step.
Another drawback of these methods is the absence of any guarantee that the communities found represent an optimal solution at the global level. More precisely, these methods suffer from the risk of community drift, in which the solution can be dragged away from an originally relevant one. Another consequence is that communities found by these algorithms at step $t$ depend on the particular sequence of previous graph modifications: the same graph produced by a different graph’s history would yield a different partition.

\textbf{\textit{Cross-time approaches:}} 
Approaches in this category do not have the problems of instability and community drift affecting the two categories of approaches mentioned above. The limitation of these approaches is that it is not possible to handle a real-time community detection since the computation of communities at each new network modification requires all the topological history of the network.

\subsection{Positioning our contributions with respect to the state of the art}\label{chp3:Discussion}

From this review of the literature, we can highlight significant research gaps that still need to be addressed on the topic of dynamic community detection, especially when dealing with social networks. 
In this section, we will show how our contributions are the answer to the issues we have observed in this review. 
To do so, we use Table \ref{tab:meth0} which summarizes the comparison between some methods of dynamic community detection. Given the large number of methods being presented in this survey, the comparison is restricted to only some representative methods from each class of approaches. The two methods we propose in this thesis are also taken into consideration for comparison (see the two last references  highlithed in bold). The following comparison criteria are used here:
\begin{itemize}
\item \textit{Approach:} indicates the class of approaches to which the method belongs: \textbf{IO} (Instant Optimal), \textbf{TTO} (Temporal Trade-off) or \textbf{CT} (CrossTime). 
\item Network model: refers to the type of the dynamic network model used by the method: \textbf{{\textcolor{red}{SN}}} (Snapshot Network), \textbf{{\textcolor{green!80!black}{TN}} }(Temporal networks).
\item \textit{Stability:} indicates if the method suffers (\xmark) or not (\cmark) from the Instability problem.
\item \textit{No drifting:} 
indicates if the method suffers (\xmark) or not (\cmark) from the community drift problem.
\item \textit{Overlapping: }indicates if the method can detect overlapping community structures (\cmark) or not (\xmark).
\item \textit{Multi-temporal scale: }indicates if the method can detect community structures at multiple temporal scales (\cmark) or only on one temporal scale (\xmark).
\end{itemize}

\begin{table}[!htbp]
\resizebox{\textwidth}{!}{%
\scriptsize
\begin{tabular}{|p{3.6cm}||p{1.4cm}|p{1.2cm}|p{1.4cm}|p{1.7cm}|p{1.7cm}|p{1.6cm}|}
\hline
\textbf{Reference} & \textbf{Approach}& \textbf{Network Model}& \textbf{Stability}& \textbf{No drifting}& \textbf{Overlapping}& \textbf{Multi-temporal scale} \\ \hline
\cite{asur2009event}& 
\textbf{IO}& 
\textbf{{\textcolor{red}{SN}}}& 
\xmark&
\cmark&
\xmark&
\xmark
\\ \hline
\cite{hopcroft2004tracking}& 
\textbf{IO}& 
\textbf{{\textcolor{red}{SN}}}& 
\cmark&
\cmark&
\xmark&
\xmark
\\ \hline
\cite{palla2007quantifying}& 
\textbf{IO}& 
\textbf{{\textcolor{red}{SN}}}& 
\cmark&
\cmark&
\cmark&
\xmark
\\ \hline
\cite{tajeuna2015tracking}& 
\textbf{IO}& 
\textbf{{\textcolor{red}{SN}}}& 
\xmark&
\cmark&
\xmark&
\xmark
\\ \hline
\cite{wang2008community}& 
\textbf{IO}& 
\textbf{{\textcolor{red}{SN}}}& 
\cmark&
\cmark&
\xmark&
\xmark
\\ \hline
\cite{chen2010detecting}& 
\textbf{IO}& 
\textbf{{\textcolor{red}{SN}}}& 
\cmark&
\cmark&
\xmark&
\xmark
\\ \hline
\cite{aynaud2010static}& 
\textbf{TTO}& 
\textbf{{\textcolor{red}{SN}}}& 
\xmark&
\xmark&
\xmark&
\xmark
\\ \hline
\cite{aktunc2015dynamic}& 
\textbf{TTO}& 
\textbf{{\textcolor{red}{SN}}}& 
\xmark&
\xmark&
\xmark&
\xmark
\\ \hline
\cite{nguyen2011adaptive}& 
\textbf{TTO}& 
\textbf{{\textcolor{green!80!black}{TN}}}& 
\xmark&
\xmark&
\xmark&
\xmark
\\ \hline
\cite{chakrabarti2006evolutionary}& 
\textbf{TTO}& 
\textbf{{\textcolor{red}{SN}}}& 
\cmark&
\xmark&
\xmark&
\xmark
\\ \hline
\cite{xu2013analyzing,xu2013community}& 
\textbf{TTO}& 
\textbf{{\textcolor{green!80!black}{TN}}}& 
\cmark&
\xmark&
\xmark&
\xmark
\\ \hline
\cite{aynaud2011multi}& 
\textbf{CT}& 
\textbf{{\textcolor{red}{SN}}}& 
\cmark&
\cmark&
\xmark&
\xmark
\\ \hline
\cite{xu2013analyzing,xu2013community}& 
\textbf{CT}& 
\textbf{{\textcolor{red}{SN}}}& 
\cmark&
\cmark&
\xmark&
\xmark
\\ \hline
\cite{ludkin2018dynamic}& 
\textbf{CT}& 
\textbf{{\textcolor{red}{SN}}}& 
\cmark&
\cmark&
\xmark&
\xmark
\\ \hline
\cite{mucha2010community}& 
\textbf{CT}& 
\textbf{{\textcolor{red}{SN}}}& 
\cmark&
\cmark&
\xmark&
\xmark
\\ \hline
\cite{jdidia2007communities}& 
\textbf{CT}& 
\textbf{{\textcolor{red}{SN}}}& 
\cmark&
\cmark&
\xmark&
\xmark
\\ \hline 
\textbf{
\cite{boudebza2018}}& 
\textbf{TTO}& 
\textbf{{\textcolor{green!80!black}{TN}}}& 
\cmark&
\cmark&
\cmark&
\xmark
\\ \hline
\textbf{
\cite{boudebza2019detecting}}& 
\textbf{CT}& 
\textbf{{\textcolor{green!80!black}{TN}}}& 
\cmark&
\cmark&
\cmark&
\cmark
\\ \hline
\end{tabular}}
\caption{Comparing dynamic community detection methods.}
\label{tab:meth0}
\end{table}

As illustrated in Table \ref{tab:meth0}, the existing dynamic community detection methods have a number of weaknesses. These limitations are to consider when designing new methods. As our interest in this thesis lies on highly dynamic social networks, the community detection in such networks has to meet some particular requirements :

\begin{itemize}
\item \textbf{Temporal networks:} Snapshot models fail to capture the full temporal evolution of networks. To deal with social networks and other highly evolving networks, it is therefore more appropriate to use a temporal network representation.
Methods working on temporal networks are few (see methods in \cite{palla2007quantifying,xu2013community,xu2013analyzing}) compared to those using snapshot models. Despite their efficiency in handling highly dynamic networks, some of these methods are computationally expensive as they require repetitive computations of communities at each network change. Therefore, the majority of recent methods adopt local computations to minimize the computational cost, like the method by \cite{nguyen2011adaptive}. 
\item \textbf{The stability of detection:} The stability is an important aspect to take into account when designing community detection algorithms for dynamic networks in general, and in particular for highly dynamic networks. The instability problem affects only the two classes of Instant-optimal approaches and Temporal Trade-off approaches like those based on Modularity \cite{asur2009event}, or Random Walk \cite{tajeuna2015tracking}.  
This problem arises from the use of non-intrinsic community definitions, i.e., when the community definition depends on the whole network, a community can be modified as a result of some changes in an unrelated part of the network and this could lead to misleading results. It may also occur when using stochastic community definitions,
i.e., when the algorithm reaches different results on the same or slightly modified network. Different solutions have been proposed to mitigate this problem: one uses sable communities \cite{hopcroft2004tracking}, other searches for community core \cite{wang2008community} or stable community core \cite{xu2013analyzing,xu2013community}, etc. 
Cross-time approaches do not suffer from the instability problem as they propose to study simultaneously all the network evolution steps. Methods using deterministic definitions resolve naturally this problem, like k-clique community definition in the CPM method by \cite{palla2007quantifying}.
\item \textbf{No community drifting}. The risk of community drift is also an important aspect to consider. The community drifting problem may affect only temporal trade-off approaches, and in particular, 
updating approaches. 
This problem occurs when using stochastic community definitions like Modularity (see methods in \cite{aynaud2010static, nguyen2011adaptive,aktunc2015dynamic}), as partitions evolve based on local optimal solutions, it is not guaranteed to yield an optimal solution at the global level.

\item \textbf{Overlapping communities:} Social networks are well known to exhibit highly overlapping community structures, i.e., nodes often belong to multiple communities at once. Most methods in literature focus on detecting disjoint communities and are no longer adapted to find overlapping communities. 
The best-known exception is the clique percolation method by \cite{palla2007quantifying}. 

\item \textbf{Multiple temporal scales of analysis :} The snapshot model often requires choosing an arbitrary temporal scale to divide the dynamic network. Communities detected at such an arbitrary temporal granularity could lead to misleading results: communities resulting from high-frequency interactions and short duration are invisible when considering large temporal scale (window time), while communities resulting from low-frequency interactions and large duration are invisible using a fine temporal granularity. A multiple temporal scale analysis of communities seems, therefore, the right solution in such a case. To the best of our knowledge, this question has not yet been studied in the literature.
\end{itemize}

The methods we propose in this thesis are meant to meet these requirements. Our first method \cite{boudebza2018} falls into Temporal-trade off approaches. It is a variant of the CPM method \cite{palla2005uncovering} working on temporal networks. The temporal complexity is reduced by adopting local updating of communities. The use of the k-clique community definition allows to naturally resolve the problems of instability and community drift and enables overlaps between communities.
Our second method \cite{boudebza2019detecting} falls into Cross-time approaches, thus it does not suffer from the instability problem and the risk of community drifting as all the network evolution steps are studied simultaneously. It also works on temporal networks. This method is the first to consider a multiple temporal scale analysis to detect community structures. Given the generality of this method, both overlapping and non-overlapping communities can be detected.

\section{Conclusion}
In this chapter, we have reviewed the literature on the topic of community detection in dynamic networks. A large amount body of research has shown a great deal of interest in this topic. Even though several approaches have been proposed for detecting communities in dynamic networks, there are still many challenges ahead, especially when dealing with dynamic social networks. As noted previously (see Section \ref{chp3:Discussion}), these challenges are mainly related to the community definition and the dynamic network representation. Most of the existing community definitions are stochastic and extrinsic, which often lead to the problems of instability and community drift. 
In addition to that, a few definitions deal with community overlaps which is a natural property of many real-world networks, especially social networks
, while the majority of them are designed for disjoint communities which are neither appropriate nor realistic to find overlapping communities. Unlike temporal networks, snapshot network models cannot capture the full temporal network evolution, so they are not suitable to model highly evolving networks which is the case for most real networks like social networks. Within this model, we often have to choose a temporal granularity to analyze communities which could lead to misleading results in community detection if the chosen scale is not good.
The methods to be presented in Chapter \ref{chp4:OLCPM} and Chapter \ref{chp:Stable} are suggested as solutions to these issues.
\clearpage
\titleformat{\chapter}[display]
  {\gdef\chapterlabel{}
   \normalfont\sffamily\Huge\bfseries\scshape}
  {\gdef\chapterlabel{\thechapter\ }}{0pt}
  {\begin{tikzpicture}[remember picture,overlay]
    \node[yshift=-5cm] (0) at (current page.north west)
      {\begin{tikzpicture}[remember picture, overlay]
        \draw[fill=grisclaire] (0,0) rectangle
          (\paperwidth,5cm);
          \tikzstyle{GN}=[circle,fill=grisfonce, draw,text=white,font=\bfseries,font=\Huge,
          minimum size=3cm]
           \tikzstyle{PN}=[circle,fill=white, draw,text=black,scale=0.4]
            \path 
 node[GN] (0) at (3.3,3) {\thechapter}
 node[PN] (1) at (5.3,3.3) {}
 node[PN] (2) at (6,2) {} 
 node[PN] (3) at (7.1,2.75) {}
 node[PN] (4) at (6.2,4.25) {}
 node[PN] (5) at (7.75,4) {} 
 node[PN] (6) at (8.2,2.5) {} 
 node[PN] (7) at (9,3.3) {} 
 node[PN] (8) at (10,4) {} 
 node[PN] (9) at (10.8,2.4) {}
 node[PN] (10) at (12.9,3.25) {}
 node[PN] (11) at (12.3,3.8) {}
 node[PN] (12) at (13.9,3) {}
 node[PN] (13) at (16,2.5) {}
 node[PN] (14) at (17,3) {}
 node[PN] (15) at (18.3,2.5) {}
 node[PN] (16) at (17.1,2.25) {};
        \node[anchor=east,xshift=.95\paperwidth,rectangle, rounded corners=16pt,inner sep=11pt,fill=grisfonce,text width=120mm,font=\Large] (17) {\color{white}#1};
              \draw[line width=0mm,   black] (0)--(1)--(2)--(3)--(4)--(1)--(3)--(6)--(7)--(5)--(4);
 \draw[line width=0mm,   black] (7)--(8)--(9)--(7)--(11)--(10)--(9)--(11)--(8);
  \draw[line width=0mm,   black] (10)--(12)--(13)--(16)--(15)--(14)--(12)--(15);
  \draw[line width=0mm,   black] (13)--(17);
       \end{tikzpicture}
      };
   \end{tikzpicture}
  }
\clearpage
\titleformat{\chapter}[display]
  {\gdef\chapterlabel{}
   \normalfont\sffamily\Huge\bfseries\scshape}
  {\gdef\chapterlabel{\thechapter\ }}{0pt}
  {\begin{tikzpicture}[remember picture,overlay]
    \node[yshift=-5cm] (0) at (current page.north west)
      {\begin{tikzpicture}[remember picture, overlay]
        \draw[fill=grisclaire] (0,0) rectangle
          (\paperwidth,5cm);
          \tikzstyle{GN}=[circle,fill=grisfonce, draw,text=white,font=\bfseries,font=\Huge,
          minimum size=3cm]
           \tikzstyle{PN}=[circle,fill=white, draw,text=black,scale=0.4]
            \path 
 node[GN] (0) at (3.3,3) {\thechapter}
 node[PN] (1) at (5.3,3.3) {}
 node[PN] (2) at (6,2) {} 
 node[PN] (3) at (7.1,2.75) {}
 node[PN] (4) at (6.2,4.25) {}
 node[PN] (5) at (7.75,4) {} 
 node[PN] (6) at (8.2,2.5) {} 
 node[PN] (7) at (9,3.3) {} 
 node[PN] (8) at (10,4) {} 
 node[PN] (9) at (10.8,2.4) {}
 node[PN] (10) at (12.9,3.25) {}
 node[PN] (11) at (12.3,3.8) {}
 node[PN] (12) at (13.9,3) {}
 node[PN] (13) at (16,2.5) {}
 node[PN] (14) at (17,3) {}
 node[PN] (15) at (18.3,2.5) {}
 node[PN] (16) at (17.1,2.25) {};
        \node[anchor=east,xshift=.9\paperwidth,rectangle,
              rounded corners=16pt,inner sep=11pt,
              fill=grisfonce,font=\Large] (17) 
              {\color{white}#1};
              \draw[line width=0mm,   black] (0)--(1)--(2)--(3)--(4)--(1)--(3)--(6)--(7)--(5)--(4);
 \draw[line width=0mm,   black] (7)--(8)--(9)--(7)--(11)--(10)--(9)--(11)--(8);
  \draw[line width=0mm,   black] (10)--(12)--(13)--(16)--(15)--(14)--(12)--(15);
  \draw[line width=0mm,   black] (13)--(17);
       \end{tikzpicture}
      };
   \end{tikzpicture}
  }
\chapter{Detecting Overlapping Communities in Dynamique Social Networks}\label{chp4:OLCPM}
\thispagestyle{empty}
\vspace{1cm}

\parindent=0em
\etocsettocstyle{\rule{\linewidth}{\tocrulewidth}\vskip0.5\baselineskip}{\rule{\linewidth}{\tocrulewidth}}
\localtableofcontents

\clearpage
\section{Introduction}\label{chp4:Introduction}
No wonder the previous chapter demonstrated a plethora of methods to help discover communities from dynamic networks, but the latter is subject to many challenges. These challenges are mainly related to the model used to represent the dynamic network or to the community definition.

The way to model the dynamic network has a direct impact on the community discovery process. 
For instance, the snapshot network model can not capture the full temporal network evolution, so it is not suitable to model highly evolving networks which is the case for most real networks like social networks. On the contrary, temporal network model is the most suitable in this case. In recent years, several authors have proposed methods allowing to work on dynamic graphs provided as a stream. In this case, there are too many modifications of the network to run a complete algorithm at each step. Therefore, these methods update communities found at previous steps based on local rules. These are some examples of such methods \cite{xie2013labelrank, nguyen2011adaptive, cazabet2011simulate, rossetti2017tiles} (for more details, see the category of updating approaches using a set of rules in Section \ref{chp3:tradeoff} ). 

The use of stochastic and non-intrinsic community definitions in these algorithms arises some weaknesses. First, the absence of any guarantee that the communities found represent an optimal solution at the global level this is because communities at each step are based on communities found in a previous step by applying a set of local rules. More precisely, these methods suffer from the risk of community drift in which the solution can be dragged away from an originally relevant solution. The second limitation is that,  communities found by these algorithms at step $t$ depend on the particular sequence of previous graph modifications: the same graph produced by a different graph's history would yield a different partition. 

Another important challenge facing dynamic community detection lies in detecting overlapping communities. Most methods are designed for disjoint communities and do not consider the overlap property between communities. This latter is considered as one of the most important properties of real-world networks, in particular for social networks: in such networks, individuals often belong to several social groups. One of the most prominent methods to reveal overlapping and evolving community structures was proposed by \cite{palla2007quantifying}. The latter falls into the category of community matching approaches. The clique percolation method (CPM) \cite{palla2005uncovering} is used to extract the community structure at each time step of an evolving network. Then, communities in consecutive time steps are matched. This method works on snapshot models and it is not suitable for highly evolving networks. Nevertheless, it provides an interesting community definition which can naturally fix the previously mentioned problems arising from stochastic and non-intrinsic community definitions.

In this chapter, we propose a 
framework for detecting overlapping and evolving communities in social networks. The approach we propose is built upon the Clique Percolation Method (CPM). It works on a fine-grained network that is capable of capturing the full dynamics of the network.
Due to the nature of the definition of communities in CPM, we are able to provide an algorithm that handles a flow of changes with local modifications, while guaranteeing that the same state of the graph will always yield the same community structure.

This chapter is organized as follows: 
We first present the rationale basis of this framework in Section \ref{chp4:Rationale}. 
In Section \ref{chp4:streamGraph}, 
we introduce our new model for representing dynamic networks. We describe in detail the proposed framework, called OLCPM, in Section \ref{chp4:OLCPMsection}. Furthermore, we discuss the obtained results of experiments in Section \ref{chp4:Experiments}. 
\section{Rationale for an online version of CPM}\label{chp4:Rationale}


The CPM method, thanks to its community definition, has interesting properties compared with other popular methods such as Louvain and Infomap \cite{blondel2008fast, rosvall2008maps}: 
\begin{itemize}
\item It is deterministic, i.e., two runs on two networks with the same topology will yield the same results.
\item Communities are defined intrinsically, i.e., each community exists independently from the rest of the network, unlike methods using a global quality function such as the \textit{modularity} \cite{girvan2002community}, that suffer from resolution limits \cite{fortunato2007resolution} binding the size of communities to the size of the network.
\item Communities can overlap, i.e., a node can be part of several communities.
\end{itemize}

These properties represent an advantage when working with social networks and with dynamic networks. In particular, a well-known problem with the discovery of evolving communities is the so-called instability of methods \cite{aynaud2010static}, which can be summarized as follows: because community detection methods are unstable, the difference observed in the partition between two consecutive periods of the network might be due either to significant changes in the network or to random perturbations introduced by the algorithm itself. This problem is due to (1) the usage of stochastic methods, as two runs on very similar (or even identical) networks can yield very different results if the algorithm reaches different local maxima, (2) non-intrinsically defined communities, as a modification of a community might be due to changes introduced in an unrelated part of the network. 

Given these observations, CPM appears as a natural candidate to be used for dynamic community detection. The method adapting CPM to the dynamic case \cite{palla2007quantifying}, however, suffers from at least two weaknesses for which we propose solutions in this chapter, one due to CPM itself, and other to its adaptation to the dynamic case:
\begin{itemize}
\item All cliques need to be discovered anew at each step, both in the new graph snapshot and in a joint graph between snapshots at $t$ and $t-1$, which is computationally expensive for networks with many steps of evolution.
\item Nodes must belong to a clique of size at least $k$ to be part of a community, and as a consequence, some nodes might not be affected to any community. As most social networks have a scale-free degree distribution, a large number of nodes remain without a community.
\end{itemize}

To circumvent these issues, we propose a new two-step framework for detecting overlapping and evolving communities in social networks. First, built upon the classical algorithm CPM, we introduce an Online CPM algorithm (OCPM) to identify the core nodes of communities in real-time. To do that, we propose to use \textit{stream graph} as a network model. At every change in the network, the community structure is updated at the local scale. This allows significant improvements in computational complexity compared with dynamic CPM \cite{palla2007quantifying}. Second, to deal with the coverage problem of CPM, we propose a label propagation post-process (OLCPM) and thus, nodes not embedded in any community will be assigned to one or more communities. As the original CPM method, the proposed framework falls into the class of Temporal Trade-off approaches, and more precisely, it is part of updating approaches using a set of rules. 

\section{Stream graph}\label{chp4:streamGraph}
In this section, we introduce our own formalism for evolving graphs, which is better suited to deal with \textit{stream graphs}, i.e., graphs whose modifications occur as a flow, not necessarily known \textit{a priori}. This formalism has the same expressivity as interval graphs.

Networks are often represented by a graph $G=(V, E)$, where $V$ is the set of nodes and $E$ is the set of edges between nodes. We represent dynamic graphs as an ordered sequence of events, which can be node addition, node removal, edge addition, or edge removal. We use the following notations:

\begin{itemize}
\item {\em Inserting or removing a node} is represented as triples $(v,e,t)$,  where $v$ is the node, $e$ is the event observed among $\{+,-\}$ (insert ($+$) or remove($-$)), and $t$ is the time when the event occurs.

\item {\em Inserting or removing an edge} is represented as quadruplets $(u,v,e,t)$, where $u$ and $v$ are endpoints of the edge, $e$ is the event observed among $\{+,-\}$ (insert ($+$) or remove($-$)), $t$ is the time when the event occurs.

\end{itemize}

Note that this formalism, for edges, is identical in nature to an interval graph, but is more convenient for stream algorithms, as new operations can be added at the end of the ordered sequence of events without affecting previous ones.

\section{OLCPM framework}\label{chp4:OLCPMsection}

Our framework comprises two main steps. First, we propose to adapt the classical algorithm CPM \cite{palla2005uncovering} for static overlapping community detection to deal with evolving networks. We propose an online version of CPM called OCPM (Online CPM). This algorithm is based on analyzing the dynamic behaviors of the network, which may arise from inserting or removing nodes or edges, i.e., every time a change is produced in the network, we update locally the community structure alongside the involved node or edge. 

As stated earlier, CPM may not cover the whole network, i.e., some nodes have no community membership. To deal with this problem, we assume that the communities corresponding to OCMP contain core nodes, and we propose a way to discover the community peripheral nodes. In the second step of our framework, we extend OCMP using label propagation method and we propose OLCPM (Online Label propagation CPM). These proposals will be presented in detail in the next section.

\subsection{OCPM: Online Clique Percolation Method}

This section proposes the first step of our framework OLCPM, an online  Clique Percolation Method (OCPM). This method takes two inputs: 
\begin{itemize}
\item $SE$, chronologically ordered sequence of events which models networks modification, following the format: $(n, e, t)$ or $(i, j, e, t)$ as defined in Section \ref{chp4:streamGraph}
\item the parameter $k$, which determines the clique size; it is an integer value greater than or equal to 3 
\end{itemize}

The OCPM method  maintains after each modification three elements: 
\begin{itemize}
	\item $G(V, E)$ the current state of the network
	\item $AC$ the set of currently Alive Communities 
	\item $DC$ the set of Dead Communities
\end{itemize}

It is therefore possible to know the community structure status at every network modification step. 

\subsubsection{Definition of the OCPM algorithm}

The core of the OCPM algorithm can be defined by an algorithm that updates the current state of all variables according to a Sequence of Events $SE$, as detailed in Algorithm \ref{algo:OCPM}. The task carried out by the algorithm depends on the type of event encountered:

\IncMargin{1em}
\begin{algorithm}[!htbp]
\SetKwData{Left}{left}
\SetKwData{This}{this}
\SetKwData{Up}{up}
\SetKwFunction{Union}{Union}
\SetKwFunction{FindCompress}{FindCompress}
\SetKwInOut{Input}{input}
\SetKwInOut{Output}{output}
\Input{$K, G, AC, DC, SE$}
\Output{Update $AC, DC, G$}
\BlankLine
\For{$ev \in SE$}{
\Switch{$ev$}{
\tcc{Add Node}
\Case{(n, +, t) }{
$V \leftarrow V \cup\{n\}$\;
break\;
}
\tcc{Add Edge}
\Case{(i, j, +, t)}{
$E \leftarrow E \cup\{(i,j)\}$\;
\eIf{ $(C_{i} \neq \varnothing)  or  (C_{j} \neq \varnothing)$ }{
\tcc{$(C_{i}$ and $C_{j}$ are respectively communities of nodes $i$ and $j$) }
$AddNonExternalEdge(i, j, t, AC, DC, G)$\;
}
{
$AddExtrenalEdge(i, j, t, AC, G)$\;
}
break\;
}
\tcc{Remove Node}
\Case{(n, -, t) }{
$V \leftarrow V \backslash \{n\}$\;
$E \leftarrow E \backslash \{\forall(i,j), i=n$ or $j=n \}$\;
\If{ $(C_{n} \neq \varnothing) $ }{
$RemoveInternalNode(n, t, AC, DC, G)$\;
}
break\;
}
\tcc{Remove Edge}
\Case{(i, j, -, t)}{
$E \leftarrow E \backslash\{(i,j)\}$\;
\If{ $(C_{i} \cap C_{j} \neq \varnothing) $ }{
$RemoveIntrnalEdge(i, j, t, AC, DC, G)$\;
}
break\;
}
}
}
\caption{Online Clique Percollation Method (OCPM)}
\label{algo:OCPM}
\end{algorithm}\DecMargin{1em}

\newpage

\begin{itemize}
\item \textbf{Add a new node}: adding an isolated node $n$ has no influence on the community partition. In this case, only $n$ is added to the graph $G$ and no other action is performed until the next event. 
\item \textbf{Add a new edge}: when a  new edge $(i,j)$ appears, we add this edge to the graph $G$. 
According to the type of edge, we distinguish two cases:

	\begin{itemize}

	\item When inserting an external edge, i.e., both its endpoints are outside any community, we check if one or more new $k$-cliques (KCliques() function Algorithm \ref{KCliques}) are created. If it is the case, we gather all adjacent $k$-cliques one to the other. Then, for each group of adjacent $k$-cliques, we create a single community. Figure \ref{fig:AddExternalEdge} shows two examples of adding external edges and the changes it brings to the community structure (see Algorithm \ref{Algo:AddExternalEdge}).
	\IncMargin{1em}
\begin{algorithm}[!htbp]
\SetKwData{Left}{left}
\SetKwData{This}{this}
\SetKwData{Up}{up}
\SetKwFunction{Union}{Union}
\SetKwFunction{FindCompress}{FindCompress}
\SetKwInOut{Input}{input}
\SetKwInOut{Output}{output}
\Input{$i, j, t, AC, G$}
\Output{Update $AC$}
\BlankLine
$KC \leftarrow KCliques(\{i,j\},G)$\;
$AKC \leftarrow AdjKCliques(KC)$\;
\For{$c \in AKC$}{
$Birth(c, t, AC)$\;
}
\caption{Add External Edge}
\label{Algo:AddExternalEdge}
\end{algorithm}\DecMargin{1em}

\begin{figure} [h!]
\centering
\begin{subfigure}{\linewidth}
    \centering
    \includegraphics{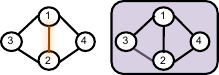} 
    \caption{Example with $k=3$}
 \end{subfigure}

\begin{subfigure}{\linewidth}
    \centering
    \includegraphics{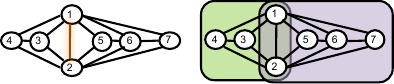} 
    \caption{Example with $k=4$}
 \end{subfigure}

\caption{Examples of adding an edge with both endpoints outside any community. $(a)$ Example for $k=3$: when the edge$(1, 2)$ is added, a new community $\{$1, 2, 3, 4$\}$ is created from two adjacent $k$-cliques $\{1, 2, 3\}$ and $\{1, 2, 4\}$. $(b)$ Example for $k=4$: the insertion of edge$(1, 2)$ leads to the creation of two communities $\{ 1, 2, 3, 4\}$ and $\{1, 2, 5, 6, 7\}$ from respectively two groups of not-adjacent $k$-cliques $\{\{1, 2, 3, 4\}\}$ and $\{\{1, 2, 5, 6\}$,$\{1, 2, 6, 7\}\}$}.
\label{fig:AddExternalEdge}
\end{figure}

	\item In all other cases, i.e., when a new edge appears with one or two internal extremities, we check all $k$-cliques created when adding this edge and not belonging to any community. Then, all adjacent $k$-cliques are grouped together and for each group, we check if there are other adjacent $k$-cliques included in any community to which belongs any node in this group. If they exist, the corresponding communities will grow with the nodes of this group and they can eventually be merged (Merge()function Algorithm \ref{Merge}). Otherwise, a new community appears containing nodes of this group. Figures \ref{fig:OneEexternalEndpoint} and \ref{fig:TwoInternalEndpoints} depict some examples of adding edges with one or two internal endpoints and the changes to the community structure (see Algorithm \ref{Add_internal_edge}).
	\end{itemize}
\IncMargin{1em}
\begin{algorithm}[!htbp]
\SetKwData{Left}{left}
\SetKwData{This}{this}
\SetKwData{Up}{up}
\SetKwFunction{Union}{Union}
\SetKwFunction{FindCompress}{FindCompress}
\SetKwInOut{Input}{input}
\SetKwInOut{Output}{output}
\Input{$i, j, t, AC, DC, G$}
\Output{Update $AC, DC$}
\BlankLine
$KC \leftarrow KCliques(\{i,j\},G)$\;
$KC \prime \leftarrow \{kcl \in KC,\forall cm \in AC, kcl \nsubseteq cm \}$\;
$AKC \leftarrow AdjKCliques(KC\prime)$\;
\For{$c \in AKC$}{
$C_{c} \leftarrow \{ \}$\;
\For{$e \in c $}{
$C_{c} \leftarrow C_{c} \cup \{ C_{e}\}$\;
}
$AdjC \leftarrow \{ \}$\;
\For{$cm \in C_{c}$}{ 
\If{$\vert cm \cap c \vert \geqslant K-1 $}{
$AdjC \leftarrow AdjC\cup\{cm\}$\;
}
}
\eIf{$AdjC\neq\varnothing$}{
\For{$cm \in AdjC$}{ 
$Growth(cm, c, AC)$\;
}
\If{$\vert AdjC\vert > 1$}{
$Merge(AdjC, t, AC, DC)$\;
}
}
{
$Birth(c, t, AC)$\;
}

}
\caption{Add Non-External Edge}
\label{Add_internal_edge}
\end{algorithm}\DecMargin{1em}
\begin{figure} [h!]

\centering
\begin{subfigure}{\linewidth}
    \centering
    \includegraphics{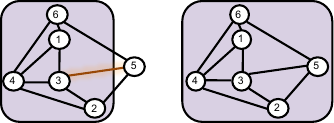} 
    \caption{Simple grow}
 \end{subfigure}

\begin{subfigure}{\linewidth}
    \centering
    \includegraphics{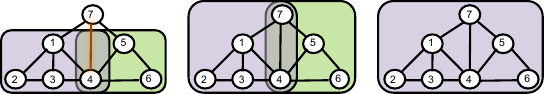}
    \caption{Grow and merge}
 \end{subfigure}

\begin{subfigure}{\linewidth}
    \centering
    \includegraphics{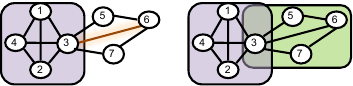} 
    \caption{New community}
 \end{subfigure}
 
\caption{Example of adding an edge with an external endpoint and internal one(for $k=3$). (a) The community $\{1, 2, 3, 4, 6 \}$ grows with node $5$ when adding edge $(3, 5)$. (b) When the edge $(4,7)$ is added, the communities $\{ 1,2,3,4\}$ and $\{4,5,6\}$ grow with node $7$, and then merged. The resulting community takes the identity of the one that contains more nodes.(c) By adding edge $(3,6)$, a new community $\{ 3,5,6,7\}$ is created.}

\label{fig:OneEexternalEndpoint}
\end{figure}

\begin{figure} [h!]

\begin{subfigure}{\linewidth}
    \centering
    \includegraphics{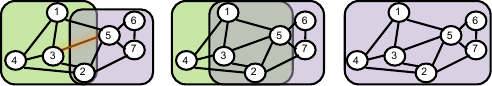}
    \caption{Grow and Merge}
 \end{subfigure}
 
 \begin{subfigure}{\linewidth}
    \centering
    \includegraphics{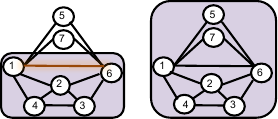} 
    \caption{Grow}
 \end{subfigure}

\caption{Examples of adding an edge with two internal endpoints(k=3). (a) The communities $\{1,2,3,4\}$ and $\{2,5,6,7\}$ grow with the nodes of adjacent $k$-cliques $\{ \{1,3,5\},\{2,3,5\}\}$ formed when adding the edge $(3,5)$, and then merged. (b) The community $\{1,2,3,4, 6\}$ grows with the nodes of adjacent $k$-cliques $\{ \{1,7,6\},\{1,5,6\},\{1,2,6\}\}$ formed when adding the edge $(1,6)$.}
\label{fig:TwoInternalEndpoints}
\end{figure}
\newpage
\item \textbf{Delete node}: In this case, we remove the node from the graph G, and all its edges are removed as well. If the node is external, i.e., it does not belong to any community, the community structure is not affected and no action is performed until the next event. When the removed node belongs to one or more communities, we check for each community to which this node belongs whether it still contains at least a $k$-clique after the node is removed. This community dies if it loses all $k$-cliques (see Figure (c) \ref{fig:DeleteInternalNode}). Otherwise, the community shrinks, i.e., it loses this node and all its associated edges. Here, we distinguish two cases:
\begin{itemize}
\item The community may remain coherent and the community structure does not change (see Figure (a) \ref{fig:DeleteInternalNode} ).
\item The community may become disconnected and therefore, it will be break up into small communities (see Figure (b) \ref{fig:DeleteInternalNode}).
\end{itemize}
The split function (see Algorithm \ref{Split}) deals with these two cases. After the community shrinking, its structure is recalculated keeping the principle of CPM -checking all maximal cliques of size not less than $k$. The resulting community having the largest number of nodes keeps the identity of the original one, where the others have new identities.

The Algorithm \ref{Remove_internal_node} describes this case.
\IncMargin{1em}
\begin{algorithm}[!htbp]
\SetKwData{Left}{left}
\SetKwData{This}{this}
\SetKwData{Up}{up}
\SetKwFunction{Union}{Union}
\SetKwFunction{FindCompress}{FindCompress}
\SetKwInOut{Input}{input}
\SetKwInOut{Output}{output}
\Input{$n, t, AC, DC, G$}
\Output{Update $AC, DC$}
\BlankLine
\For{$c \in C_{n}$}{
$KC \leftarrow KCliques(c,G)$\;
\eIf{$KC = \varnothing$}{
$Death(c, t, AC, DC)$\;
}
{
$Shrink(c,n, AC)$\;
$Split(c, t, AC, G)$\;
}
}
\caption{Remove Internal Node}
\label{Remove_internal_node}
\end{algorithm}\DecMargin{1em}
\begin{figure} [h!]
\centering

\begin{subfigure}{\linewidth}
    \centering
    \includegraphics[width=.3\linewidth]{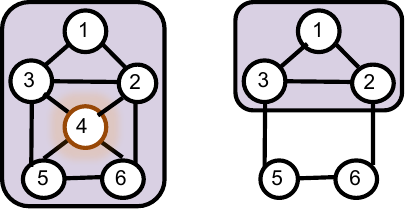}
    \caption{Shrink}
 \end{subfigure}
 
 \begin{subfigure}{\linewidth}
    \centering
    \includegraphics[width=.4\linewidth]{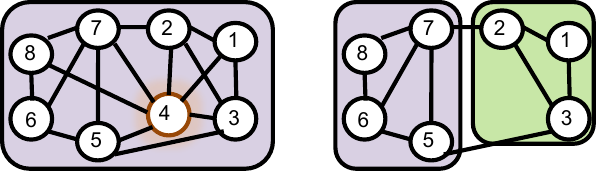} 
    \caption{Shrink and Split}
 \end{subfigure}
 
 \begin{subfigure}{\linewidth}
    \centering
    \includegraphics{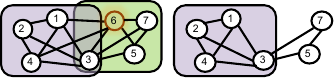} 
    \caption{Death}
 \end{subfigure}
 
\caption{Example of removing internal node (k=3 for (a) and (b), $k=4$ for (c)). (a) When removing the node $4$, the members $\{4,5,6\}$ leaves out the community $\{1,2,3,4,5,6\}$.(b) When removing the node $4$, the community $\{1,2,3,4,5,6,7,8\}$ shrinks, i.e., it loses this node and all its edges, and then splits into two communities: $\{5,6,7,8\}$ and $\{1,2,3\}$. (c)By removing the node $6$, the community $\{1,2,3,4\}$ shrinks and the community $\{3,5,6,7\}$ dies}
\label{fig:DeleteInternalNode}
\end{figure}
\newpage
\item \textbf{Delete edge}: First, we remove the edge from the graph G. The removal of an edge with two endpoints belonging to the same community(ies) (called internal edge) follows the same mechanism as internal node removal: the communities to which belong the two extremities of this edge may split or die. For each of them, we check whether it still contains $k$-cliques. If so, we use the function Split (Algorithm \ref{Split}) to check whether or not the community is divided into smaller parts. Otherwise, this community dies (see Algorithm \ref{Remove_internal_edge}). Figure \ref{fig:RemoveInternalEdge} shows two examples of removing internal Edge and the changes that it brings to the community structure.

For all other types of edges, the community structure does not change. 
\IncMargin{1em}
\begin{algorithm}[!htbp]
\SetKwData{Left}{left}
\SetKwData{This}{this}
\SetKwData{Up}{up}
\SetKwFunction{Union}{Union}
\SetKwFunction{FindCompress}{FindCompress}
\SetKwInOut{Input}{input}
\SetKwInOut{Output}{output}
\Input{$i, j, t, AC, DC, G$}
\Output{Update $AC, DC$}
\BlankLine
\For{$c \in C_{i j}=\{\forall cm, cm \in (C_{i}\cap C_{j})\}$}{ 
$KC \leftarrow KCliques(c,G)$\;
\eIf{$KC = \varnothing$}{
$Death(c, t, AC, DC)$\;
}
{
$Split(c, t, AC, G)$\;
}
}
\caption{Remove Internal edge}
\label{Remove_internal_edge}
\end{algorithm}\DecMargin{1em}
\begin{figure} [h!]

 \begin{subfigure}{\linewidth}
    \centering
    \includegraphics{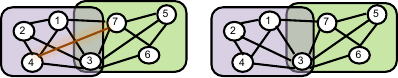}
    \caption{No change in the community structure}
 \end{subfigure}
 
  \begin{subfigure}{\linewidth}
    \centering
    \includegraphics{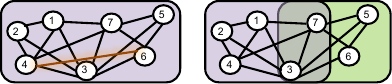}
    \caption{Community split}
 \end{subfigure}
 
\caption{Examples of removing internal edge (k=4). (a) The community structure does not change when removing the edge $(4, 7)$. (b) When removing the edge $(4, 6)$, the community splits into two small communities, each of which contains a group of adjacent $k$-cliques in the original community.}
\label{fig:RemoveInternalEdge}
\end{figure}

\end{itemize}
\newpage

Here, we detail some functions used in our algorithm:

\begin{itemize}

\item \textbf{Kcliques}(): (Algorithm \ref{KCliques}) This function takes a set of nodes SN as input parameter and returns all maximal cliques of size not less than $k$ containing this set. In order to optimize the performance of our algorithm, $k$-cliques are locally launched in the subgraph including the set $SN$ and all common neighbors among its members. 
\IncMargin{1em}
\begin{algorithm} [!h]
\SetKwData{Left}{left}
\SetKwData{This}{this}
\SetKwData{Up}{up}
\SetKwFunction{Union}{Union}
\SetKwFunction{FindCompress}{FindCompress}
\SetKwInOut{Input}{input}
\SetKwInOut{Output}{output}
\Input{$ SN$:Set of nodes,$G$}
\Output{$ SKC$: Set of Set of nodes}
\BlankLine
\For{$n \in SN$}{
$N \leftarrow Neighbors(n,G)$\;
$L \leftarrow L\cup \{N\}$\;
}
$CN \leftarrow \cap_{l\in L} l$\;
$SCL \leftarrow MaximalCliques(CN, K)$\;
\caption{KCliques}
\label{KCliques}
\end{algorithm}\DecMargin{1em}
\item \textbf{Merge}(): (Algorithm \ref{Merge}) This function is used for merging adjacent communities. The resulting community takes the identity of the one with the highest number of nodes. 
\IncMargin{1em}
\begin{algorithm} [!h]
\SetKwData{Left}{left}
\SetKwData{This}{this}
\SetKwData{Up}{up}
\SetKwFunction{Union}{Union}
\SetKwFunction{FindCompress}{FindCompress}
\SetKwInOut{Input}{input}
\SetKwInOut{Output}{output}
\Input{$ Adjc, t, AC, DC$}
\Output{Update$  AC,DC$}
\BlankLine
$mc \leftarrow c, |c|= max_{x \in Adjc} |x| $\;
$Adjc \leftarrow Adjc \backslash \{mc\}$\;
$mc \leftarrow \cup_{x \in Adjc}$ $x$\;
\For{$x \in Adjc$}{
$Death(c, t, AC, DC)$\;
}
\caption{Merge}
\label{Merge}
\end{algorithm}\DecMargin{1em}

\item \textbf{Split}(): (Algorithm \ref{Split}) This function is used for splitting a community if possible. It takes as input a community and creates from it one or more communities.  We proceed as follows: first, we identify all maximal cliques of size not less than $k$ in this community and we aggregate adjacent $k$-cliques with each other. Then, for each of the aggregated $k$-cliques, we create a new community. The community which has the largest number of nodes keeps the identity of the original one.
\end{itemize}
\IncMargin{1em}
\begin{algorithm}[!htbp]
\SetKwData{Left}{left}
\SetKwData{This}{this}
\SetKwData{Up}{up}
\SetKwFunction{Union}{Union}
\SetKwFunction{FindCompress}{FindCompress}
\SetKwInOut{Input}{input}
\SetKwInOut{Output}{output}
\Input{$ c, t, AC, G$}
\Output{Update$  AC$}
\BlankLine
$KCc \leftarrow KCliques(c, G)$\;
$Adjc \leftarrow AKCliques(KCc)$\;
$c \leftarrow mc, |mc|= max_{x \in Adjc} |x| $\;
$Adjc \leftarrow Adjc \backslash \{c\}$\;
\For{$cm \in Adjc$}{
$Birth(cm, t, AC)$\;
}
\caption{Split}
\label{Split}
\end{algorithm}\DecMargin{1em}
\newpage
In Table \ref{tab:Actions} we summarize the actions which can be carried out by OCPM according to graph events.

\begin{table}[!h]
\centering
\begin{tabular}{|l|l||c|}
\hline
\multicolumn{2}{|c||}{\textbf{Event}}     & \multicolumn{1}{l|}{\textbf{Actions}} \\ \hline
\multicolumn{2}{|l||}{Add new node}       & -                                     \\ \hline
\multirow{2}{*}{Add new edge} & External & Birth                                 \\ \cline{2-3} 
                              & Other    & Grow+{[}Merge{]}, Birth                \\ \hline
\multirow{2}{*}{Delete Node}  & External & -                                     \\ \cline{2-3} 
                              & Internal & Shrink+{[}Split{]}, Death              \\ \hline
\multirow{2}{*}{Delete Edge}  & Internal & Split, Death                           \\ \cline{2-3} 
                              & Other    & -                                     \\ \hline
\end{tabular}
\caption{Actions that can be performed according to graph events. Brackets denotes events that can only follow the preceding community event.}\label{tab:Actions}
\end{table}

\subsubsection{Complexity of the algorithm}

Instead of computing all $k$-cliques for the whole network at each event occurring in the network, OCPM updates the community structure on the local scale, and thus only the community structure alongside the node or the edge involved in the event is recomputed. For certain events, like adding or deleting an isolated node or deleting an external edge, the community structure does not change and hence, the computational time saving reaches its maximum. For instance, if we have $n$ $k$-cliques when such an event is produced, the computational time savings will be $n$ times the average time for calculating $k$-cliques. For other events, the computational time saving is also significant. See Section \ref{chp4:empComplexity} for an empirical evaluation of time complexity.

\subsubsection{Community tracking process}
One of the difficulties when tracking the evolution of communities is to decide which community is a continuation of which. Our framework allows a trivial matching in the case of \textit{continuation} (no merge or split) of communities. In the case of merge and split, deciding which community keeps the original identity is a well-known problem with no consensus in the literature \cite{cazabet2014dynamic}. In OCPM, we took the simple yet reasonable decision to consider that the \textit{largest} community involved in a merge or split have the same identifier as the merged/split one. This strategy can be replaced without altering the algorithm logic. A more advanced process could be added to solve problems of \textit{instability}, e.g. communities merging and quickly splitting back to their original state.

\subsection{OLCPM: Online Label propagation CPM}

This section describes the second step of our framework. A post-processing based on label propagation is set out on the output communities of OCPM to discover the peripheral nodes. This module is called OLCPM (Online Label propagation CPM).  

There is a twofold reason for using a post-process extending core-communities found by OCPM:
\begin{itemize}
	\item In a network evolving at fast path, one can update core-communities efficiently after each event, and run the post-process only when the current state of communities needs to be known, thus saving computation time
	\item It is known that the periphery of communities is often not well defined and unstable. As seen earlier, and because OCPM is deterministic and it searches for core-communities, it reduces this instability problem. By using the label propagation mechanism only as a post-process for analysis, communities at $t$ do not depend on the periphery of communities that might have been computed at $t-1$, but only on the stable part found by OCPM.
\end{itemize}

\subsubsection{OLCPM algorithm}

\begin{figure*}[h!]
    \centering
     \begin{subfigure}{\linewidth}
     \centering
    \includegraphics{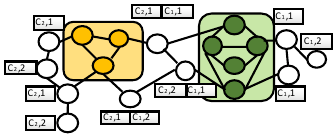}
    \caption{Label spreading step}
 \end{subfigure}
  \begin{subfigure}{\linewidth}
  \centering
    \includegraphics{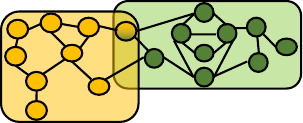}
    \caption{Community structure after label analysis (k=3)}
 \end{subfigure}
\caption{Peripheral community updates by OLCPM. (a) Label spreading step. (b) Community structure after label analyses (for $k=3$). Green nodes are members of the community $C1$; Yellow nodes are  members of the community $C2$; uncolored nodes have no affiliation.}
\label{fig:OLCPM}
\end{figure*}

First, each core-community (community found by OCPM) spreads to neighboring peripheral nodes (nodes not covered by OCPM) a label containing its identity and a weight representing the geodesic distance (the length of the shortest path) between this neighboring node and any other node in the core-community. Each peripheral node has a local memory allowing the storage of many labels. The label propagation process is based on breadth-first search (BFS). When all labels have been shared, nodes are associated with all communities with which they have the shortest geodesic distance. Note that nodes can, therefore, belong to several communities, if they are at the same distance of community found by OCPM. This algorithm is defined formally in Algorithm \ref{algo:OLCPM}.
Figure \ref{fig:OLCPM} presents an illustration of this process.
\IncMargin{1em}
\begin{algorithm} [!h] 
\SetKwData{Left}{left}
\SetKwData{This}{this}
\SetKwData{Up}{up}
\SetKwFunction{Union}{Union}
\SetKwFunction{FindCompress}{FindCompress}
\SetKwInOut{Input}{input}
\SetKwInOut{Output}{output}
\Input{$ AC, G$}
\Output{Update$  AC$}
\BlankLine
$PN \leftarrow \{ n, n \in c \forall c\in AC \}$\;
//Label spreading\\

\For{$c \in AC$}
{
$d \leftarrow 1$\;
$S \leftarrow c$\;
$\textbf{x:} N \leftarrow Nieghbors(S)$\;
$N \leftarrow N \cap PN$\;
\If{$N \neq\ \varnothing$}
{
\For{$n \in N$}
{
$Label(n,c.id,d)$\;
}
$S \leftarrow N$\;
$d \leftarrow d+1$\;
\textbf{goto x:}
}
}
//label Analyses\\

\For{$n \in PN$}{
$idc \leftarrow LabelAnalysis(n.label)$\;
$Growth(idc, n)$\;
}
\caption{OLCPM}
\label{algo:OLCPM}

\end{algorithm}\DecMargin{1em}

\newpage

\section{Experiments}\label{chp4:Experiments}
In this section, we begin by evaluating the effectiveness of OCPM algorithm. Thus, we compare the time complexity (running time) of OCPM with the dynamic version of CPM \cite{palla2007quantifying}. Second, we are interested in the quality of the communities that OLCPM is able to find, considering both synthetic and real-world networks.

Note: the full code of the proposed framework OLCPM and the datasets used in the experiments are available at the following web site: \url{http://olcpm.sci-web.net}

\subsection{Measuring OCPM time complexity gain for highly dynamic networks}
\label{chp4:empComplexity}
In this section, we compare the empirical time complexity of the original dynamic version of CPM (hereafter, DyCPM)\cite{palla2007quantifying} and our proposed version (OCPM). We generate synthetic dynamic networks and compare how the running time of both algorithms varies with the properties of the network and of its dynamic. Note that we compare OCPM only with CPM because both algorithms try to solve the \textit{same problem}, i.e, they have the same definition of communities. Other streaming algorithms like \cite{xie2013labelrank, nguyen2011adaptive, cazabet2011simulate, rossetti2017tiles} have an \textit{ad hoc} definition of communities introduced together with the method and does not have the same properties, such as being deterministic and not being dependent on the network history. Their time complexity is, in theory, similar to the one of OCPM (local updates at each modification).

\subsubsection{Generation of dynamic networks with community structure}
We propose a simple process to generate dynamic networks with realistic community structures. First, a static network is generated using the LFR benchmark \cite{lancichinetti2009benchmarks}, the most used benchmark for community detection. Then, for this network, we generate a step-by-step evolution. In order to conserve the network properties (community structure, size, density), we define an \textit{atomic modification} as the following process:

\begin{enumerate}
\item Choose randomly a planted community as provided by LFR
\item Select an existing edge in this community
\item Select a pair of nodes without edges in this community
\item Replace the selected existing edge with the selected not-existing one.
\end{enumerate}

We define a step of evolution as the combination of $a$ atomic modifications. In order to test the influence of the number of modifications between steps, we test different values of $a$.

Note that we use synthetic networks instead of real networks at this step since: 
\begin{itemize}
	\item We are only interested in measuring the time complexity of algorithms. Synthetic networks are mostly criticized for having unrealistic community structures, while here we are mainly interested in the size and rate of evolution of the networks.
	\item It allows controlled experiments. With real evolving networks, changes in the structure/size of the network could affect computation time at each step, and we could not control the number of modifications between snapshots, or vary the size of networks while keeping constant properties.
\end{itemize}

\subsubsection{Experimental process}

The LFR benchmark \cite{lancichinetti2009benchmarks} is, as of today, one of the most widely used benchmark to evaluate community detection methods. It is known to generate realistic networks with heterogeneous degrees and community sizes.

It has the following parameters : $N$ is the network size, $k$ is the average degree of nodes, $kmax$ the maximum degree, $t1$ and $t2$ are power-law distribution coefficients for the degree of nodes and the size of community respectively, $\mu$ is the mixing parameter which represents the ratio between the external degree of the node with respect to its community and the total degree of the node, $minc$ and $maxc$ are the minimum and maximum community size respectively, $On$ is the number of overlapping nodes , $Om$ is the number of community memberships of each overlapping node. 

In order to obtain realistic networks, we first generate an original network with $n$ nodes using the LFR benchmark, with fixed parameters $k=7$, $maxk=15$, and $\mu=0.4$. Other parameters stay at their default values. In order to test the influence of the network size, we test different values of $n$.

\begin{figure} [h!]
\begin{center}
\includegraphics[width=0.8\linewidth]{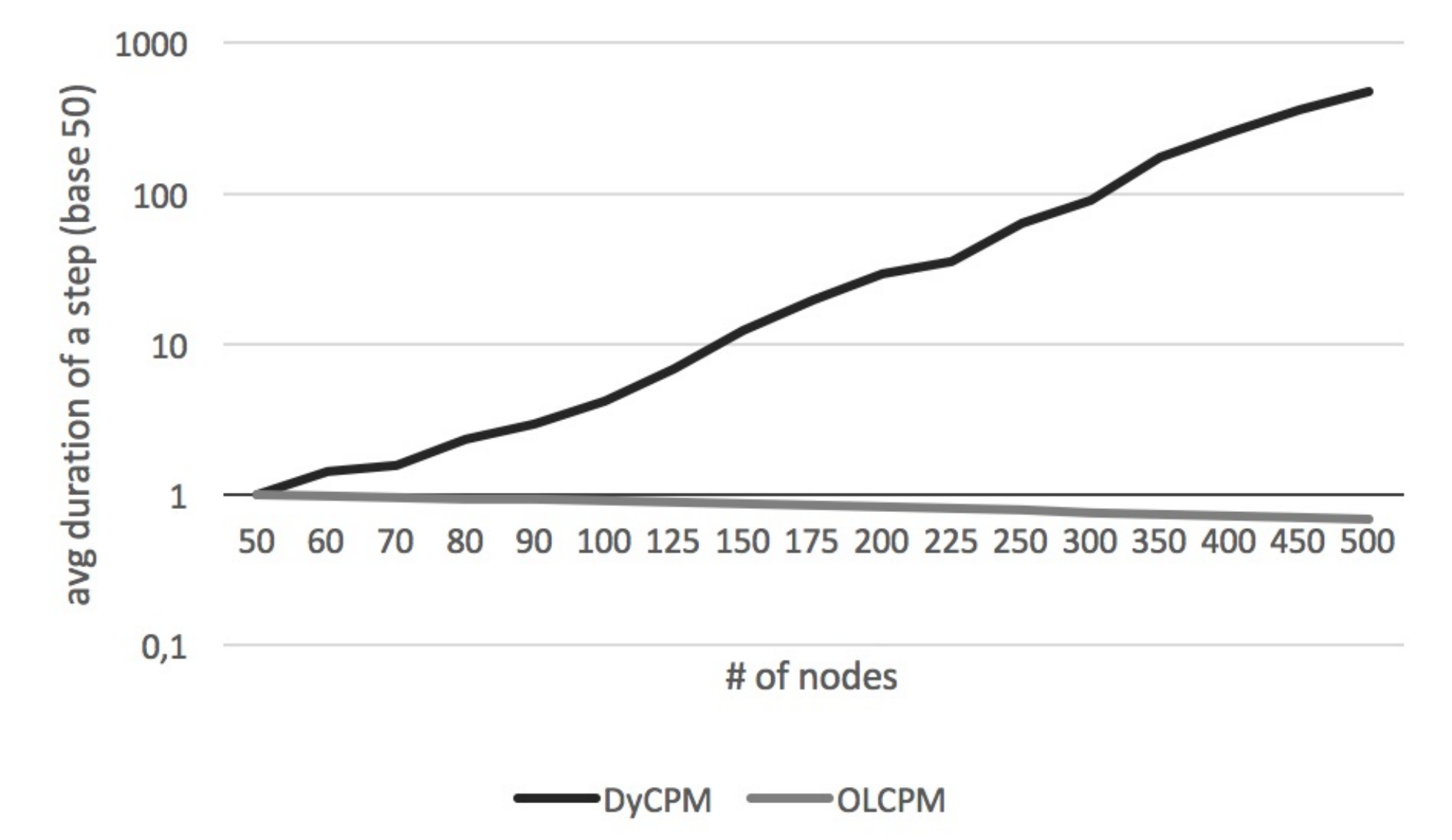}
\end{center}
\caption{Evolution of time complexity when varying the size of the network (number of nodes), and keeping other parameters constant (average node degree, community, size, etc.). DyCPM complexity increases exponentially with the size of the network, while OLCPM one stays constant or slightly decreases. Expressed in base 50, i.e, 10 on the vertical axis means 10 times slower than with 50 nodes.}
\label{fig:time1}
\end{figure}

\begin{figure}[!ht]
\begin{center}
\includegraphics[width=0.8\linewidth]{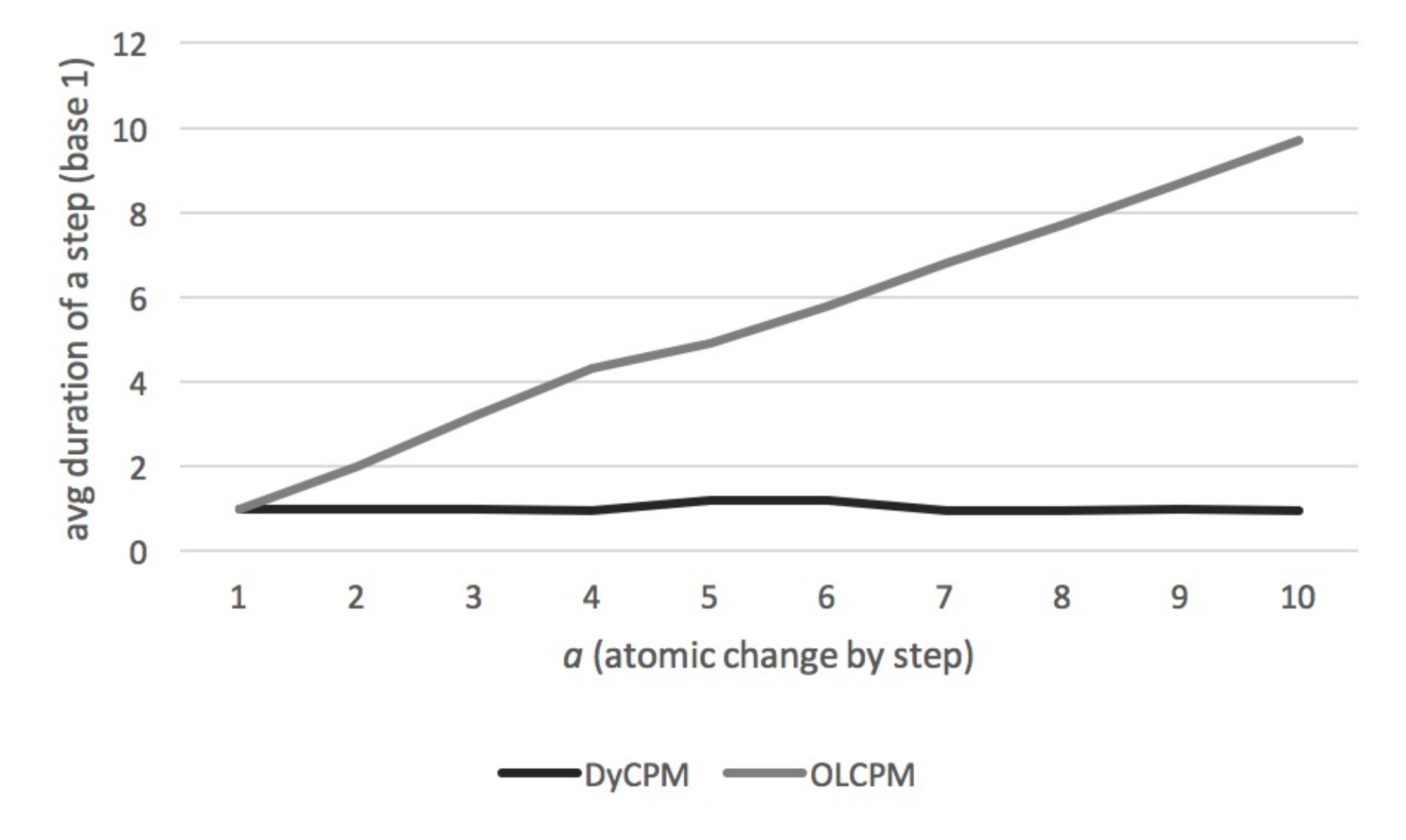}
\end{center}
\caption{Evolution of time complexity when varying the number of atomic changes by step. DyCPM complexity is independent relatively to $a$ while OLCPM's complexity increases linearly with $a$ Time.}
\label{fig:time2}
\end{figure}

As can be seen in Figures \ref{fig:time1} and \ref{fig:time2}, the time complexity of both algorithms depends on very different parameters. With OLCPM, the time needed to update communities after a modification step does not increase proportionally to the size of the network at any given time, but increases linearly with the number of atomic modifications. 

On the contrary, the time complexity of DyCPM depends on the properties of the static network, but not on the number of atomic modifications between steps.

As expected, OLCPM is appropriate to deal with stream graphs, in which modifications are known at a fine granularity, as the cost of each update is low. On the contrary, DyCPM is appropriate to deal with network snapshots, i.e., a dynamic network composed of a few observations collected at regular intervals.

\subsection{Measuring OLCPM communities quality}

To quantify the quality of communities detected by OLCPM framework, we used both synthetic and real-world networks with ground truth community structure. We remind the reader that communities found by DyCPM and OCPM are identical, the difference lies only in the label propagation post-process of OLCPM.

Normalized Mutual Information (NMI) is used as the measurement criterion. This measure is borrowed from information theory \cite{danon2005comparing} and widely adopted for evaluating community detection algorithms. It measures the similarity between a ground truth partition and the one delivered by an algorithm. As the original definition is only well defined for \textit{partitions} (each node belong to one and only one community), a variant of the NMI adapted for \textit{covers} (nodes can belong to zero, one or more communities) have been introduced by \cite{lancichinetti2009compare}. This variant is the most used in the literature for comparing overlapping communities. We used the original implementation by the authors \footnote{\url{https://sites.google.com/site/andrealancichinetti/software}}. The NMI value is defined between 0 and 1, with a higher value meaning higher similarity.

\subsubsection{Static synthetic networks}

We use the LFR benchmark \cite{lancichinetti2009benchmarks} to generate realistic artificial networks.  

We use two different network sizes, \textit{small networks} (1000 nodes) and \textit{large networks} (5000 nodes), and for a given size we use two ranges for community size: \textit{small communities}, having between $10$ and $50$ nodes and \textit{large communities}, having between $20$ and $100$ nodes. We generate eight groups of LFR networks. 

In the first four networks, $\mu$ ranges from $0$ to $0.5$ (steps of $0.1$) while $Om$ is set to $100$ for small networks and $500$ for large networks ($5000$ nodes). In the other networks, $\mu$ is fixed to $0.1$ and $On$ ranges from $0$ to $500$ (steps of $100$) for small networks  and from $0$ to $2000$ (steps of $500$) for large networks. All these networks share the common parameters: $k = 10$, $maxk = 30$, $t1 = 2$, $t2 = 1$, $On = 2$. The parameter settings are shown in Table \ref{tab:LFRParm}. 

\begin{table} [!h] 
\centering
\begin{tabular}{|c|c|c|c|c|c|c|}

        \hline
        \textbf{Network group ID} & \textbf{N} & \textbf{minc} &	\textbf{maxc} & \textbf{$\mu$} & \textbf{On}\\ 
        \hline
       N1 & 1000 & 10 & 50 & 0-0.5 & 100 \\ 
       \hline
       N2 & 1000 & 20 & 100 & 0-0.5 & 100 \\ 
       \hline
       N3 & 5000 & 10 & 50 & 0-0.5 & 500 \\ 
       \hline
       N4 & 5000 & 20 & 100 & 0-0.5 & 500 \\ 
       \hline
       N5 & 1000 & 10 & 50 & 0.1 & 0-500 \\ 
       \hline
       N6 & 1000 & 20 & 100 & 0.1 & 0-500 \\ 
       \hline
       N7 & 5000 & 10 & 50 & 0.1 & 0-2000 \\ 
       \hline
       N8 & 5000 & 20 & 100 & 0.1 & 0-2000 \\ 
       \hline

\end{tabular}
\caption{LFR parameter setting}\label{tab:LFRParm}

\end{table}

\begin{figure*}[!h]
\centering     
\begin{subfigure}{0.4\linewidth}
    \includegraphics[width=\linewidth]{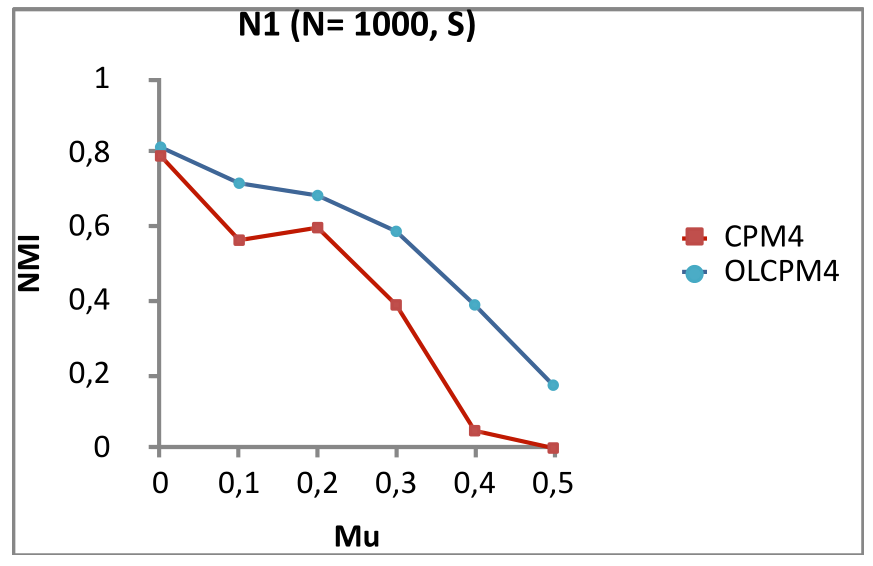}
 \end{subfigure}
     \begin{subfigure}{0.4\linewidth}
    \includegraphics[width=\linewidth]{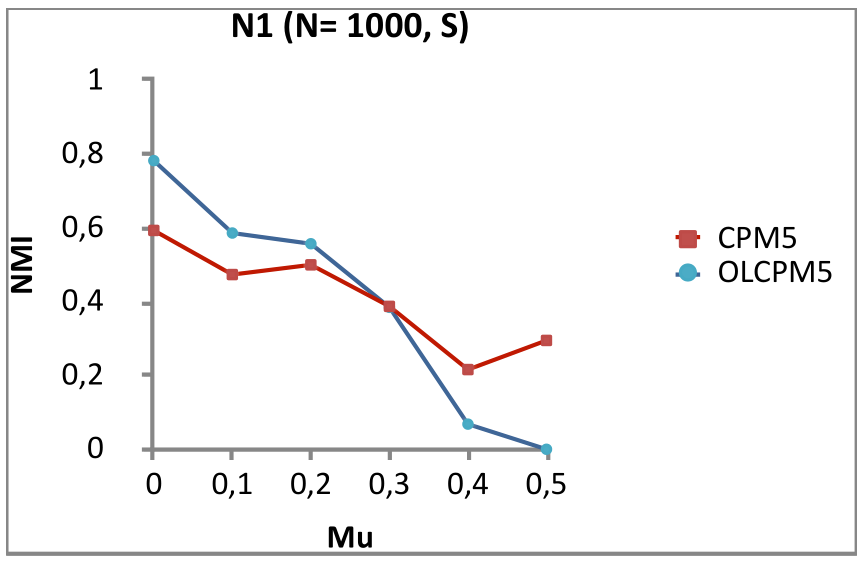}
 \end{subfigure}
 
     \begin{subfigure}{0.4\linewidth}
     \includegraphics[width=\linewidth]{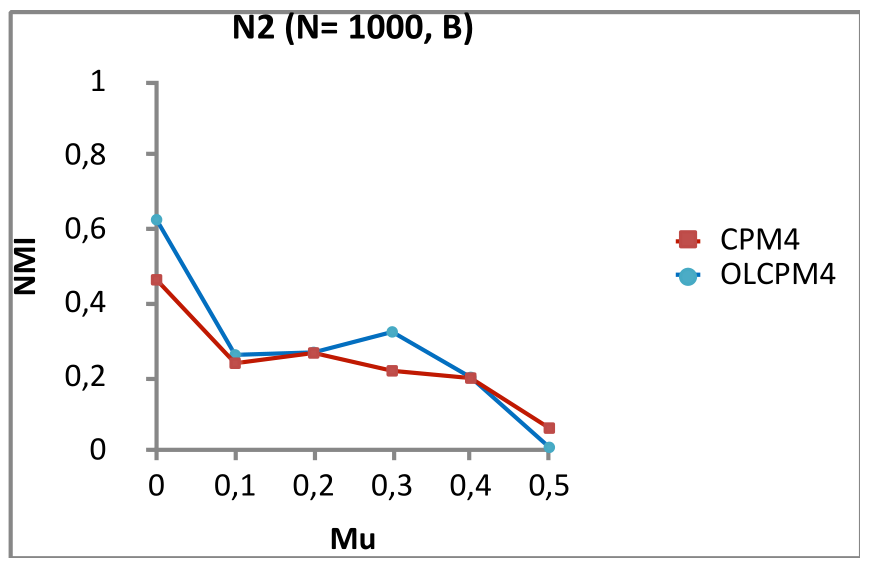}
 \end{subfigure}
     \begin{subfigure}{0.4\linewidth}
    \includegraphics[width=\linewidth]{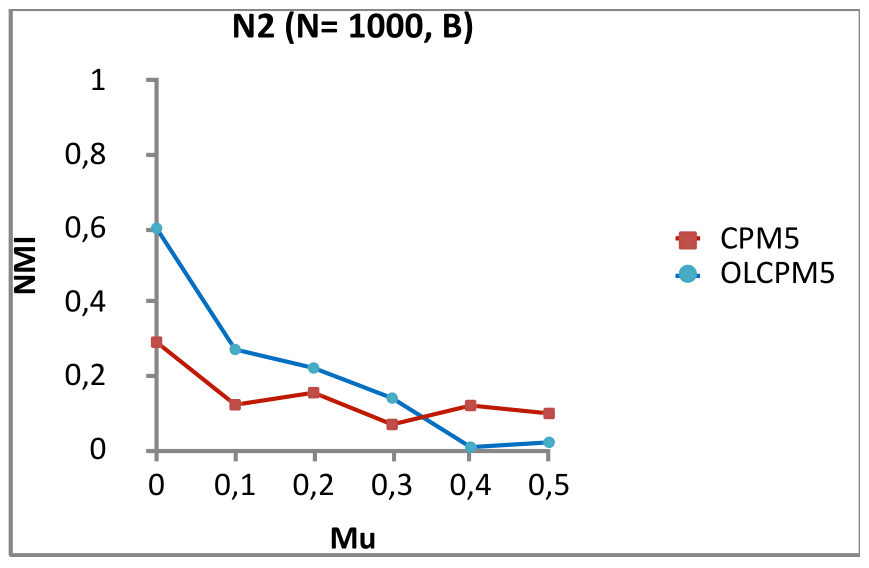}
 \end{subfigure}
 
     \begin{subfigure}{0.4\linewidth}
    \includegraphics[width=\linewidth]{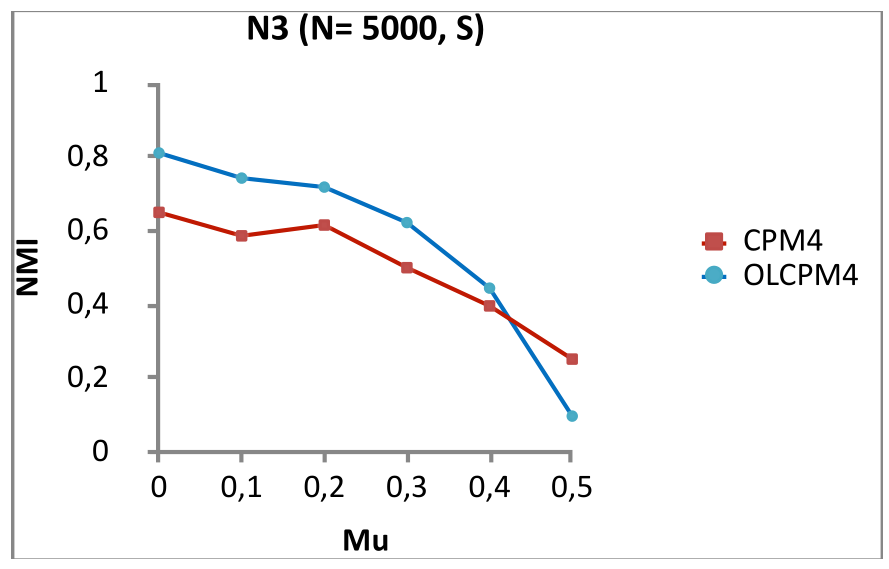}
 \end{subfigure}
     \begin{subfigure}{0.4\linewidth}
    \includegraphics[width=\linewidth]{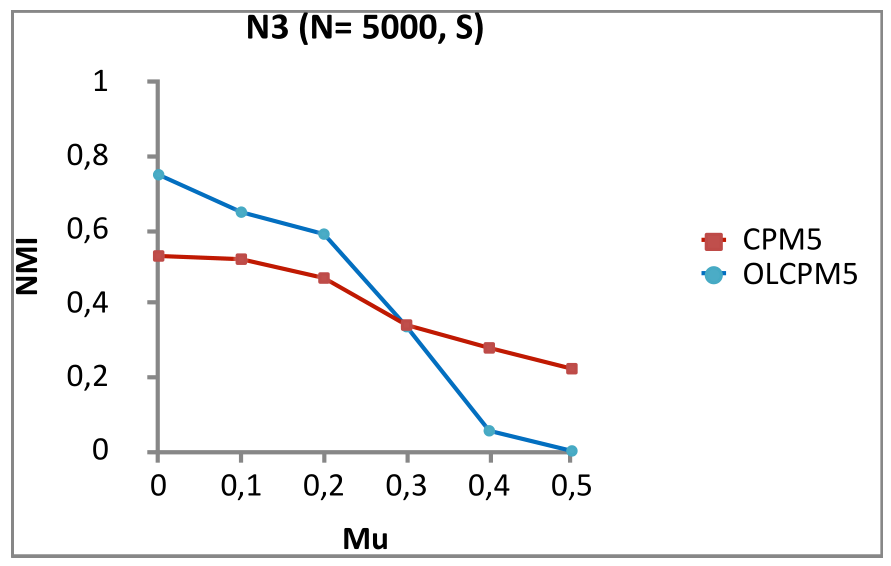}
 \end{subfigure}
     \begin{subfigure}{0.4\linewidth}
    \includegraphics[width=\linewidth]{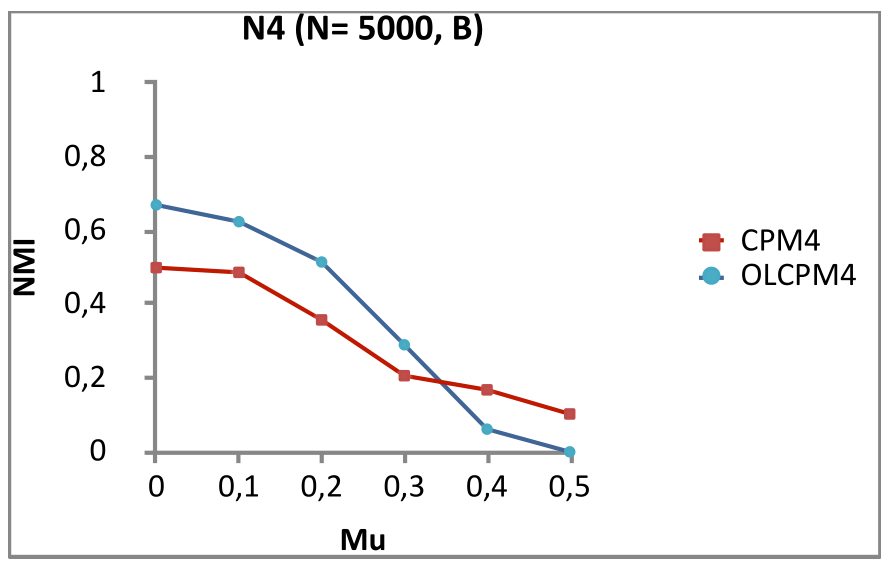}
 \end{subfigure}
     \begin{subfigure}{0.4\linewidth}
    \includegraphics[width=\linewidth]{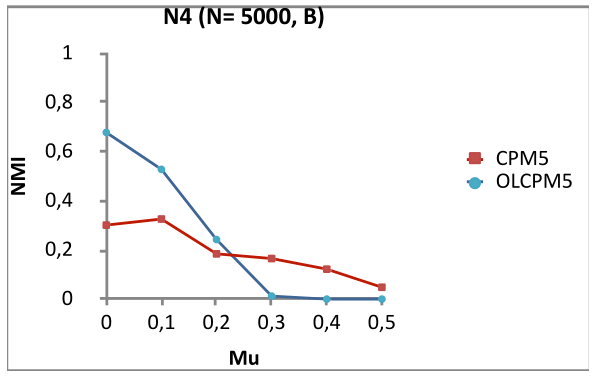}
 \end{subfigure}
\caption{The NMI scores of OCPM and OLCPM for $k=4$ and $k=5$ on the LFR benchmark networks as a function  of the mixing parameter $\mu$ for different network sizes (small networks in the upper half plots and large networks in the lower half plots) and different community sizes ($(S)$ ranges from $10$ to $50$ and $(B)$ ranges from $20$ to $100$).}
\label{fig:LFRRes1}
\end{figure*}

\begin{figure*}[!h]
\centering     
\begin{subfigure}{0.4\linewidth}
    \includegraphics[width=\linewidth]{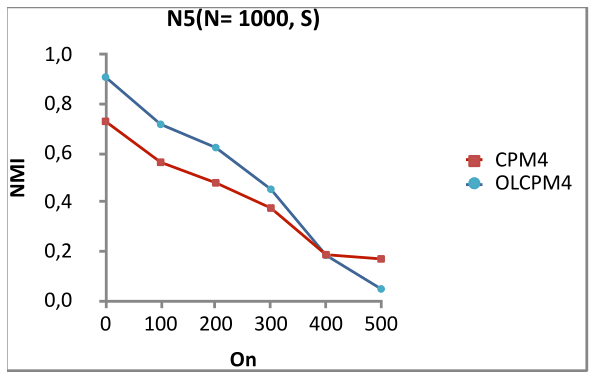}
 \end{subfigure}
     \begin{subfigure}{0.4\linewidth}
    \includegraphics[width=\linewidth]{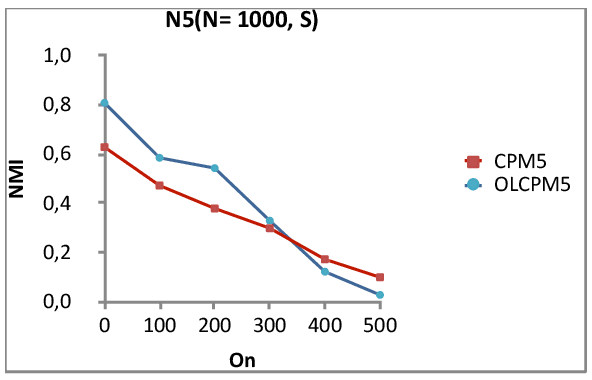}
 \end{subfigure}
 
     \begin{subfigure}{0.4\linewidth}
    \includegraphics[width=\linewidth]{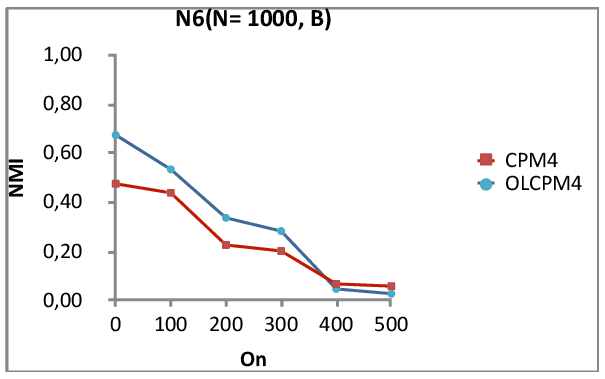}
 \end{subfigure}
     \begin{subfigure}{0.4\linewidth}
    \includegraphics[width=\linewidth]{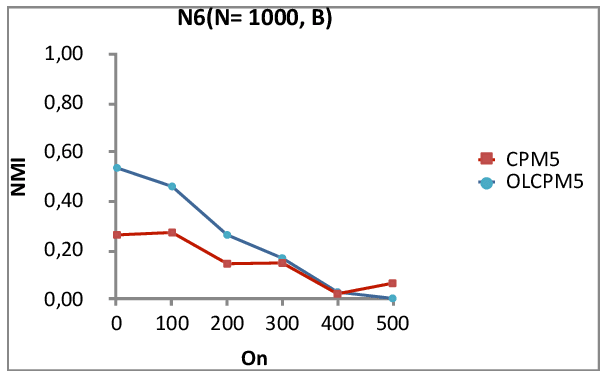}
 \end{subfigure}
 
     \begin{subfigure}{0.4\linewidth}
    \includegraphics[width=\linewidth]{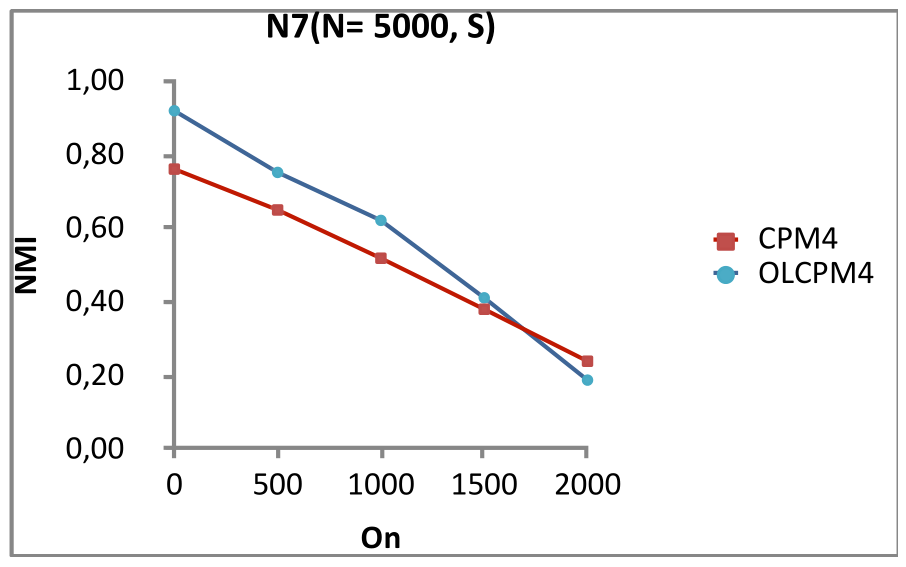}
 \end{subfigure}
     \begin{subfigure}{0.4\linewidth}
    \includegraphics[width=\linewidth]{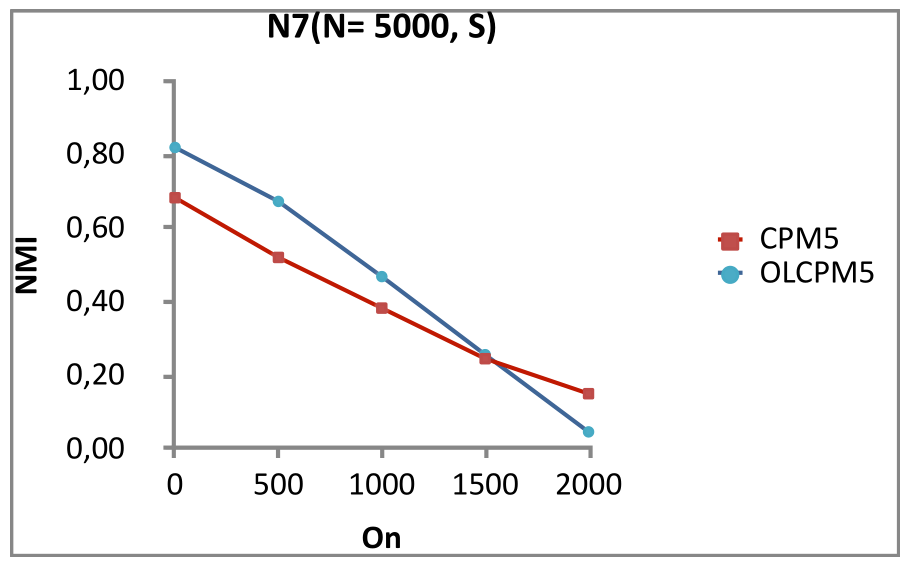}
 \end{subfigure}
     \begin{subfigure}{0.4\linewidth}
    \includegraphics[width=\linewidth]{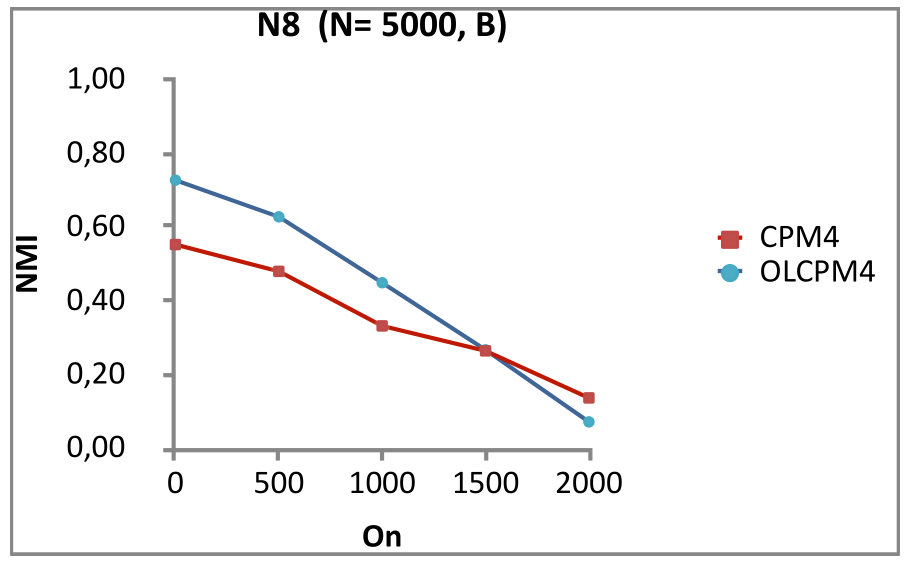}
 \end{subfigure}
     \begin{subfigure}{0.4\linewidth}
    \includegraphics[width=\linewidth]{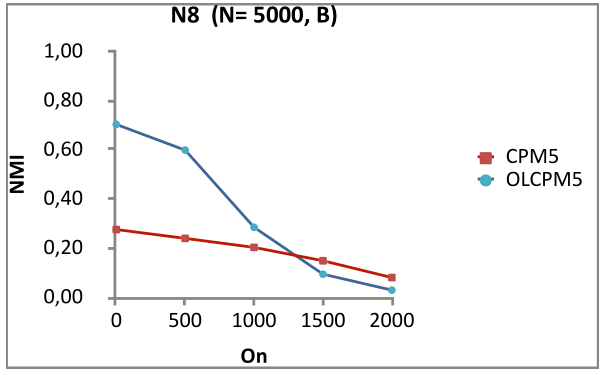}
 \end{subfigure}
\caption{The NMI scores of OCPM and OLCPM for $k=4$ and $k=5$ on the LFR benchmark networks as a function of the number of overlapping nodes $On$ for different network sizes (small networks in the upper half plots and large networks in the lower half plots) and different community sizes ($(S)$ ranges from $10$ to $50$ and $(B)$ ranges from $20$ to $100$).}
\label{fig:LFRRes2}
\end{figure*}

CPM and OLCPM are run for $k=4$ and for $k=5$. The NMI values of communities detected by CPM and OLCPM are depicted in Figure \ref{fig:LFRRes1} and Figure \ref{fig:LFRRes2}. Note that communities found by CPM and OCPM are identical, therefore the observed differences are only due to the post process.




As can be seen from Figure \ref{fig:LFRRes1} and Figure \ref{fig:LFRRes2}, the NMI scores of OCPM and OLCPM are substantially equivalent for $k=4$ and $k=5$. In most cases, OLCPM achieves the highest results, except for the two cases where: (1)  the community structure becomes very fuzzy ( $On >= 400$ for small networks or $On >=1500$ for large networks) or (2) the value of $\mu$ is large (greater than $0.3$). In these cases, OLCPM performs similar or slightly worse than CPM- it depends on the value of $k$. When the community structure becomes too fuzzy for CPM, the irrelevant core-communities provided are probably worsened by the post-process. 
For $k=5$ achieves the highest results in most cases, except for the two cases where: (1)  the community structure becomes very fuzzy ( $On >= 400$ for small networks or $On >=1500$ for large networks) or (2) the value of $\mu$ is large (greater than $0.3$). In these cases, OLCPM performs similar or worse than CPM, especially for higher value of $k$. When the community structure becomes too fuzzy for CPM, the irrelevant core-communities provided are probably worsened by the post-process. 
As a conclusion, we can consider that in situations in which CPM finds meaningful communities in a network, the proposed post-process improves the solution. 

\subsubsection{Dynamic real-world networks}

In order to evaluate the community detection results of our framework OLCPM on real temporal networks, we leverage a high-resolution time-varying network describing contact patterns among high school students in Marseilles, France \cite{fournet2014contact}. The dataset was collected by the SocioPatterns collaboration using wearable sensors, able to capture proximity between individuals wearing them. The dataset was gathered during nine days (Monday to Tuesday) in November 2012. Data collection involved 180 students from five classes. Proximity relations are detected over 20-second intervals. Data collection involved students' classes corresponding to different specializations: 'MP' classes focus more on mathematics and physics, 'PC' classes on physics and chemistry and 'PSI' classes on engineering studies.  These classes represent the expected ground-truth community structure.

We construct a dynamic network composed of 216 snapshots, each corresponding to 1 hour of data. Nodes correspond to students, and there is an edge between two nodes in a snapshot if the corresponding students have been observed in interaction at least once during the corresponding period. (Please refer to the original paper \cite{fournet2014contact} for details about the meaning of \textit{interaction}. To sum up, two students are in interaction if they stand face-to-face at a distance between 1 and 1.5 meters.)

The constructed dynamic network is used into two sets of experiments. The first experiments compares the two algorithms of our framework:  OLCPM and OCPM, while the second compares the OLCPM with other methods.

\textbf{Comparing OCPM and OLCPM. }

\begin{figure*}[!h]
\begin{center}
\includegraphics[width=\linewidth]{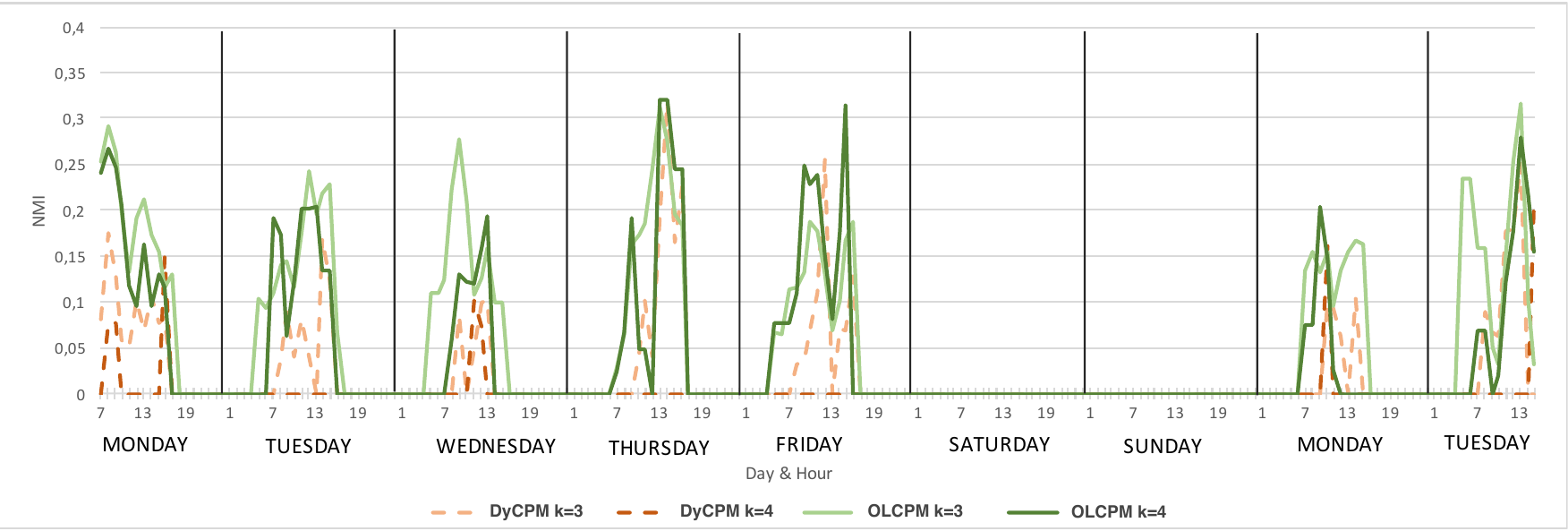}
\end{center}
\caption{NMI values of OLCPM and DyCPM \cite{palla2007quantifying} for $k=3$ and $k=4$ on SocioPatterns collaboration networks \cite{fournet2014contact}. }
\label{fig:NMI}
\end{figure*}

In this set of experiments, we compute the communities at each step using both DyCPM and OLCPM (Communities yielded by DyCPM and OCPM are identical). Then, for each snapshot, we compute the NMI according to \cite{lancichinetti2009compare}.
Results are displayed in Figure \ref{fig:NMI}. We show results for $k=3$ and $k=4$, which yield the best results.

The average NMI over all snapshots is provided in Table \ref{tab:ANMI}.
\begin{table} [!h] 
\resizebox{\textwidth}{!}{%
\small
\begin{tabular}{|c||c|c|c|c|}
        \hline
        \textbf{Algorithm} & DyCPM $k=3$ & DyCPM $k=4$ & OLCPM $k=3$ & OLCPM $k=4$\\ 
        \hline
       \textbf{ Average NMI} & 0.024	 & 0.004  & 0.059  & 0.044 \\ 
       \hline
\end{tabular}} \\  
      
        \caption{Average NMI scores of OLCPM and DyCPM \cite{palla2007quantifying} for $k=3$ and $k=4$ on SocioPatterns collaboration networks \cite{fournet2014contact}.}\label{tab:ANMI}
\end{table}

We can observe that the average NMI of OLCPM is higher than the original DyCPM, and that values of NMI are also higher for most snapshots.

The longitudinal visualization of Figure \ref{fig:NMI} illustrates the relevance of studying the evolution of a network with a fine granularity: only looking at this plot, we can see that the class structure is not always present in the data. For instance, we can observe that there is no community structure during evenings and weekends, or that the community structure is less observable during several days around lunchtime (Thursday, Friday, second Monday). One can then look in more detail to the communities found and their evolution to interpret these observations.
In this example, we were able to run DyCPM because of the small size of the network, the restriction to one-hour interval, and the limitation to 9 days of data, but, as shown previously, it would not be possible to extend this analysis to a much larger number of steps due to the increase in complexity.

\textbf{Comparing OLCPM with other methods. }

In this set of experiments, we compare our framework against three state-of-the-art community detection methods : 

\begin{itemize}
    \item The method by \cite{greene2010tracking}: We used a costume implementation which uses the Louvain method for community detection, and  the Jaccard coefficient to match between communities in consecutive snapshots, with a minimal similarity threshold of $0.3$.
    \item The method by \cite{falkowski2006mining}: the implementation used for this method run the Louvain algorithm in each snapshot, then uses  the Jaccard coefficient with a minimal similarity threshold of 0.3 to match any community with any other one in any other snapshot, constituting a survival graph. Louvain algorithm is then applied on this survival graph, yielding dynamic communities.
    \item The method by \cite{guo2014evolutionary}: we use a naive implementation of this method which creates, at each snapshot, a new graph  combining the graph at this step and a graph in which edges are present between any two nodes belonging to the same community in the previous step. The method has is a parameter to tune how important is the weight of the current topology compared with previous partition. This parameter is set to $0.9$.
\end{itemize}

\begin{figure*}[!htbp]
    \centering
 \begin{subfigure}{\linewidth}
  \centering
    \includegraphics[scale = 0.3]{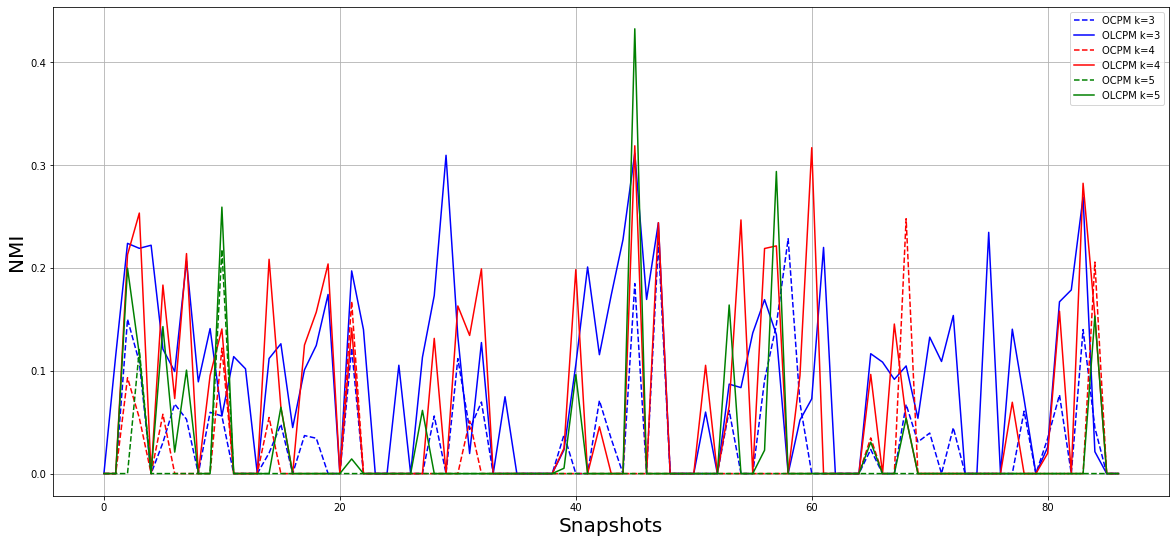}
 \end{subfigure}
\caption{NMI values for OCPM and OLCPM for different values of $k$ ($k=3$, $k=4$ and $k=5$) on SocioPatterns collaboration networks \cite{fournet2014contact}.}
\label{fig:COMP1}
\end{figure*}

\begin{figure*}[!htbp]
    \centering
     \begin{subfigure}{\linewidth}
     \centering
    \includegraphics[scale = 0.3]{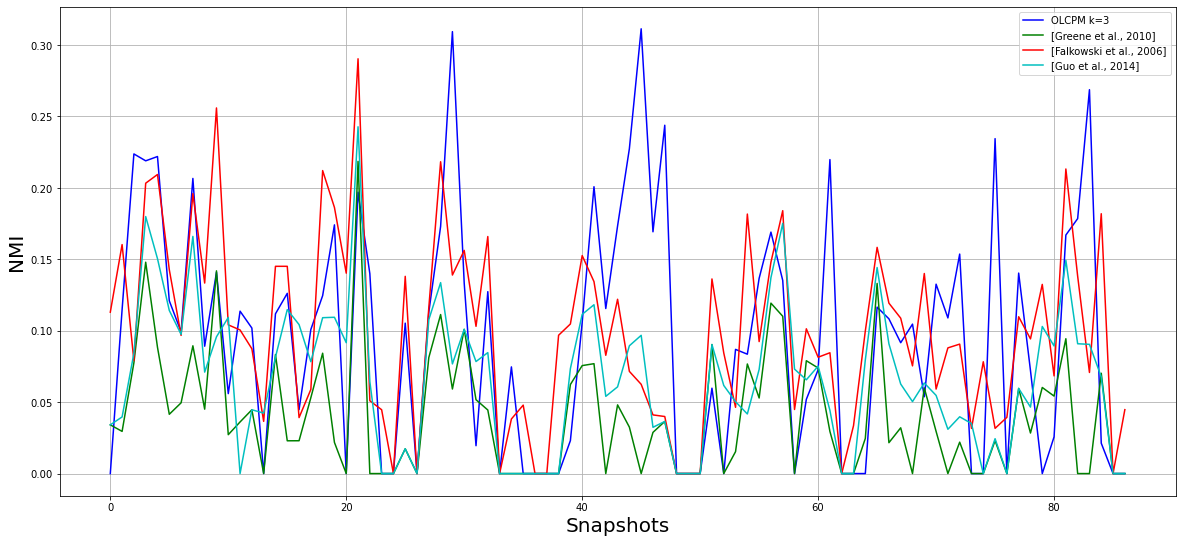}
 \end{subfigure}
\caption{Comparaison of NMI scores obtained  for OLCPM ($k=3$) and other methods on SocioPatterns collaboration networks \cite{fournet2014contact}.}
\label{fig:COMP2}
\end{figure*}

For each method, we compute at each step (snapshot) the similarity between the obtained communities and the ground truth. 
The Normalized Mutual Information (NMI) \cite{lancichinetti2009compare} is used as similarity measure. 
To enhance legibility, we kept out only non-empty snapshots (87 in total).
Results are displayed in Figure \ref{fig:COMP1} and Figure \ref{fig:COMP2}. 
In Table \ref{tab:NMI2}, we summarize the average NMI over all time steps for each method.


As can be seen, the average NMI for OCPM for the three values of $k$ is lower than for all other methods. The results are significantly improved by applying the post processing (OLCPM)(see Figure \ref{fig:COMP1}). For $k=3$, OLCPM reached a score equivalent to the best one obtained in the method by \cite{falkowski2006mining}(see Figure \ref{fig:COMP2}). 
\begin{table}[!h]
\begin{tabular}{|l||l|}
\hline
\textbf{Method} & \textbf{Average NMI} 
\\ \hline
OCPM ($k=3$)     & 0.031 
\\ \hline
OLCMP ($k=3$)     & \textbf{0.095} 
\\ \hline
OCPM ($k=4$)     & 0.012
\\ \hline
OLCPM ($k=4$)     & 0.066 
\\ \hline
OCPM ($k=5$)     & 0.003 
\\ \hline
OLCPM ($k=5$)     & 0.025
\\ \hline
\cite{greene2010tracking}               & 0.040                 \\ \hline
\cite{falkowski2006mining}               & \textbf{0.099}                
\\ \hline
\cite{guo2014evolutionary}               & 0.066                \\ \hline
\end{tabular}
\caption{Average NMI of OLCPM (for different values of $k$) and other methods on SocioPatterns collaboration netwoks \cite{fournet2014contact}.}
\label{tab:NMI2}
\end{table}
\newpage
\section{Conclusion}
This chapter introduced a novel framework for the problem of detecting overlapping dynamic community structures within social networks. The proposed framework \cite{boudebza2018}, meets three requirements: it is built on a deterministic and intrinsic community definition; it is capable of detecting overlapping communities; and it works on steam graphs which are suitable for modeling social networks.  
We proposed OCPM, an online version of the Clique percolation method (CPM) \cite{palla2005uncovering}, working on a fully dynamic network model, i.e., described as ﬂows of events, where nodes or edges can be added or removed over time. Instead of calculating all k-cliques for the whole network at each event occurring in the network, our method updates only the community structure alongside the node or the edge involved in the event. This local update of the community structure provides a signiﬁcant improvement in computational time. To cope with the covering problem of CPM, we proposed OLCPM algorithm, a post-process on OCPM based on label propagation, applied on peripheral nodes, i.e., nodes that do not belong to OCPM communities. The experimental results of our framework in both artiﬁcial and real-world networks exhibit good performance in both computing time and quality detection.
In Chapter \ref{chp:Stable}, we will address another problem of dynamic community detection related to the use of snapshots models. We will also propose a solution to avoid the problem of determining the right temporal scale within snapshot models.

\chapter{Temporal Multi-Scale Community Detection}\label{chp:Stable}
\thispagestyle{empty}
\vspace{1cm}

\parindent=0em
\etocsettocstyle{\rule{\linewidth}{\tocrulewidth}\vskip0.5\baselineskip}{\rule{\linewidth}{\tocrulewidth}}
\localtableofcontents

\clearpage

\section{Introduction}
In recent years, studying interactions over time has witnessed a growing interest in a wide range of fields, such as sociology, biology, physics, etc. Such dynamic interactions are often represented using the snapshot model: the network is divided into a sequence of static networks, i.e., snapshots, aggregating all contacts occurring in a given time window. The main drawback of this model is that it often requires to choose arbitrarily a temporal scale of analysis. The link stream model \cite{Latapy2018} is a more effective way for representing interactions over time, that can fully capture the underlying temporal information. Under this model, the main challenge is to mine more efficiently and smoothly both temporal and topological structures. 

Real-world networks evolve frequently at many different time scales. Fluctuations in such networks can be observed at yearly, monthly, daily, hourly, or even smaller scales.  For instance, if one were to look at interactions among workers in a company or laboratory, one could expect to discover clusters of people corresponding to \textit{meetings} and/or \textit{coffee breaks}, interacting at \textbf{high frequency} (e.g., every few seconds) for \textbf{short periods} (e.g., few minutes), \textit{project members} interacting at \textbf{medium frequency} (e.g., once a day) for \textbf{medium period}s (e.g., a few months), \textit{coordination groups} interacting at \textbf{low frequency} (e.g., once a month) for \textbf{longer periods} (e.g., a few years), etc.

An analysis of communities found at an arbitrarily chosen scale would necessarily miss some of these communities: low latency ones are invisible using short aggregation windows, while high-frequency ones are lost in the noise for long aggregation windows. A multiple temporal scale analysis of communities seems therefore the right solution to study networks of interactions represented as link streams.

To the best of our knowledge, no such method exists in the literature. In this chapter, we propose a method having roots both in the literature on change-point detection and in dynamic community detection and more precisely in streaming methods. It detects what we call \textbf{stable communities}, i.e., \textit{groups of nodes} forming a \textit{coherent community} throughout a \textit{period of time}, at a given \textit{temporal scale}. The proposed method falls into the class of Cross-time approaches. 


The remainder of this chapter is organized as follows. In Section \ref{related_work}, we present the roots for the method we propose namely streaming methods and change-point detection. 
Then, we describe the proposed framework in detail in Section \ref{framework}. We experimentally evaluate the proposed method on both synthetic and real-world networks in Section \ref{experiments}.

\section{Origin of the method}
\label{related_work}
Our contribution in this chapter relates to two active body of research: $i)$ streaming methods for dynamic community detection and $ii)$ change-point detection. The aim of the former is to discover groups in link streams, while the objective of the latter is to detect changes in the overall structure of a dynamic network. This section briefly presents both categories which are the basis of our proposal. 

\subsection{Streaming Methods}
As highlighted in our review about dynamic community detection, most methods consider that the studied dynamic networks are represented as sequences of snapshots, with each snapshot being a well-formed graph with meaningful community structure, see for instance \cite{mucha2010community,greene2010tracking}. Some other methods work with interval graphs and update the community structure at each network change, e.g., \cite{rossetti2017tiles, cazabet2011simulate}. However, those methods are not adapted to deal with link streams, for which the network is usually not well-formed at any given time. Using them on such a network would require to first aggregate the links of the stream by choosing arbitrarily a temporal scale (aggregation window). 

The little research that exists to handle link streams has nonetheless some limitations. For instance, due to the usage of Stochastic Block Model, the method by \cite{Viard:2016:CMC:2853249.2853730} (see Section \ref{chpt2:DCDM} for further detail) provides only a single partition of the nodes that is considered constant through time. It is therefore impossible for nodes to switch between communities, appear, disappear or change behaviors. Another interesting method is the one introduced by \cite{Viard:2016:CMC:2853249.2853730}. It is not dedicated to the detection of communities but rather to maximal $\Delta$ cliques (an extreme case of communities), i.e., groups of nodes having at least one interaction between all of them during a user-defined period $\Delta$. This method requires to fix a unique granularity using a time period $\Delta$. 

Compared to those methods, the solution we propose is able to: $i)$ Discover communities at multiple temporal scales without redundancy and $ii)$ Allow nodes to belong to several communities at different periods and different temporal scales.

\subsection{Change-point detection}

Our work is also related to research conducted on change-point detection considering community structures in dynamic networks. 
In these approaches, given a sequence of snapshots, one wants to detect periods during which the network organization and/or the community structure remains stable. In what follows we introduce examples of such methods.

The work by \cite{Peel204} introduced the first change-point detection method for evolving networks that uses generative network models and statistical hypothesis testing. The Generalized Hierarchical Random Graph (GHRG) model is used to define a parametric probability distribution over network snapshots and to compactly model nested community structure. Then, two models are inferred (a model for representing the change at time $t$ in the snapshot, and a model for representing the null hypothesis of no change over the entire snapshot). Bayesian hypothesis testing is used to choose the best model. This method can detect when do the change occurs, and the shape of the change. The main issue of this method is its scalability on large networks. 

The authors in \cite{Wang2017} proposed a method to detect local and global change points in the community structure. Each snapshot is contracted into a weighted hypergraph, in which hyper-nodes are the communities detected at this snapshot. Then, local changes are detected by measuring similarities between hyper-nodes (communities) in successive hyper-networks, while global changes are detected by running community detection algorithms on successive hyper-graphs. 

Similarly, \cite{Zhu2018} formulated the problem of change-point detection as clustering of hyper-networks. 
First, they construct a weighted hyper-network, where nodes represent the snapshots and weights represent similarities between snapshots, based on structural feature and similarity measurements (relative importance of nodes). Then, community detection is run on this hyper-graph. At final, they serialize the community detection results in chronological order, where each snapshot is labeled by its community identity. This method can detect $i)$ local changes, $ii)$ global changes and $iii)$ isomorphic changes.

We also note that our work is related to the one by \cite{Masuda2019}. The authors seek to identify sequences of system states in data streams. The idea is to $i)$ transform a data stream into a sequence of snapshots (non-overlapping windows of size $W$), $ii)$ measure the pairwise distance between snapshots, $iii)$ run a clustering algorithm on the matrix distance and categorize snapshots into discrete states.

From those methods, our proposal keeps the principle of stable periods delimited by change points, and the idea of detecting changes at local and global scales. But our method differs in two directions: $i)$ we are searching for stable individual communities instead of stable graph periods, and $ii)$ we search for stable structures at multiple levels of temporal granularity.

\section{Method}
\label{framework}
The goal of our proposed method is $i)$ to detect stable communities $ii)$ at multiple scales without redundancy and $iii)$ to do so efficiently. Thus, we adopt an iterative approach, searching communities from the coarser to the more detailed temporal scales. At each temporal scale, we use a three-step process:
\begin{enumerate}
    \item \textbf{Seed Discovery}, to find relevant community seeds at this temporal scale.
    \item \textbf{Seed Pruning}, to remove seeds that are redundant with communities found at higher scales.
    \item \textbf{Seed Expansion}, expanding seeds in time to discover stable communities.
\end{enumerate}

We start by presenting each of these three steps, and then we describe the method used to iterate through the different scales in Section \ref{iterative}.

Our work aims to provide a general framework that could serve as a baseline for further work in this field. We define three generic functions that can be set according to the user needs:

\begin{itemize}
    \item \textbf{CD($g$)}, a static community detection algorithm on a graph $g$.
    \item \textbf{QC($N,g$)}, a function to assess the quality of a community defined by the set of nodes $N$ on a graph $g$.
    \item \textbf{CSS($N_1$,$N_2$)}, a function to assess the similarity of two sets of nodes $N_1$ and $N_2$.
\end{itemize}

See Section \ref{parameters} on how to choose proper functions for those tasks.

We define a stable dynamic community $c$ as a triplet $c=(N,p,\gamma)$, with $c.N$ the list of nodes in the community, $c.p$ its period of existence defined as an interval, e.g., $c.p=[t_1,t_2[$\footnote{We use right open intervals such as a community starting at $t_x$ and another one ending at the same $t_x$ have an empty intersection, which is necessary to have coherent results when handling discrete time steps.} means that the community $c$ exists from $t_1$ to $t_2$, and $c.\gamma$ the temporal granularity at which $c$ has been discovered. 

We denote the set of all stable dynamic communities $\mathcal{D}$.

\subsection{Seed discovery}
For each temporal scale, we first search for interesting seeds. A temporal scale is defined by a granularity $\gamma$,  expressed as a period of time (e.g.; 20 minutes, 1 hour, 2 weeks, etc).We use this granularity as a window size, and, starting from a time $t_0$ --by default, the date of the first observed interaction-- we create a cumulative graph (snapshot) for every period $[t_0,t_0+\gamma[,[t_0+\gamma,t_0+2\gamma[,[t_0+2\gamma,t_0+3\gamma[,etc.$, until all interactions belong to a cumulative graph. This process yields a sequence of static graphs, such as $G_{t_0,\gamma}$ is a cumulated snapshot of link stream $G$ for the period starting at $t_0$ and of duration $\gamma$. $G_{\gamma}$ is the list of all such graphs.

Given a static community detection algorithm $CD$ yielding a set of communities, and a function to assess the quality of communities $QC$, we 
apply $CD$ on each snapshot and filter promising seeds, i.e., high quality communities,
using $QC$. The set of valid seeds $\mathcal{S}$ is therefore defined as: 
 
\begin{equation} 
\mathcal{S} = \{QC(s,g)>\theta_q, \forall g \in G_{\gamma}, \forall s \in CD(g)\},
\end{equation}
with $\theta_q$ a threshold of community quality.  

Since community detection at each step is independent, we can run it in parallel on all steps, this is an important aspect for scalability. 

\subsection{Seed pruning}
The seed pruning step has a twofold objective: $i)$ reducing redundancy and $ii)$ speed up the multi-scale community detection process. Given a measure of structural similarity $CSS$, we prune the less interesting seeds, such as the set of filtered seeds $\mathcal{FS}$ is defined as:

\begin{equation} 
\mathcal{FS} = \{(CSS(s.N, c.N) \leq \theta_s) \vee (s.p \cap c.p = \emptyset), \forall s \in \mathcal{S}, \forall c \in \mathcal{D}\},
\end{equation}

where $\mathcal{D}$ is the set of stable communities discovered at coarser (or similar, see next section) scales, $s.p$ is the interval corresponding to the snapshot at which this seed has been discovered, and $\theta_s$ is a threshold of similarity.

Said otherwise, we keep as interesting seeds those that are topologically not redundant, i.e., having similar structure(nodes/edges), OR temporally not redundant, i.e., appearing at several scales. In other words, a seed is kept if it corresponds to a situation never seen before. 

\subsection{Seed expansion}
The aim of this step is to assess whether a seed corresponds to a \textit{stable} dynamic community.

Most static algorithms suffer from a major drawback when dealing with temporal networks: the \textit{instability} problem. This problem has been identified since the early stages of the dynamic community detection field \cite{aynaud2010static}. In few words, the same algorithm ran twice on the same network after introducing minor random modifications might yield very different results. As a consequence, one cannot know if the differences observed between the community structure found at $t$ and $t+1$ are due to structural changes or to the instability of the algorithm. This problem is usually solved by introducing smoothing techniques \cite{rossetti2018community}. Our method uses a similar approach, but instead of comparing communities found at step $t$ and $t-1$, we check whether a community found at $t$ is still relevant in previous and following steps, recursively.

More formally, for each seed $s \in S$ found on the graph $G_{t,\gamma}$, we iteratively expand the duration of the seed $s.d=[t,t+\gamma[$ (where $t$ is the time start of this duration) at each step $t_i$ in both temporal directions ($t_i \in (...[t-2\gamma,t-\gamma[,[t-\gamma,t]; [t+\gamma,t+2\gamma[,[t+2\gamma,t+3\gamma]...))$ as long as the quality $QC(s.N,G_{t_i,\gamma})$ of the community defined by the nodes $s.N$ on the graph at $G_{t_i,\gamma}$ is good enough.
Here, we use the same similarity threshold $\theta_s$ as in the seed pruning step. If the final duration period $|s.p|$ of the expanded seed is higher than a duration $\theta_p \gamma$, with $\theta_p$ a threshold of stability, the expanded seed is added to the list of stable communities, otherwise, it is discarded. This step is formalized in Algorithm \ref{alg:extend}.

\begin{algorithm}[H]
\SetAlgoLined
\KwIn{$s, \gamma,\theta_p,\theta_s 
$}
 $t \gets t^{start} | s.p = [t^{start},t^{end}[$ \;
 $g \gets G_{t,\gamma}$\;
 $p \gets [t,t+\gamma[$\;
 \While{$QC(s.N,g)>\theta_s$}{
  $s.p \gets s.p \cup p$\;
  $t \gets t+\gamma $\;
  $p \gets [t,t+\gamma[$\;
  $g \gets G_{t,\gamma}$\;
  }
  \If{$|s.p| \geq \theta_p \gamma$}{
    $\mathcal{D} \gets  \mathcal{D} \cup \{s$\}\;
   }

 \caption{\textbf{Forward seed expansion}. 
 Forward temporal expansion of a seed $s$ found at time $t$ of granularity  $\gamma$. The reciprocal algorithm is used for \textit{backward} expansion: $t+1$ becomes $t-1$.}
 \label{alg:extend}
\end{algorithm}

In order to select the most relevant stable communities, we consider seeds in descending order of their $QC$ score, i.e., the seeds of higher quality scores are considered first.
Due to the pruning strategy, a community of the lowest quality might be pruned by a community of the highest quality at the same granularity $\gamma$.

\subsection{Multi-scale iterative process}
\label{iterative}
Until then, we have seen how communities are found for a particular time scale. In order to detect communities at multiple scales, we first define the ordered list of studied scales $\Gamma$. The largest scale is defined as $\gamma^{max}=|G.d|/\theta_p$, with $|G.d|$ the total duration of the dynamic graph. Since we need to observe at least $\theta_p$ successive steps to consider the community stable, $\gamma^{max}$ is the largest scale at which communities can be found.

We then define $\Gamma$ as the ordered list: 

\begin{equation}
    \Gamma=[\gamma^{max}, \gamma^{max}/2^1, \gamma^{max}/2^2, \gamma^{max}/2^3,..., \gamma^{max}/2^k ], 
\end{equation}
with $k$ such as $\gamma^{max}/2^k>\theta_{\gamma}>= \gamma^{max}/2^{k+1} $, $\theta_{\gamma}$ being a parameter corresponding to the finest temporal granularity to evaluate, which is necessarily data-dependant (if time is represented as a continuous property, this value can be fixed at least at the sampling rate of data collection).

This exponential reduction in the studied scale guarantees a limited number of scales to study.

The process to find seeds and extend them into communities is then summarized in Algorithm \ref{alg:iterative}.

\begin{algorithm}[H]
\SetAlgoLined
\KwIn{$G,\theta_q,\theta_s,\theta_p,\theta_{\gamma}$}
$\mathcal{D} \gets \{\emptyset\}$\;
$\Gamma \gets $studied\_scales($G,\theta_{\gamma})$ \;
 \For{$\gamma \in \Gamma$}{
    $\mathcal{S} \gets $ Seed\_Discovery($\gamma
    ,CD,QC,\theta_q$)\;
    $\mathcal{FS} \gets $Seed\_Pruning($\mathcal{S},CSS,\theta_s$)\;
    \For{$s \in \mathcal{S}$}{
        Seed\_Expansion($s,\gamma,\theta_p,\theta_s$)\;
     }
  }

 \caption{\textbf{Multi-temporal-scale stable communities finding}. Summary of the proposed method. See corresponding sections for the details of each step. $G$ is the link streams to analyze, $\theta_q,\theta_s,\theta_p,\theta_{\gamma}$ are threshold parameters.
}
 \label{alg:iterative}
\end{algorithm}

\subsection{Choosing functions}
\label{parameters}
The proposed method is a general framework that can be implemented using different functions for $CD, QC$ and $CSS$. 
This section provides explicit guidance for selecting each function, and introduces the choices we make for the experimental section. 

\subsubsection{Community Detection - CD}
Any algorithm for community detection could be used, including overlapping methods, since each community is considered as an independent seed. Following literature consensus, we use the Louvain method \cite{blondel2008fast}, which yields non-overlapping communities using a greedy modularity-maximization method. The Louvain method performs well on static networks, it is in particular among the fastest and most efficient methods.
Note that it would be meaningful to adopt an algorithm yielding communities of good quality according to the chosen $QC$, which is not the case in our experiments, as we wanted to use the most standard algorithms and quality functions in order to show the genericity of our approach.

\subsubsection{Quality of Communities - QC}
The $QC$ quality function must express the quality of a set of nodes w.r.t a given network, unlike functions such as modularity, which express the quality of a whole partition w.r.t a given network. Many such functions exist, like \textit{Link Density} or \textit{Scaled Density} \cite{Labatut2017}, but the most studied one is probably the \textit{Conductance} \cite{leskovec2009community}. Conductance takes into account both internal density and out-going edges, and is defined as the ratio of $i)$ the number of edges between nodes inside the community and nodes outside the community, and $ii)$ the sum of degrees of nodes inside the community (or outside if this value is larger). Its value ranges from 0 (Best, all edges starting from nodes of the community are internal) to 1 (Worst, no edges between this community and the rest of the network). 
Since our generic framework expects good communities to have $QC$ scores higher than the threshold $\theta_q$, we adopt the definition $QC$=1-conductance. 

\subsubsection{Community Seed Similarity - CSS}
This function takes as input two sets of nodes and returns their similarity. Such a function is often used in dynamic community detection to assess the similarity between communities found in different time steps. Following \cite{greene2010tracking}, we choose as a reference function the Jaccard Index, a measure of similarity between pairs of sample sets. It is defined as the size of the intersection divided by the size of the union of the sample sets. 
Given two sets A and B, it is defined as: $J(A,B) = \frac{|A \cap B|}{ |A \cup B|}$

\subsection{Choosing parameters}
The algorithm has four parameters, $\theta_{\gamma},\theta_q,\theta_s,\theta_p$, defining different thresholds. We explicit them and provide the values used in the experiments.  
\begin{enumerate}
    \item $\theta_{\gamma}$ is data-dependant. It corresponds to the smallest temporal scale that will be studied and should be set at least at the collection rate. 
    For synthetic networks, it is set at 1 (the smallest temporal unit needed to generate a new stream), while, for SocioPatterns dataset, it is set to 20 seconds (the minimum length of time required to capture a contact). 
    \item $\theta_q$ determines the minimal quality a seed must have to be preserved and expanded. The higher this value, the more strict we are on the quality of communities. We set $\theta_q=0.7$ in all experiments. It is dependent on the choice of the $QC$ function.
    \item $\theta_s$ determines the threshold above which two communities are considered redundant. The higher this value, the more communities will be obtained. We set $\theta_s=0.3$ in all experiments. It is dependent on the choice of the $CSS$ function.
    \item $\theta_p$ is the minimum number of consecutive periods a seed must be 
    expanded in order to be considered as stable community. We set $\theta_s=3$ in all experiments. The value should not be lower in order to avoid spurious detections due to pure chance. Higher values could be used to limit the number of results.
\end{enumerate}

\section{Experiments and results}
\begin{figure}[!h]
\centering
\begin{subfigure}{0.8\textwidth}
    \includegraphics[width=\textwidth]{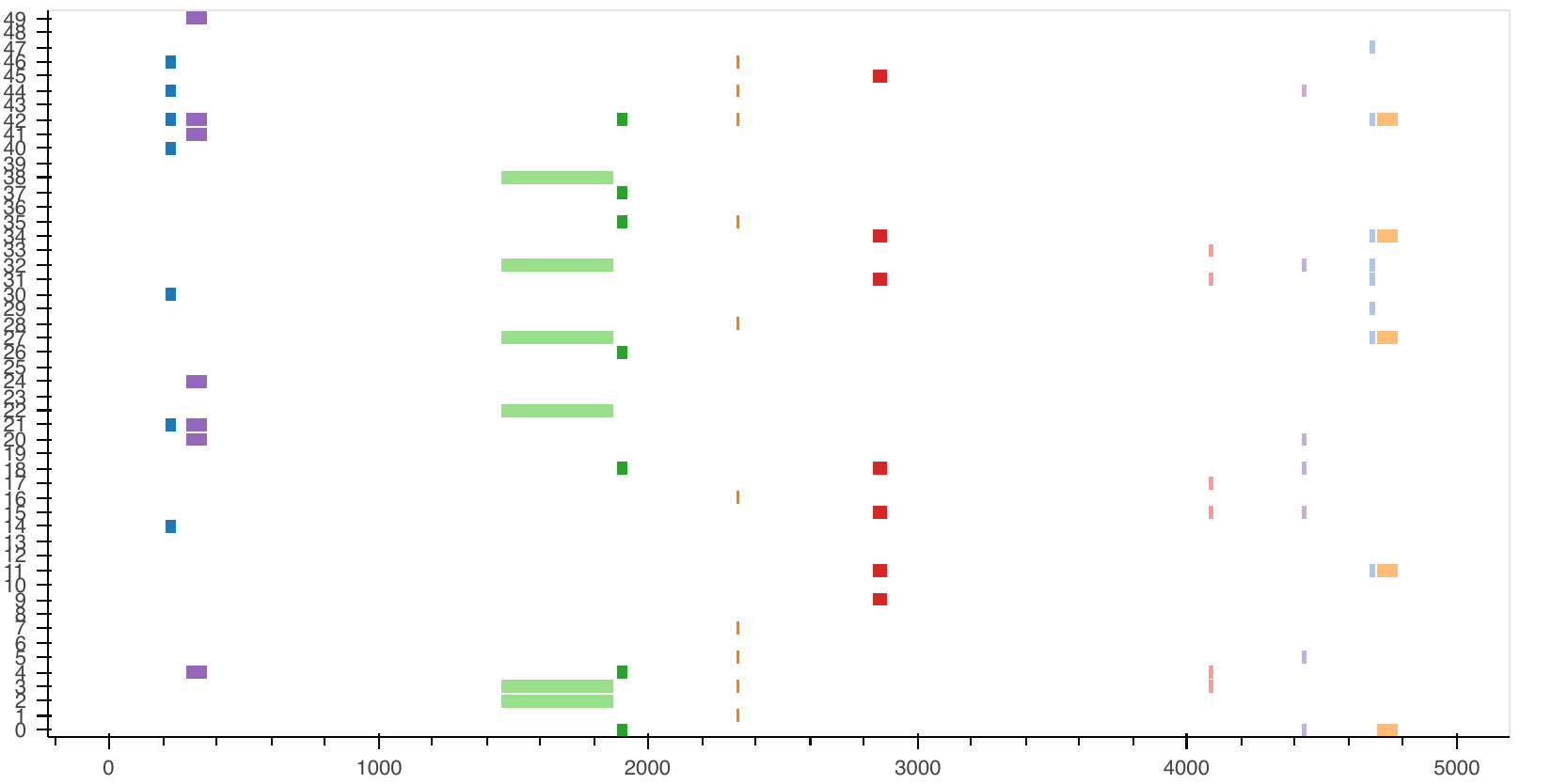}
    \caption{Stable communities produced by the generator.}
\end{subfigure}

\begin{subfigure}{0.8\textwidth}
    \includegraphics[width=\textwidth]{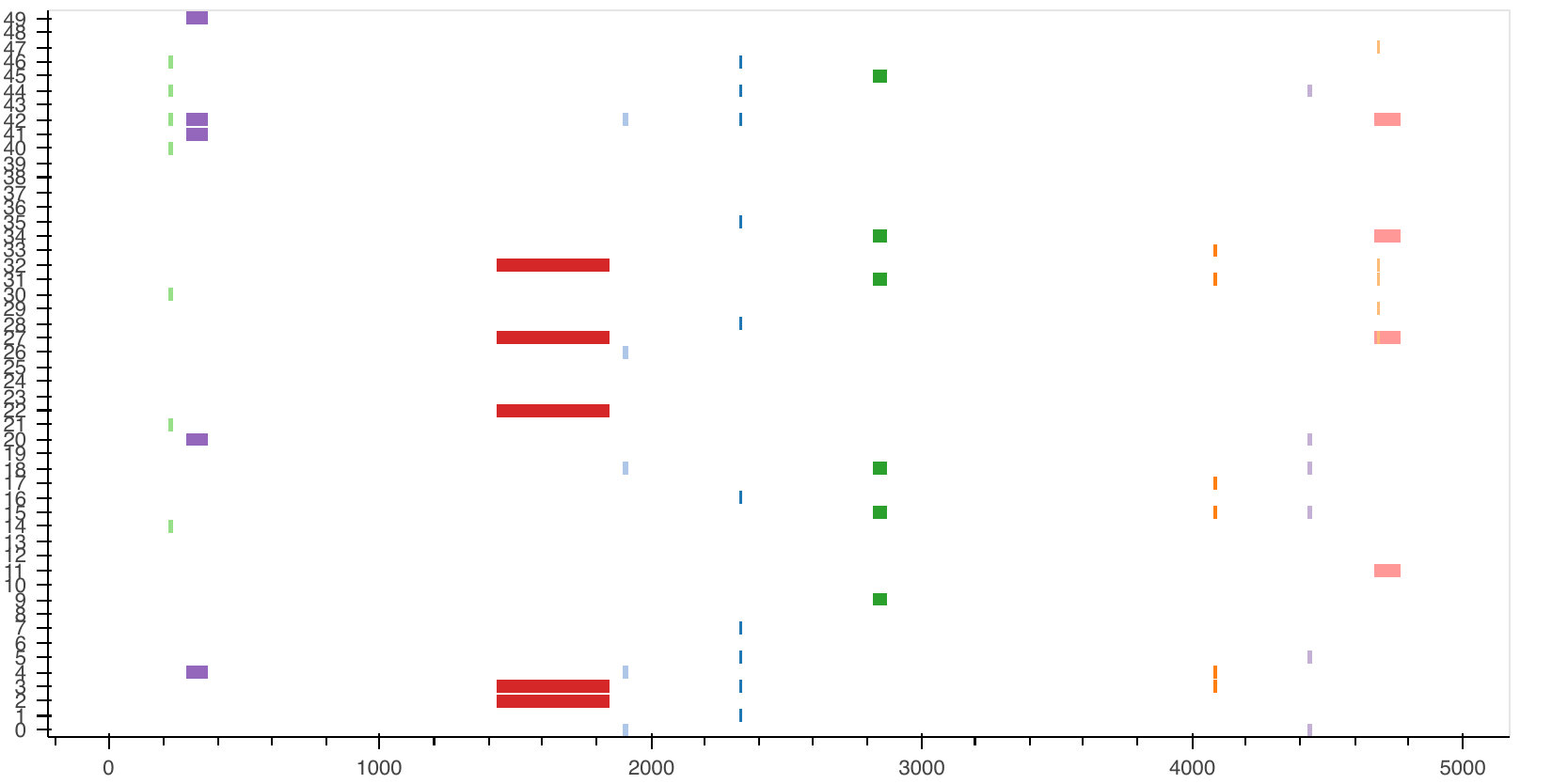}
        \caption{Stable communities discovered by the proposed method.}
\end{subfigure}

	\caption{Visual comparison between planted and discovered communities. Time steps on the horizontal axis, nodes on the vertical axis.  Colors correspond to communities and are randomly assigned. We can observe that most communities are correctly discovered, both in terms of nodes and duration.}
	\label{fig:synthetic}	
\end{figure}

\label{experiments}
The evaluation of community detection algorithms is a difficult task, and, to the best of our knowledge, we are the first to have to evaluate multi-scale dynamic community detection. 

The validation, we propose here, encompass 
three main aspects: $i)$ the validity of communities found, and $ii)$ the multi-scale aspect of our method, $iii)$ its scalability. We conduct two kinds of experiments: on synthetic data, on which we use planted \textit{ground-truth} to quantitatively compare our results, and on real networks, on which we use both qualitative and quantitative evaluation to validate our method.  

\subsection{Validation on synthetic data}
To the best of our knowledge, no existing network generator allows to generate dynamic communities at multiple temporal scales. We, therefore, introduce a simple solution to do so.
Let us consider a dynamic network composed of $T$ steps and $N$ different nodes. We start by adding some random noise: at each step, an Erdos-Renyi random graph\cite{erdos1959} is generated, with a probability of edge presence equal to $p$. We then add a number $SC$ of random stable communities. For each community, we attribute randomly a set of $n\in [4,N/4]$ nodes, a duration $d \in [10,T/4]$ and a starting date $s\in [0,T-d]$. $n$ and $d$ are chosen using a logarithmic probability, in order to increase variability. The temporal scale of the community is determined by the probability of observing an edge between any two of its nodes during the period of its existence, set as $10/d$. As a consequence, a community of duration 10 will have edges between all of its nodes at every step of its existence, while a community of length 100 will have an edge between any two of its nodes only every 10 steps on average.

Since no algorithm exists to detect communities at multiple temporal scales, we compare our solution to a baseline: communities found by a static algorithm on each window, for different window sizes. It corresponds to \textit{detect \& match}
methods for dynamic community detection such as \cite{greene2010tracking}. We then compare the results by computing the overlapping NMI as defined in \cite{lancichinetti2009detecting}, at each step. For those experiments, we set $T=5000,N=100,p=10/N$. We vary the number of communities $SC$.

\begin{figure}[!h]
\centering
\begin{subfigure}{0.7\textwidth}
    \includegraphics[width=\textwidth]{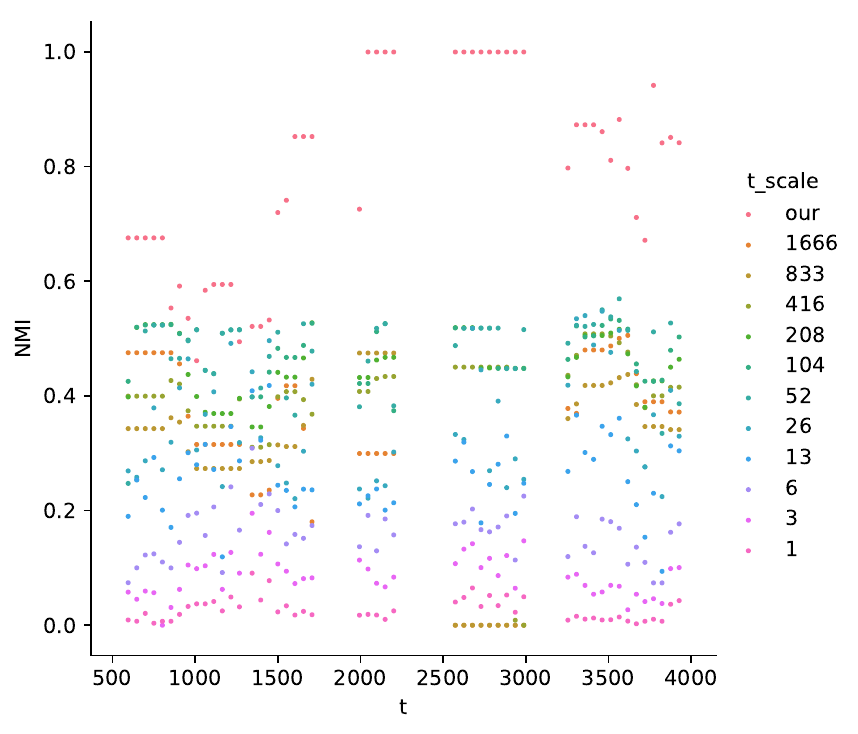}
    \caption{NMI at each step, for one run, $CD=10$}
\end{subfigure}

\begin{subtable}{0.7\textwidth}
\begin{tabular*}{\textwidth}{c @{\extracolsep{\fill}}clcccccc}
 t\_scale ($\gamma$) &    5 &   10 &   20 &   30 &   40 &   50 \\ \hline
     \textit{Proposed} & \textbf{0.91} & \textbf{0.78} & \textbf{0.69} & \textbf{0.69} & \textbf{0.62} & \textbf{0.54} \\
    1666 & 0.41 & 0.32 & 0.24 & 0.23 & 0.15 & 0.19 \\
     833 & 0.36 & 0.30 & 0.29 & 0.27 & 0.23 & 0.25 \\
    416 & 0.39 & 0.40 & 0.36 & 0.34 & 0.32 & 0.33 \\
    208 & 0.46 & 0.45 & 0.40 & 0.42 & 0.41 & 0.37 \\
     104 & 0.47 & 0.48 & 0.44 & 0.46 & 0.45 & 0.42 \\
   52 & 0.45 & 0.47 & 0.45 & 0.47 & 0.47 & 0.45 \\
    26 & 0.35 & 0.35 & 0.38 & 0.42 & 0.42 & 0.41 \\
     13 & 0.28 & 0.26 & 0.30 & 0.31 & 0.32 & 0.31 \\
     6 & 0.17 & 0.16 & 0.19 & 0.19 & 0.20 & 0.19 \\
      3 & 0.12 & 0.09 & 0.11 & 0.10 & 0.12 & 0.11 \\
      1 & 0.05 & 0.03 & 0.04 & 0.03 & 0.05 & 0.04 \\
\end{tabular*}
\caption{Average NMI. For each $\gamma$, average over 10 runs.}

\end{subtable}
	\caption{Comparison of NMI scores (over 10 runs) obtained for the proposed method (\textit{Proposed}) and for each of the temporal scales ($\gamma \in \Gamma$) used by the proposed method, taken independently.}
	\label{fig:res_synthetic}	
\end{figure}

Figure \ref{fig:synthetic} represents the synthetic communities to find for $SC=10$, and the communities discovered by the proposed method. We can observe a good match, with communities discovered throughout multiple scales (short-lasting and long-lasting ones).
We report the results of the comparison with baselines in Figure. \ref{fig:res_synthetic}. We can observe that the proposed method outperforms the baseline at every scale in all cases in terms of average NMI, but also for most timesteps in NMI computed at that step. 

The important implication is that the problem of dynamic community detection is not only a question of choosing the right scale through a window size but that if the network contains communities at multiple temporal scales, one needs to use an adapted method to discover them.

\subsection{Validation on real datasets}

\begin{figure}[!h]
\centering
    \begin{subfigure}{0.7\textwidth}
        \includegraphics[width=\textwidth]{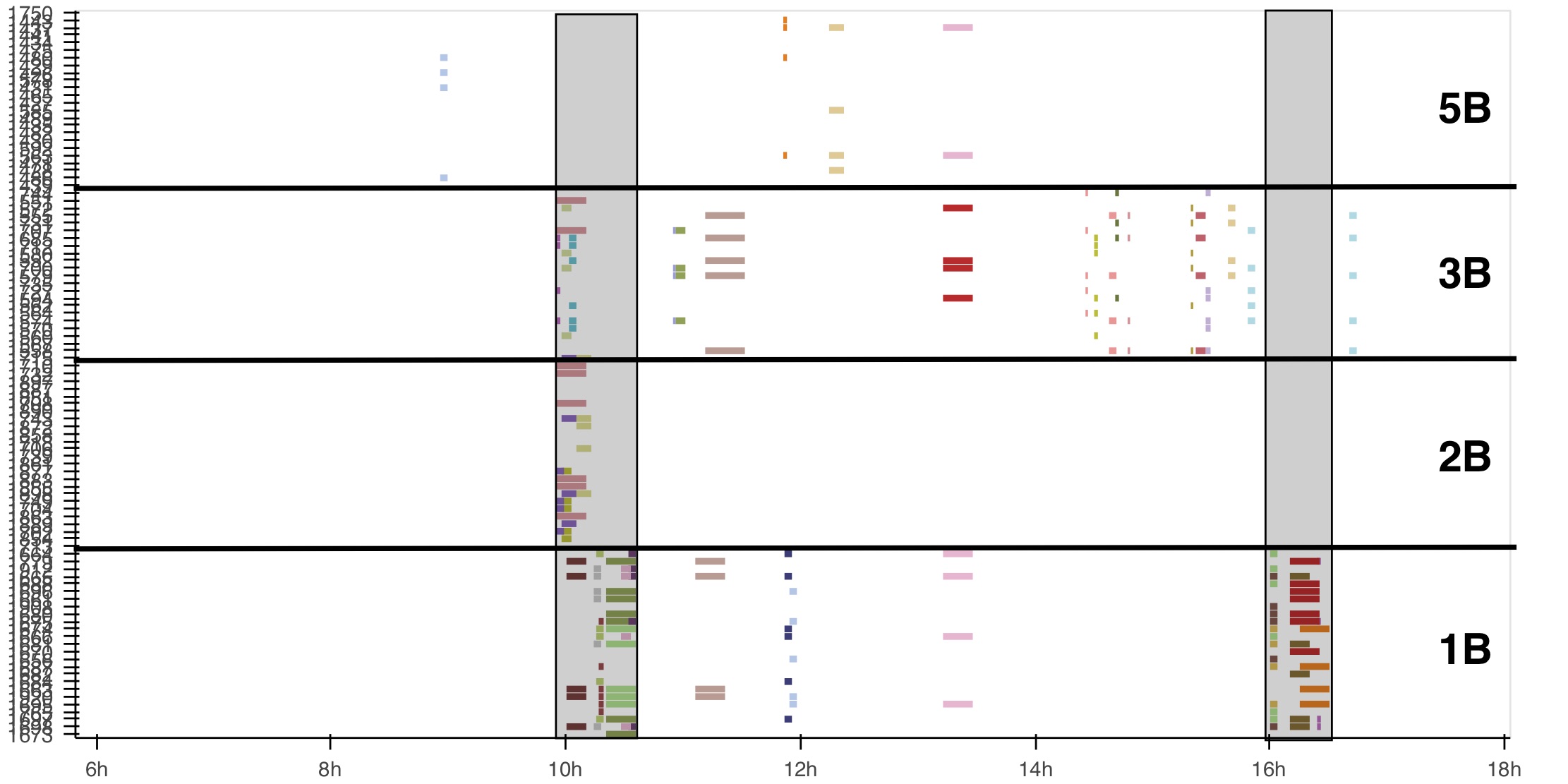}
        \caption{Second day, length$<$30min. Grey vertical areas correspond to most likely break periods.}
        \label{fig:AllComs}
    \end{subfigure}
    
    \begin{subfigure}{0.7\textwidth}
        \includegraphics[width=\textwidth]{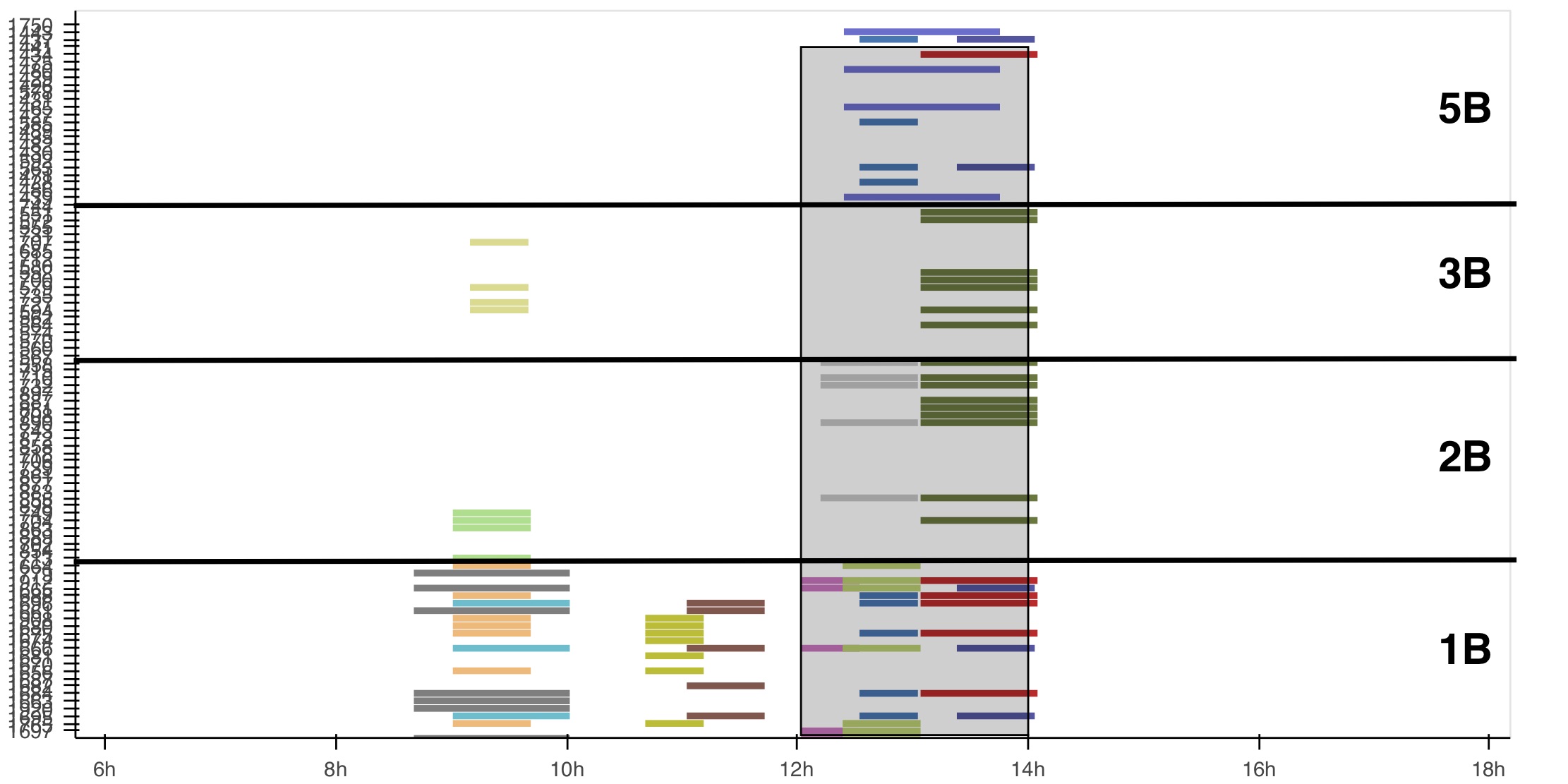}
        \caption{Second day, 30min.$>$length$>$2hours. Grey vertical area corresponds to the lunch break}
        \label{fig:TwentyComs}
    \end{subfigure}
    
    \begin{subfigure}{0.7\textwidth}
        \includegraphics[width=\textwidth]{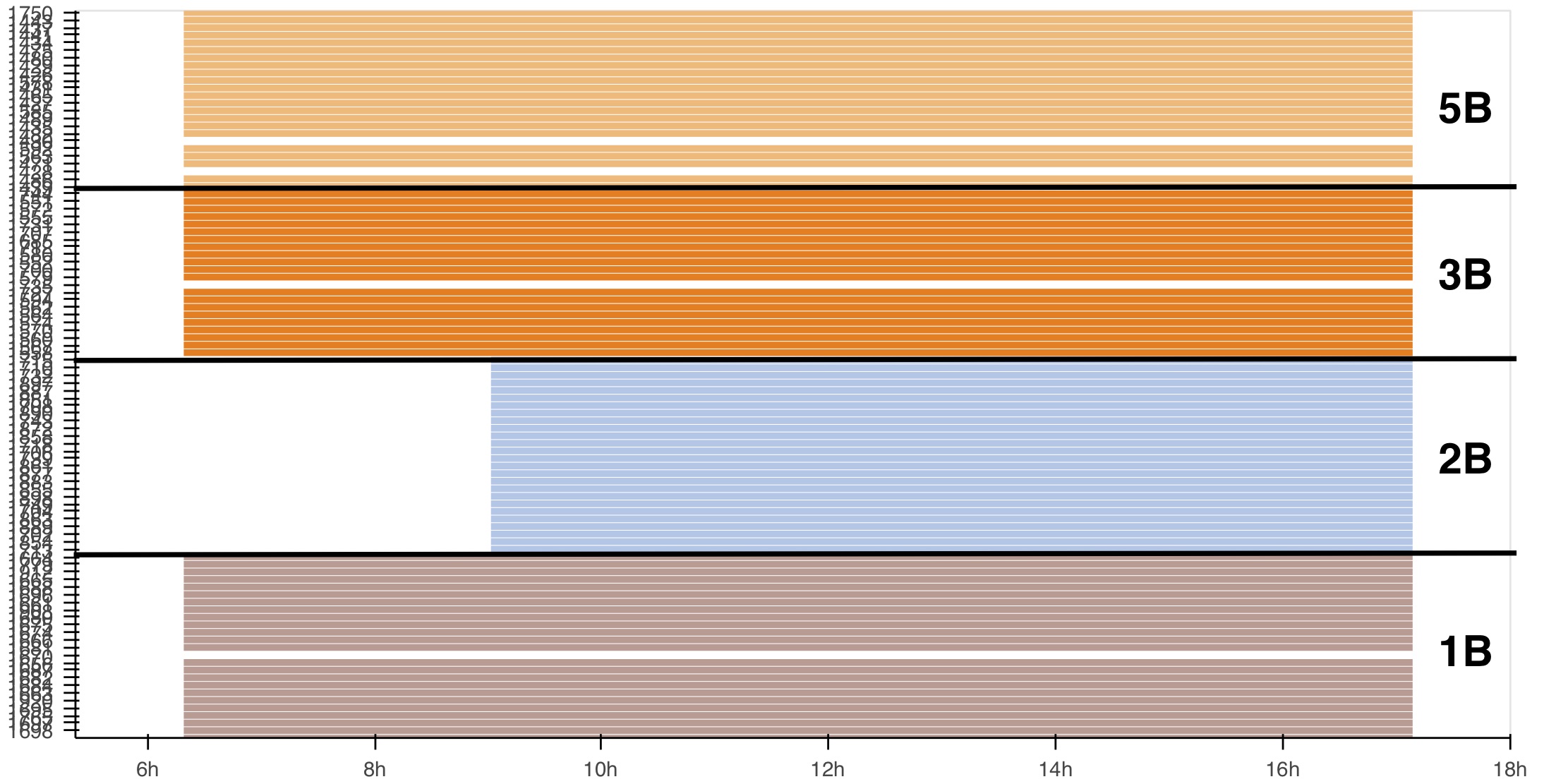}
        \caption{Second day, length$>$2hours}
        \label{fig:SecondDayComs}
    \end{subfigure}    
	\caption{Stable communities of different lengths on the SocioPatterns Primary School Dataset. Time on the horizontal axis, children on the vertical axis. Colors are attributed randomly.}
	\label{fig:PrimarySchool}	
\end{figure} 


We validate our approach by applying it to two 
real datasets. Because no ground truth data exist to compare our results with, we validate our method by using both quantitative and qualitative evaluation. We use the quantitative approach to analyze the scalability of the method and the characteristics of communities discovered compared with other existing algorithms. We use the qualitative approach to show that the communities found are meaningful and could allow an analyst to uncover interesting patterns in dynamic datasets.

The datasets used are the following:
\begin{itemize}
    \item \textbf{SocioPatterns} primary school data\cite{sociopattrens2011}, face-to-face interactions between children in a school (323 nodes, 125 773 interaction). 
    \item \textbf{Math overflow} stack exchange interaction dataset \cite{paranjape2017motifs}, a larger network to evaluate scalability (24 818 nodes, 506 550 interactions).
\end{itemize}

\subsubsection{Qualitative evaluation}
For the qualitative evaluation, we used the primary school data\cite{sociopattrens2011} collected by the SocioPatterns collaboration \footnote{www.sociopatterns.org} using RFID devices. They capture face-to-face proximity of individuals wearing them, at a rate of one capture every 20 seconds. The dataset contains face-to-face interactions between 323 children and 10 teachers collected over two consecutive days in October 2009 days (Thursday, October 1st and Friday, October 2nd, 2009) from 8.45 am to 5.20 pm on the first day, and from 8.30 am to 5.05 pm on the second day. 
This school has 5 levels, each level is divided into 2 classes(A and B), for a total of 10 classes.

No community ground truth data exists to validate quantitatively our findings. We, therefore, focus on the descriptive information highlighted in the SocioPatterns study \cite{sociopattrens2011}, and we show how the results yielded by our method match the course of the day as recorded by the authors in this study.

In order to make an accurate analysis of our results, the visualization has been reduced to one day (the second day), and we limited ourselves to 4 classes (1B, 2B, 3B, 5B)
\footnote{Note that full results can be explored online using the provided notebook (see conclusion section \ref{conclusion})}. 
120 communities are discovered in total on this dataset. We created three different figures, corresponding to communities of length respectively i)less than half an hour, ii) between half an hour and 2 hours, iii) more than 2 hours. Figure \ref{fig:PrimarySchool} depicts the results. Nodes affiliations are ordered by class, as marked on the right side of the figure. The following observations can be made:
\begin{itemize}
    \item Communities having the longest period of existence clearly correspond to the class structure. Similar communities had been found by the authors of the original study using aggregated networks per day. 
    \item Most communities of the shorter duration are detected during what are probably breaks between classes. In the original study, it had been noted that break periods are marked by the highest interaction rates (measured as the density of the aggregated graphs, for 20 minutes time windows in the original study). We know from the data description that classes have 20/30 minutes breaks and that those breaks are not necessarily synchronized between classes. This is compatible with observation, in particular with communities found between 10:00 and 10:30 in the morning, and between 4:00 and 4:30 in the afternoon.
    \item Most communities of medium duration occur during the lunch break. We can also observe that most communities are separated into two intervals, 12:00-13:00 and 13:00-14:00.  This can be explained by the fact that children have a common canteen and a shared playground. As the playground and the canteen do not have enough capacity to host all the students at the same time, only two or three classes have breaks at the same time, and lunches are taken in two consecutive turns of one hour. Some children do not belong to any communities during the lunch period, which matches the information that about half of the children come back home for lunch \cite{sociopattrens2011}.
    \item During lunch breaks and class breaks, some communities involve children from different classes, see the community with dark-green color during lunchtime (medium duration figure) or the pink community around 10:00 for short communities, when classes 2B and 3B are probably in a break at the same time. This confirms that an analysis at the coarser scales only can be misleading, as it leads only to the detection of the stronger class structure, ignoring that communities exist between classes too, during shorter periods.
\end{itemize}

\subsubsection{Quantitative evaluation}
In this section, we compare our proposition with other methods on two aspects: scalability, and aggregated properties of communities found. The methods we compare ourselves to are: 
\begin{itemize}
    \item An Identify and Match framework proposed by \cite{greene2010tracking}. We implement it using the Louvain method for community detection, and the  \textit{Jaccard coefficient} to match communities, with a minimal similarity threshold of 0.7. We used a custom implementation, sharing the community detection phase with our method.
    \item The multislice method introduced by \cite{mucha2010community}. We used the authors' implementation, with interslice coupling $\omega=0.5$.
    \item The dynamic clique percolation method (DyCPM) introduced by \cite{palla2007quantifying}. We used a custom implementation, the detection in each snapshot is done using the implementation in the networkx library \cite{hagberg2008exploring}.
\end{itemize}
For the methods Identify and Match, DyCPM, and our approach, the community detection phase is performed in parallel for all snapshots. This is not possible for Mucha et al., since the method is performed on all snapshots simultaneously. On the other hand, DyCPM and Identify and Match are methods with no dynamic smoothing.

\begin{figure}[!h]
\centering
    \includegraphics[width=0.7\textwidth]{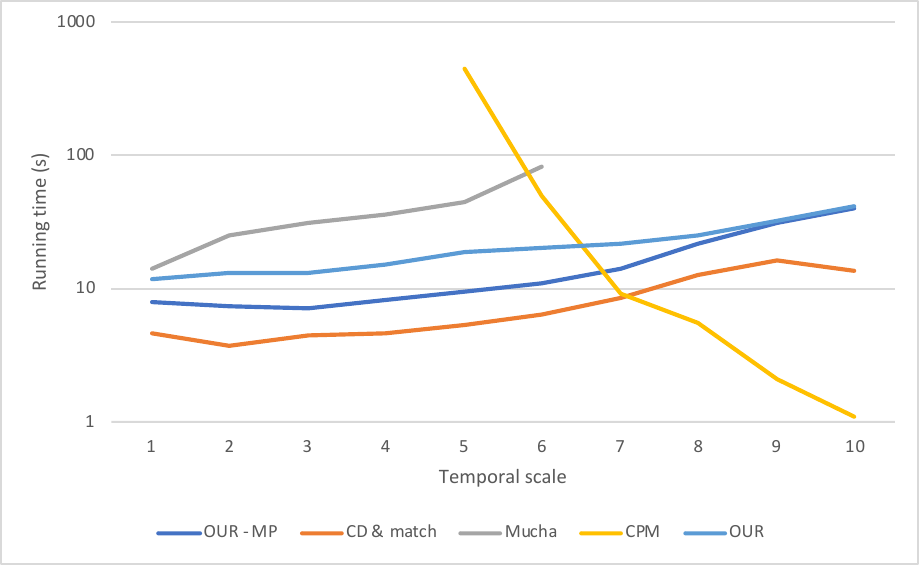}

	\caption{Speed of several dynamic community detection methods for several temporal granularities, on the Math Overflow dataset. Missing points correspond to computation time above 1000s. Temporal scales correspond to window sizes and are divided by 2 at every level, from 1=67 681 200s (about 2 years) to 10=132 189s (about 36h). OUR and OUR-MP corresponds to our method using or not multiprocessing (4 cores)}
	\label{fig:scalability}	
\end{figure}

Figure \ref{fig:scalability} presents the time taken by those methods and our proposition, for each temporal granularity, on the Math Overflow network. The task accomplished by our method is, of course, not comparable, since it must not only discover communities but also avoid redundancy between communities in different temporal scales, while other methods yield redundant communities in different levels. Nevertheless, we can observe that the method is scalable to networks with tens of thousands of nodes and hundreds of thousands of interactions. It is slower than the Identify and Match(CD\&Match) approach but does not suffer from the scalability problems as f the two other ones(DyCPM and Mucha et al.,). In particular, the clique percolation method is not scalable to large and dense networks, a known problem due to the exponential growth in the number of cliques to find. For the method by Mucha et al., the scalability issue is due to the memory representation of a single modularity matrix for all snapshots.




In Table \ref{tab:carac}, we summarize the number of communities found by each method, their persistence, size, stability, density, and conductance. It is not possible to formally rank those methods based on these values only, that correspond to vastly different scenarios. What we can observe is that existing methods yield much more communities than the method we propose, usually at the cost of lower overall quality. 
When digging into the results, it is clear that other methods yield many noisy communities, either found on a single snapshot for methods without smoothing, unstable for the smoothed Mucha method, and often with low density or Q.

\begin{table}[!h]
\resizebox{\textwidth}{!}{%
\small
\begin{tabular}{lllllll}
Method   & \#Communities & Persistance & Size  & Stability & Density & Q \\
\hline
OUR      & 179 & 3.44        & 10.89 & 1.00      & 0.50    & 0.91        \\
CD\&MATCH & 29846 & 1.21        & 5.50  & 0.97      & 0.42    & 0.96        \\
CPM     & 3259  & 1.87        & 5.37  & 0.51      & 0.01    & 0.53        \\
MUCHA   & 1097  & 15.48       & 9.72  & 0.62      & 0.38    & 0.85       
\end{tabular}}
\caption{Average properties of communities found by each method (independently of their temporal granularity). \#Communities: number of communities found. Persistence: number of consecutive snapshots. Size: number of nodes. Stability: average Jaccard coefficient between nodes of the same community in successive snapshots. Density: average degree/size-1. Q: 1-Conductance (higher is better)}
\label{tab:carac}
\end{table}

\section{Discussions}

We have shown, using real and synthetic data, the added value of our method compared with existing ones, that consider only structures at an arbitrarily chosen temporal scale. This method is, to the best of our knowledge, the first to tackle the problem of multiple temporal scales. As an exploratory work, it has limitations on which we will discuss in this section, that offer interesting extension possibilities. We will start with the computational complexity, then we discuss the difficulty in interpreting some results.

\subsection{Computational complexity}
We proposed several mechanisms to make the complexity of the method tractable: $i)$A pruning mechanism limits the number of seeds to consider, $ii)$an exponential decrease of the studied scales limits the number of temporal scales to consider. 
The main bottleneck is the large number of communities found that need to be performed at the finer temporal scales. Although each detection is fast, because graphs at fine scales are smaller and sparser compared with coarser scales, the number of steps can still be prohibitive for networks studied for a long period at a small minimal scale. 

This problem is common with all dynamic community detection methods based on snapshots. As an initial solution, we proposed to perform community detection in parallel for all snapshots. A future direction of research that could help to solve this problem is to implement targeted community detection. Since most communities found are discarded anyway, either due to the quality threshold or to the pruning process, heuristics or statistical selection procedures could be implemented to target only the time periods with the highest chances of finding new interesting seeds.

\subsection{Results interpretation}
Another limit of the method is that it yields a large number of stable communities, which can represent a challenge for interpretation. The problem is mainly due to communities found at the finest temporal scale and is even worst with other dynamic community detection methods such as \cite{greene2010tracking} that suffer from the instability problem. We nevertheless propose future directions of research that could mitigate this difficulty: 
\begin{enumerate}
\item \textbf{Stable intermittent communities}: in this work, we consider two stable communities composed of similar nodes at different non-overlapping periods as distinct: one stable community corresponds, for instance, to one particular meeting. Considering such communities as being part of the same stable intermittent community could greatly reduce the complexity of the result, at the cost of introducing approximations, as the question of \textit{nearly-similar} communities will arise.
\item \textbf{Hierarchical organization of stable communities}: another way to simplify the results would be to take into account the hierarchical relations --both temporal and structural-- between communities. For instance, a group of twenty individuals 
belonging to the same group forms a long lasting stable community, but some members of this group might be involved in a large number of shorter stable communities, e.g., subgroup meetings, lunch breaks, etc. Considering those as subgroups of the larger one in a hierarchical ordering would greatly simplify the method outcomes.
\end{enumerate}

\section{Conclusion}\label{conclusion}
To conclude, this work only scratches the surface of the possibilities of multiple-temporal-scale community detection. We have proposed the first method for the detection of such structures, that we validated on both synthetic and real-world networks, highlighting the interest of such an approach. The method is proposed as a general, extensible framework, and its code is available \footnote{The full code is available at \url{https://github.com/Yquetzal/ECML_PKDD_2019}}\footnote{An online notebook to test the method is available at \url{https://colab.research.google.com/github/Yquetzal/ECML_PKDD_2019/blob/master/simple_demo.ipynb}}as an easy to use library, for replications, applications, and extensions.
\chapter{Conclusion and Future Work}\label{chp:conclusionfuturework}
\section{Conclusion}\label{sec:conclusion}

This thesis tackles one of the most important problems in social network analysis, which is community detection. In particular, we focused on dynamic community detection which takes into account the evolutionary nature of social networks.

Despite the considerable body of the existing literature on dynamic community detection, significant research gaps remain. As  stated previously in Chapter \ref{chp3:BackgroundCD}, these research questions are related to the way we define a dynamic community and how to model the dynamic network. 
Most existing definitions of dynamic communities are stochastic and extrinsic, which, therefore, pose the so-called instability problem \cite{aynaud2010static}. Furthermore, most community definitions do not consider the overlapping property which is a natural common property in social groups. The dynamic network model has a great impact on the community detection. 
Unlike snapshot models, temporal network models are suitable to deal with highly evolving networks (like most real world networks), but their higher complexity to analyze remains problematic.
Snapshot models often require an arbitrary chosen \textit{temporal scale} of analysis which could lead to misleading results: communities of short duration are lost in the noise for large temporal scale, while communities of large duration are invisible using a fine temporal granularity. 

To address these concerns, two frameworks 
have been developed. The first framework proposed a method to detect dynamic communities, while $i)$ considering certain specificities related to the highly dynamic nature of social networks and to the overlapping property of social groups; $ii)$ avoiding the instability problem; and  $iii)$ achieving a good computational complexity. 
The second framework proposes a first method to detect stable communities through a multiple temporal scale analysis. The method is capable of discovering communities of very different lengths (duration). 
For both frameworks, we conducted experiments with both synthetic and real-world data sets to assess their applicability. 
The experiments have shown that the algorithms achieve the goals they are designed for. What follows is a quick summary of each of these contributions.



\textbf{OLCPM framework to discover overlapping and evolving communities in social networks}

In Chapter \ref{chp4:OLCPM}, we presented our framework for detecting overlapping  dynamic community structures within social networks, defined by OLCPM \cite{boudebza2018}. OLCPM is based on clique percolation and label propagation methods. At first, we proposed OCPM, a dynamic version of the clique percolation method CPM by \cite{palla2005uncovering}, intended to work on fine grained dynamic networks. The community structure is updated at each event occurring in the network, i.e., addition or removal of nodes or edges. 
Thanks to the deterministic and intrinsic nature of clique based communities, we were able to adopt a local update strategy, thus only communities whose nodes or edges are involved in the event are recomputed. This strategy was very useful, it significantly enhanced the efficiency of our method.
At second, we extended our method using label propagation method and we proposed OLCPM to deal with the limitation of CPM method concerning the covering problem, i.e., some nodes do not belong to any community. Communities found by OCPM are considered as core nodes. We proposed a label propagation  method to discover the peripheral nodes of these core-communities. We have made the discovery of the periphery of OCPM communities as a post process to $i)$ save computational time, and $ii)$ to reduce the instability problem. We conducted experiments on both synthetic and real-world social networks in order to assess the effectiveness of our framework. The results revealed a high performance of our method in terms of 
time complexity and quality detection.

\textbf{Detecting stable communities in link streams at multiple temporal scales}

In Chapter \ref{chp:Stable}, we 
proposed an original method for discovering stable communities at multiple temporal scales \cite{boudebza2019detecting}. 
From a temporal network (Link streams), we created snapshots with exponentially decreasing window sizes, what we called temporal scales. At each temporal scale (starting from the coarser), we used a three step process: $i)$ Seed Discovery: at this step we detect communities at each snapshot , then only those interesting (with high quality) are selected; $ii)$ Seed pruning: redundant communities at lower scales are pruned according to those found at larger ones; $iii)$ Seed expansion: here, we search for stable communities, i.e., we check whether a seed stays relevant in adjacent snapshots. The proposed method is general, thus, any static community detection algorithm can be used to find communities at each snapshot, any  local quality function can be used to find the best seeds in static snapshots and  any similarity measure can be used to avoid community redundancy. The validation of our method was a difficult task because we were the first to propose a multi-scale dynamic community detection.  
We conducted two sets of experiments. The first set of experiments was conducted on synthetic networks. A network generator is proposed to produce a network with dynamic communities at different scales. Then we compared communities found by our method with the planted one and also with the communities found by running a baseline dynamic algorithm \cite{greene2010tracking} at the different scales.  The results showed the efficiency of  our method in detecting stable communities at multiple scales.
The second set of experiments was conducted on real world networks.  We tested the accuracy of our method on SocioPatterns dataset, and since the absence of ground truth data, we used knowledge highlighted on their study to validate our finding.   We also tested our method on Math overflow dataset, we compared our proposition with other methods on two aspects: scalability, and aggregated properties of communities found. The results obtained were very promising.  
 
\section{Future work}\label{sec:futurework}
Encouraged by our results which indicate that the proposed approaches are appropriate to efficiently analyze the dynamics of community structures within social network, future work could be for example to improve and to extend the proposed methods or to explore other important challenges in the field. In what follows, we will briefly discuss examples for future work in the two directions.

The OLCPM  method has some drawbacks, some of which are related to CPM itself, like the dependency of the parameter $k$ (clique size). We intend to propose a heuristic for finding appropriate values of k. Currently, the post-process is run from scratch at each step, and although it is not as costly as a clique-finding problem, running it at each step for a large network can become very costly. For future research, it will be interesting to extend OLCPM by developing an online version of the post-process.

The work carried out to uncover communities at multiple temporal scales is 
novel. In this regard, further investigations and improvements are needed. We have already proposed many strategies to reduce the complexity of our method. In the pruning step, we proposed the exponential decreases of the temporal scales and the parallelization of community detection to find the initial communities. Other methods like heuristics or statistical selection procedures could be implemented to target only the periods where there is high probability to find interesting seeds and this could reduce the number of runs for community detection. 
Another limitation of our method lies in the large number of stable communities it yields, this makes interpretation of results very difficult. Hierarchical organization of relations –both temporal and structural–between communities could greatly simplify the methods outcomes.

An essential challenge in dynamic community detection concerns the characterization of complex events during the community evolution \cite{cazabet2019challenges}. As stated earlier (see Section \ref{chpt3:Communityevolution}), the community evolution is characterized in eight events: birth, death, growth, contraction, merge, split, continue and resurgence. Most existing community detection algorithms handle simple community evolution scenarios, such as : birth, shrink and  death, in snapshots network models, and they are not suitable for handling more complex scenarios in fine grained networks, like link streams models. In real scenarios, communities are susceptible to evolve gradually. For example, a shrink event can be expressed in different ways: a node moving to another community, a node leaving the network, or a newborn community with a subset of its nodes and maybe of other nodes. Therefore, a potentially interesting avenue for future research is to explore formalism to represent such complex community evolution scenarios.

Another important challenge is related to the evaluation of dynamic communities. This is still an extremely difficult task since there is no universally accepted definition of a community. The evaluation is often performed using synthetic networks with planted community structures, where the communities found by an algorithm on the generated network are compared with the ground truth communities on the same network. Since there are many well established benchmarks for evaluating static community detection methods, such are GN \cite{girvan2002community} and LFR \cite{lancichinetti2008benchmark} benchmarks, these static generators are largely adopted for the dynamic case by generating a series of static networks with planted communities, each representing a step of the network evolution. The point here, however, is that there is no possibility to assess the smoothness of communities since each network is generated randomly. Even though there have been a few benchmarks which are directly designed for the dynamic case, the evaluation of dynamic communities remains an open and complex problem.

In this thesis we have proposed methods to detect community structures while considering the temporal evolution of networks. However, many valuable network data-sets contain information about nodes and links.  The integration of the semantic dimension of a social network could provide valuable information when characterizing the community structure. Some of the existing community detection solutions use attributed graphs to represent the network semantic. Other representations, like ontologies defined within the semantic web framework \cite{boudebza2015ontology} can provide rich description of social networks, and thus may offer new opportunities for studying community detection.

\titleformat{\chapter}[display]
  {\gdef\chapterlabel{}
   \normalfont\sffamily\Huge\bfseries\scshape}
  {\gdef\chapterlabel{\thechapter\ }}{0pt}
  {\begin{tikzpicture}[remember picture,overlay]
    \node[yshift=-5cm] (0) at (current page.north west)
      {\begin{tikzpicture}[remember picture, overlay]
        \draw[fill=grisclaire] (0,0) rectangle
          (\paperwidth,5cm);
        \node[anchor=east,xshift=.9\paperwidth,rectangle,
              rounded corners=16pt,inner sep=11pt,
              fill=grisfonce,font=\large] (17) 
              {\color{white}#1};
       \end{tikzpicture}
      };
   \end{tikzpicture}
  }
\clearpage\phantomsection\addcontentsline{toc}{chapter}{Bibliography}
\bibliographystyle{apalike}
\bibliography{bibliography}
\clearpage\phantomsection\addcontentsline{toc}{chapter}{Publications}
\chapter*{Publications}
\vspace{2cm}

The work presented in this thesis has been published in two papers:

We also list two other publications that we carried out during the thesis work, but not part of the manuscript. The first one is related to social network analysis.
The second one addresses the topic of evaluation of dynamic community detection methods.

\end{document}